\documentclass[11pt]{article}

\usepackage[english]{babel}
\usepackage[T1]{fontenc}
\usepackage[utf8]{inputenc}
\usepackage[margin=1in]{geometry}
\usepackage{lmodern}
\usepackage{microtype}
\usepackage{setspace}
\usepackage{hyperref}
\usepackage{booktabs}
\usepackage{longtable}
\usepackage{tabularx}
\usepackage{array}
\usepackage{enumitem}
\usepackage{graphicx}
\usepackage{float}
\usepackage{xcolor}
\usepackage{mdframed}
\usepackage{amsmath}
\usepackage{subcaption}
\usepackage{multicol}
\usepackage{tikz}
\usetikzlibrary{arrows.meta, positioning, fit}
\definecolor{haasnavy}{RGB}{22,44,74}
\definecolor{haasblue}{RGB}{73,120,196}
\definecolor{notebg}{RGB}{246,246,246}
\definecolor{noteline}{RGB}{180,180,180}
\definecolor{haasgreen}{RGB}{69,151,115}
\definecolor{haasorange}{RGB}{219,143,54}
\definecolor{haasred}{RGB}{190,73,73}

\hypersetup{
  colorlinks=true,
  linkcolor=haasnavy,
  citecolor=haasblue,
  urlcolor=haasblue,
  pdftitle={HAAS Studio: A Tool for Simulating and Governing Human-AI Work Allocation},
  pdfauthor={},
}

\setlist[itemize]{leftmargin=1.2em,itemsep=0.1em}
\setlist[enumerate]{leftmargin=1.2em,itemsep=0.1em}

\newmdenv[
  linewidth=0.6pt, linecolor=noteline,
  backgroundcolor=notebg,
  roundcorner=3pt,
  innerleftmargin=9pt, innerrightmargin=9pt,
  innertopmargin=7pt, innerbottommargin=7pt,
]{infobox}

\newmdenv[
  linewidth=0.6pt, linecolor=noteline,
  backgroundcolor=notebg,
  roundcorner=3pt,
  innerleftmargin=9pt, innerrightmargin=9pt,
  innertopmargin=7pt, innerbottommargin=7pt,
]{examplebox}

\newmdenv[
  linewidth=0.8pt, linecolor=haasblue,
  backgroundcolor=haasblue!4,
  roundcorner=3pt,
  innerleftmargin=9pt, innerrightmargin=9pt,
  innertopmargin=7pt, innerbottommargin=7pt,
]{capturenotebox}

\DeclareRobustCommand{\ui}[1]{\texttt{#1}}

\DeclareRobustCommand{\term}[1]{\textit{#1}}
\newcommand{\sectionrule}{}
\newcolumntype{Y}{>{\raggedright\arraybackslash}X}
\newif\ifdraftnotes
\draftnotestrue

\newcommand{\screenshot}[3][0.97\linewidth]{%
  \IfFileExists{#2}{%
    \begin{figure}[htbp]
      \centering
      \fbox{\includegraphics[width=\dimexpr #1 - 2\fboxsep - 2\fboxrule\relax,
            height=0.80\textheight, keepaspectratio]{#2}}%
      \caption{#3}
    \end{figure}%
  }{%
    \ifdraftnotes
    \begin{figure}[htbp]
      \centering
      \fbox{\parbox[c][2.4cm][c]{0.7\linewidth}{%
        \centering\itshape Screenshot pending\\%
        \scriptsize see corresponding capture note%
      }}%
      \caption{#3}
    \end{figure}%
    \fi
  }%
}

\title{%
  \textbf{HAAS Studio: A Tool for Simulating, Benchmarking,\\
  and Governing Human-AI Work Allocation}%
}
\author{%
  Vicente Pelechano\\[0.3em]
  {\small Valencian Research Institute for Artificial Intelligence (VRAIN)}\\
  {\small Universitat Polit{\`e}cnica de Val{\`e}ncia}\\
  {\small \href{mailto:pele@vrain.upv.es}{pele@vrain.upv.es}}
}
\date{}

\begin{document}

\maketitle

\begin{abstract}

We present \textbf{HAAS Studio}, a simulation and decision-support tool
implementing the HAAS framework for policy-aware adaptive task allocation
between humans and AI systems~\cite{pelechano2026haaspolicyawareframeworkadaptive}.
HAAS Studio operationalises that framework as an interactive tool for
examining how work can be allocated between humans and AI systems. The tool
targets a practical question that is still poorly supported by current software:
before deploying AI in a real workflow, how can a team compare candidate
allocation strategies, inspect governance trade-offs, and derive a defensible
operating model at task level? HAAS Studio combines a five-dimensional cognitive
representation of subtasks, a five-mode collaboration spectrum, adaptive
allocation with multi-armed bandits (UCB1, Discounted-UCB, LinUCB, Thompson
Sampling), oracle counterfactual regret analysis, contract-based governance with
four independent guards, and a multi-criteria decision-support layer with a
\emph{responsible screen} that separates efficiency winners from deployable
options. The tool additionally models Human-AI co-evolution across six layers
(L1--L6), monitors deskilling risk through sliding-window exposure metrics and
dedicated benchmark runners, and supports persistent worker modelling through
\term{Live Twin} and forward-looking \term{Planning} modules. Three domain packs are included (software engineering, manufacturing, and
healthcare), each with its own task catalog, worker profiles, and KPI
vocabulary; the architecture is designed so that new domains can be added
without modifying the simulation core. The release also ships 16 company
profiles and six governance benchmark suites. This document focuses on the tool
itself: its modeling assumptions, layered architecture, interaction workflow,
and built-in evidence assets. Eleven task-oriented recipes covering the full operational workflow are included,
together with five case-study protocols; a compact demonstration snapshot on two
representative scenarios illustrates the tool's key outputs in reproducible
form. A dedicated decision-guidance layer translates benchmark outputs into
deployment decisions through structured patterns, heuristics, and a decision
matrix.

\end{abstract}


\clearpage
\tableofcontents
\clearpage

\vspace{0.5em}
\sectionrule

\section{Introduction}
\label{sec:intro}

Organizations are no longer asking whether AI should be used at work, but
\emph{where}, \emph{how much}, and \emph{under which constraints}. Those choices
are not binary. A team may want AI autonomy for repetitive tasks, human control
for ambiguous tasks, and shared execution for safety-critical tasks. The main
problem is that such decisions are often made with fragmented evidence: one
dashboard for productivity, another for quality, and no integrated view of
human sustainability, governance risk, or deployment transitions.

Classic work on human-automation interaction already showed that automation is a
spectrum rather than a switch~\cite{parasuraman2000}, and situation-awareness
research demonstrated that inappropriate automation degrades operator
competence over time~\cite{endsley1996}. Later human-centered AI frameworks
emphasized transparency, oversight, and responsibility as first-class design
goals~\cite{shneiderman2022}, and the human-machine teaming literature has
repeatedly shown that the best allocation is neither full-human nor full-AI but
context-dependent~\cite{cummings2014}. In parallel, adaptive allocation methods
based on bandit learning provide an attractive mechanism for sequential
decision-making under uncertainty~\cite{banditsurvey}. Regulatory frameworks
such as ISO/IEC~42001~\cite{iso42001} and the NIST AI Risk Management
Framework~\cite{nistairmf} further require that AI deployments remain governed
and auditable. What is still missing in many environments is a usable tool that
brings these ideas together in one artifact: not only a simulator, but also a
workflow for configuration, comparison, recommendation, and governance review.

HAAS Studio was built to address that gap. It is a Python and Streamlit
application that implements the HAAS policy-aware framework
\cite{pelechano2026haaspolicyawareframeworkadaptive} for simulating
Human-AI work allocation at the level of subtasks, comparing strategies
under configurable business contracts, and translating benchmark traces
into deployment-oriented recommendations. The broader project
documentation currently includes a detailed system report and a manual with
step-by-step study protocols; this paper condenses those materials into a
tool-focused description of the artifact and its guided use.

The paper covers six practical areas:
\begin{itemize}
  \item it presents the conceptual foundations of the tool: the stable
        vocabulary of eleven operational terms, the five-dimensional cognitive
        subtask representation, the five-mode collaboration spectrum, the
        governed adaptation loop, and the oracle counterfactual regret
        analysis;
  \item it describes the system architecture, including the governed
        allocation loop, three domain packs (software, manufacturing,
        healthcare), the six-layer Human-AI co-evolution model, and the
        deskilling and worker sustainability monitoring mechanisms;
  \item it documents the user-facing interaction model, from the setup
        wizard and company profiles through benchmark interpretation,
        \term{Live Twin}, and \term{Planning}, including four task-oriented
        operational recipes;
  \item it provides a decision framework for converting tool outputs into
        deployment decisions, covering five recurring user intents, a
        five-step signal-to-action filter, common interpretation errors,
        and a decision matrix;
  \item it presents the built-in evidence assets: five representative
        case-study protocols and six governance benchmark suites that
        support reproducible use across domains and allocators; and
  \item it covers the operational and scope information needed to use the
        tool correctly: installation, layered architecture, a pre- and
        post-run checklist, and an explicit statement of scope limitations.
\end{itemize}

Table~\ref{tab:positioning} positions HAAS Studio relative to two common
alternatives practitioners encounter: traditional manual planning and
unconstrained simulation.

\begin{table}[H]
\centering
\small
\caption{HAAS Studio positioned against common alternatives.}
\label{tab:positioning}
\begin{tabularx}{\linewidth}{@{}>{\raggedright\arraybackslash}lYYY@{}}
\toprule
\textbf{Feature} & \textbf{Manual planning} & \textbf{Unconstrained simulation} & \textbf{HAAS Studio} \\
\midrule
Allocation granularity    & Job-level           & Configurable            & Subtask-level with cognitive profile \\
Collaboration model       & Binary (human / AI) & Binary or single-mode   & Five-mode spectrum \\
Adaptive learning         & No                  & Optional                & Bandit allocators (UCB1, LinUCB, TS, Discounted-UCB) \\
Governance constraints    & Manual policy       & None                    & Contract guards: fatigue, deskilling, trust, RLHF \\
Oracle counterfactual     & No                  & No                      & Yes, per subtask \\
Co-evolution monitoring   & No                  & No                      & Six-layer model (L1--L6) \\
Deskilling protection     & No                  & No                      & Sliding-window exposure guard \\
Persistent worker state   & External tools      & None                    & \term{Live Twin} across sessions \\
Decision recommendation   & No                  & No                      & Responsible screen and transition plan \\
\bottomrule
\end{tabularx}
\end{table}

This is intentionally a \textbf{tool-oriented paper}. Its goal is not to claim a
universal optimal allocation policy, but to show how the tool structures the
problem, what kinds of evidence it produces, and how it can support
practitioner-facing operating decisions through a guided workflow.

\section{Quick Start}
\label{sec:quickstart}

\sectionrule

For a first contact with HAAS Studio, the fastest useful path is:
\begin{enumerate}
  \item open the setup wizard and select \ui{Simulation + Benchmark};
  \item choose \ui{Software / Maintenance} or \ui{Software / Standard Sprint}
        as the scenario;
  \item select a mid-level or senior worker profile;
  \item use \ui{UCB1} as the allocator and a balanced company profile such as
        \ui{Balanced\_Wellbeing} or a quality-oriented profile such as
        \ui{Software\_Quality\_Architecture};
  \item keep a modest run length (10--12 cycles) and default exploration;
  \item set three explicit contract constraints: a quality floor, a fatigue
        cap, and a cost target;
  \item run the simulation and benchmark; then read the outputs in this order:
        \ui{Run overview}, \ui{Operational impact}, \ui{Decision support},
        \ui{History}.
\end{enumerate}

The first question to answer is not \emph{which strategy ranked first}, but
\emph{did any strategy remain deployable under the configured contract?}
If the answer is yes, continue with the task-level transition plan. If the
answer is no, use the corrective actions and watchlist views to see whether the
configuration failed because of quality, cost, fatigue, or deskilling exposure.

\begin{infobox}
\textbf{First-run recommendation.} New users should prefer
\ui{Simulation + Benchmark} over pure \ui{Simulation}. It is the shortest path
to seeing both the operational trace of one configuration and the comparative
decision layer that separates benchmark winners from deployable options.
\end{infobox}

\section{Core Concepts}
\label{sec:concepts}

\sectionrule

The interface and the analytical views repeatedly use a stable vocabulary.
Reading the rest of the document is much easier if these concepts are fixed
early:
\begin{itemize}
  \item \textbf{Subtask}: the atomic decision unit used by the simulator. Work
        is not assigned monolithically; each scenario is decomposed into
        subtasks with a cognitive profile.
  \item \textbf{Collaboration mode}: the operational relationship between
        human and AI for a subtask. HAAS Studio uses five modes:
        \ui{HUMAN\_ONLY}, \ui{COPILOT}, \ui{PEER}, \ui{SUPERVISED},
        \ui{AUTONOMOUS}.
  \item \textbf{Allocator}: the adaptive decision policy that proposes
        collaboration modes from observed outcomes. The current tool exposes
        \ui{UCB1}, \ui{Discounted-UCB}, \ui{LinUCB}, and
        \ui{Thompson Sampling}.
  \item \textbf{Contract}: the active operational constraints configured by the
        user, typically including quality floor, cost target, time or fatigue
        bounds, and any hard restrictions on task ownership.
  \item \textbf{Guardrails}: the governance layer that can demote autonomy
        when the run drifts into unacceptable fatigue, deskilling, trust, or
        Reinforcement Learning from Human Feedback (RLHF) health conditions.
  \item \textbf{Feasible set}: the subset of benchmarked strategies that satisfy
        the active contract.
  \item \textbf{Responsible screen}: a second-stage filter applied after
        feasibility. A strategy may remain feasible yet fail the responsible
        screen if it still creates excessive deskilling exposure or unsafe
        co-evolution dynamics.
  \item \textbf{Deployment decision}: the tool's operational verdict on whether
        a strategy is deployable, conditional, or non-deployable under the
        current decision frame.
  \item \textbf{Transition plan}: the task-level recommendation that says how
        specific task types should move across collaboration modes.
  \item \textbf{Oracle counterfactual}: the non-operational ceiling used for
        diagnosis. It shows what would have happened if each subtask had been
        assigned with perfect hindsight.
  \item \textbf{Live Twin}: the persistent state of one worker across sessions,
        including fatigue, skill evolution, trust, and accumulated learning
        traces.
  \item \textbf{Planning}: the forward-looking module that simulates future
        trajectories from the current persistent worker state.
\end{itemize}

\section{Tool Scope and Modeling Choices}
\label{sec:scope}

\sectionrule

\subsection{Subtasks as the unit of decision}

HAAS Studio does not allocate entire jobs monolithically. Its basic decision
unit is the \textbf{subtask}. This is important because the same workflow may
contain highly automatable activities and strongly human-centered activities.
For example, in software engineering, documentation or regression testing may be
good candidates for AI-led execution, while architecture design or incident
triage may require stronger human control. The same principle appears in
manufacturing and healthcare.

Each subtask is represented by five normalized cognitive dimensions repeatedly
used throughout the tool:
\begin{enumerate}
  \item \textbf{Repetitiveness}
  \item \textbf{Technical depth}
  \item \textbf{Creativity}
  \item \textbf{Ambiguity}
  \item \textbf{Human interaction}
\end{enumerate}

These dimensions are used to estimate the initial fit between a subtask and each
party. They are also the context features used by the contextual allocator
\ui{LinUCB}, and they provide a common language for the analytics tabs. The
default AI-affinity scoring function is:
\[
  A_{\mathrm{AI}}(s)=0.35r_s + 0.25t_s + 0.20(1-c_s) + 0.10(1-a_s) + 0.10(1-h_s)
\]
where $r,t,c,a,h$ denote the five dimensions of subtask $s$. The weights are
explicit, interpretable, and calibratable rather than hidden inside the model.

\begin{infobox}
\textbf{Design rationale.} The purpose of the cognitive model is not to provide
a universal ontology of work. Its purpose is to create a stable,
cross-domain representation that remains simple enough for users to interpret,
yet structured enough for adaptive allocation and sensitivity analysis. This
choice follows the project documentation, which consistently treats the five
dimensions as the bridge between scenario design, allocation, and explanation.
\end{infobox}

\subsection{A five-mode collaboration spectrum}

The tool also avoids a binary \emph{human vs.\ AI} view. It models five
operational collaboration modes, summarized in Table~\ref{tab:modes}.

\begin{table}[H]
\centering
\small
\caption{Canonical collaboration modes used by HAAS Studio.}
\label{tab:modes}
\begin{tabularx}{\linewidth}{@{}>{\bfseries}lccY@{}}
\toprule
Mode & Human share & AI share & Execution pattern \\
\midrule
\ui{HUMAN\_ONLY} & 1.0 & 0.0 & Human executes alone. \\
\ui{COPILOT} & $>$ AI & $<$ human & Human leads, AI assists. \\
\ui{PEER} & $\approx 0.5$ & $\approx 0.5$ & Human and AI split the work equally. \\
\ui{SUPERVISED} & $<$ AI & $>$ human & AI leads, human validates. \\
\ui{AUTONOMOUS} & 0.0 & 1.0 & AI executes alone. \\
\bottomrule
\end{tabularx}
\end{table}

This spectrum operationalizes the idea that delegation should be gradual and
governed, not all-or-nothing~\cite{parasuraman2000}. It also matters for
measurement: the same subtask executed in \ui{PEER} or \ui{SUPERVISED} mode may
have similar productivity but very different implications for fatigue, trust,
learning, and deskilling exposure.

\subsection{Governed adaptation instead of unconstrained optimization}

When HAAS Studio allocates a subtask, the decision is adaptive but not fully
free. Four learning algorithms are available in the current version:
\ui{UCB1}~\cite{auer2002}, \ui{Discounted-UCB}~\cite{garivier2011},
\ui{LinUCB}~\cite{li2010}, and \ui{Thompson Sampling}~\cite{thompson1933}.
They learn from accumulated outcomes, but their choices are constrained by a
governance layer that operates through four independent guards:
\begin{itemize}
  \item \textbf{Fatigue guard}: caps delegation when projected fatigue exceeds
        the configured threshold, preventing unsustainable workloads~\cite{wickens2021}.
  \item \textbf{Deskilling guard}: limits AI share when recent exposure is too
        high, protecting human skill maintenance over time.
  \item \textbf{Trust guard}: reduces AI autonomy after repeated quality
        failures, requiring human oversight to restore confidence~\cite{endsley1996}.
  \item \textbf{RLHF guard}: applies an autonomy cap when the AI model is in an
        unstable exploration phase (low \ui{rlhf\_health\_index}).
\end{itemize}

This is a central modeling decision of the tool. In HAAS Studio, the best
average reward is not automatically the best deployment option. The tool
separates \emph{what wins the benchmark} from \emph{what remains acceptable
under the active contract and governance constraints}~\cite{iso42001,nistairmf}.
That distinction shapes both the internals of the simulator and the structure
of the results UI.

\paragraph{Reward function.}
The quantity the allocators optimize is the per-subtask reward $R_t$, computed
by the engine's reward module. In the comprehensive four-outcome mode it is a
weighted combination of four outcome dimensions:
\[
  R_t = w_q\,q_t + w_\tau\,\tau_t + w_\gamma\,\gamma_t + w_{\!hs}\,\mathit{HS}_t
\]
where $q_t\in[0,1]$ is execution quality, $\tau_t$ is a soft time score, $\gamma_t$
is a soft cost score (both using an exponential penalty $e^{-\ln 4\cdot\text{ratio}}$
so that on-budget performance scores $0.5$), and $\mathit{HS}_t\in[0,1]$ is the
human sustainability component. Default weights are $w_q=0.30$, $w_\tau=0.20$,
$w_\gamma=0.10$, $w_{\!hs}=0.40$; they are exposed in the setup wizard and
directly calibratable through company profiles.

The sustainability term decomposes human-centered risk into five sub-signals:
\[
  \mathit{HS}_t = 1
    - \omega_f F_t
    - \omega_m M_t
    - \omega_d D_t
    - \omega_e E_t
    + \omega_r \Gamma_t
\]
where $F_t$ is the normalized fatigue increment, $M_t = r_s\cdot\lambda_s$ is a
monotony burden (repetitiveness $\times$ human share), $D_t$ is the per-subtask
deskilling exposure (defined in Section~\ref{subsec:deskilling}), $E_t$ is a
cognitive exclusion term, and $\Gamma_t$ is a fatigue-relief bonus. The
four guards described above act on this signal and on accumulated sliding-window
statistics, ensuring that high $R_t$ cannot be achieved by systematically
depleting the worker.

\subsection{Oracle counterfactual and regret}
\label{subsec:oracle}

A distinctive analytical component of HAAS Studio is the \textbf{oracle
counterfactual}. Before each real allocation decision, the engine silently
simulates two mirror executions without committing agent state: one with the
human agent and one with the AI. This produces a theoretical upper bound for
each subtask decision. The \textbf{regret} of the actual choice is then:
\[
  \text{regret}_s = \max\bigl(r_s^{H},\, r_s^{A}\bigr) - r_s^{\text{real}}
\]
where $r_s^H$ and $r_s^A$ are the estimated rewards for human and AI
execution respectively, and $r_s^{\text{real}}$ is the reward of the selected
mode. Cumulative regret $R_C = \sum_s \text{regret}_s$ tracks how much value
was left on the table by suboptimal assignments. A decreasing regret curve is
direct evidence that the allocator is learning. The normalized form
$\bar{R}_C = R_C / R_{\text{oracle}}$ enables comparison across runs of
different length. In the benchmark configurations documented in this manual,
adaptive allocators often approach $\bar{R}_C < 0.10$ by cycle 5--6, meaning
that most of the theoretical value available under the current model has
already been captured.

\begin{figure}[H]
\centering
\resizebox{\linewidth}{!}{%
\begin{tikzpicture}[
  font=\small,
  node distance=1.1cm and 1.25cm,
  box/.style={draw=haasnavy, rounded corners=2pt, thick, align=center,
    text width=3.0cm, minimum height=1.05cm, fill=haasblue!6},
  altbox/.style={draw=haasgreen, rounded corners=2pt, thick, align=center,
    text width=2.8cm, minimum height=1.05cm, fill=haasgreen!8},
  realbox/.style={draw=haasorange, rounded corners=2pt, thick, align=center,
    text width=2.8cm, minimum height=1.05cm, fill=haasorange!10},
  outbox/.style={draw=haasnavy, rounded corners=2pt, thick, align=center,
    text width=3.1cm, minimum height=1.05cm, fill=notebg},
  arrow/.style={-{Latex[length=2.2mm]}, thick, draw=haasnavy},
  dashedarrow/.style={-{Latex[length=2.2mm]}, thick, dashed, draw=haasgreen}
]
\node[box] (subtask) {Current subtask\\and cognitive profile};
\node[altbox, right=of subtask, yshift=0.85cm] (human) {Mirror execution\\human only\\$r_s^{H}$};
\node[realbox, right=of subtask] (real) {Actual allocation\\selected mode\\$r_s^{\mathrm{real}}$};
\node[altbox, right=of subtask, yshift=-0.85cm] (ai) {Mirror execution\\AI only\\$r_s^{A}$};
\node[outbox, right=1.45cm of real] (oracle) {Oracle ceiling\\$\max(r_s^{H},r_s^{A})$};
\node[outbox, right=1.25cm of oracle] (regret) {Stored trace\\$\text{regret}_s$\\and $R_C$};
\draw[dashedarrow] (subtask) -- (human);
\draw[arrow] (subtask) -- (real);
\draw[dashedarrow] (subtask) -- (ai);
\draw[dashedarrow] (human) -- (oracle);
\draw[arrow] (real) -- (oracle);
\draw[dashedarrow] (ai) -- (oracle);
\draw[arrow] (oracle) -- (regret);
\node[align=center, font=\scriptsize, text width=3.2cm, below=0.15cm of subtask]
  {Mirror executions do not commit state.};
\node[align=center, font=\scriptsize, text width=3.6cm, below=0.15cm of regret]
  {The trace feeds trends, replay, ranking, and recommendation.};
\end{tikzpicture}%
}
\caption{Oracle counterfactual mechanism. Each real allocation decision is
compared with human-only and AI-only mirror executions; the gap to the best
mirror outcome becomes per-subtask regret and accumulates into the learning
diagnostic used throughout the tool.\label{fig:oracle-counterfactual-diagram}}
\end{figure}
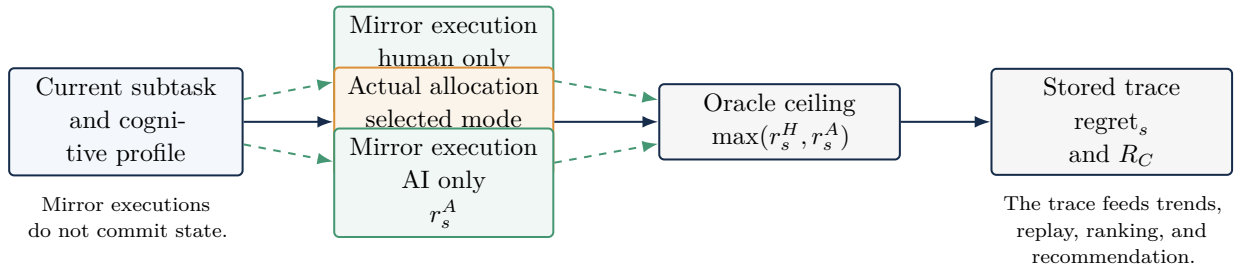

This is not a post-hoc metric. The counterfactual trace is stored per subtask
and feeds the temporal analysis surfaces, the recommendation layer, and the
local \term{replay} capability that lets users re-examine specific decisions.
The oracle also appears as an explicit benchmark baseline
(\term{Oracle counterfactual baseline}) that assigns every subtask to the
theoretically best agent, serving as a performance ceiling for strategy
comparison~\cite{cummings2014}. Figure~\ref{fig:oracle-counterfactual-diagram}
illustrates this mechanism.

\section{System Overview}
\label{sec:overview}

\sectionrule

HAAS Studio is a web application implemented in Streamlit over a layered Python
engine.

\subsection{Governed allocation loop}

The allocation loop is the operational core of HAAS Studio. Instead of making
one coarse decision for an entire scenario, the system breaks each cycle into
concrete subtasks and decides, for each one, how much work should be handled by
the human and how much by the AI. The same simple sequence is repeated for every
subtask:
\begin{enumerate}
  \item the scenario selects the next subtask and describes its cognitive
        dimensions: repetitiveness, technical depth, creativity, ambiguity, and
        human interaction;
  \item the policy layer first applies non-negotiable rules, for example tasks
        that must remain \ui{human-only} or tasks whose criticality limits AI
        autonomy;
  \item when no rule blocks the choice, the allocator recommends one
        collaboration mode: human-only, copilot, peer, supervised, or
        autonomous;
  \item the execution engine simulates what happens when the subtask is
        performed under that collaboration mode, estimating the resulting
        quality, time, cost, fatigue, trust, and other state changes;
  \item governance guards then check whether the proposed autonomy is still
        acceptable; if fatigue, low trust, or deskilling risk is too high, the
        system reduces autonomy before recording the result;
  \item the observed outcome is stored in the run trace and used as feedback for
        the allocator, so later decisions can adapt to what happened;
  \item the accumulated trace is then turned into decision-support outputs: it
        ranks allocation strategies, flags subtasks or profiles that need
        attention, shows how performance evolves over time, and supports
        transition plans for increasing or reducing autonomy.
\end{enumerate}

\begin{figure}[H]
\centering
\resizebox{\linewidth}{!}{%
\begin{tikzpicture}[
  font=\small,
  stage/.style={draw=haasnavy, rounded corners=2pt, thick, align=center,
    text width=2.75cm, minimum height=0.95cm, fill=haasblue!6},
  guard/.style={draw=haasred, rounded corners=2pt, thick, align=center,
    text width=2.75cm, minimum height=0.95cm, fill=haasred!7},
  store/.style={draw=haasgreen, rounded corners=2pt, thick, align=center,
    text width=2.75cm, minimum height=0.95cm, fill=haasgreen!8},
  arrow/.style={-{Latex[length=2.2mm]}, thick, draw=haasnavy},
  redarrow/.style={-{Latex[length=2.2mm]}, thick, draw=haasred},
  feedback/.style={-{Latex[length=2.2mm]}, thick, dashed, draw=haasnavy}
]
\node[stage] (scenario) at (0,0) {Scenario\\subtask stream};
\node[stage] (policy) at (3.25,0) {Policy rules\\hard restrictions};
\node[stage] (allocator) at (6.5,0) {Allocator\\UCB1, Disc-UCB,\\LinUCB, TS};
\node[stage] (decision) at (9.75,0) {Collaboration\\mode decision};
\node[stage] (execute) at (13.0,0) {Execution engine\\quality, cost, fatigue, trust};
\node[guard] (guards) at (9.75,-2.0) {Governance guards\\demote or cap autonomy};
\node[store] (trace) at (6.5,-2.0) {Run trace\\outcomes, regret,\\worker state};
\node[store] (decisionlayer) at (3.25,-2.0) {Decision support\\ranking, watchlist,\\transition plan};
\draw[arrow] (scenario) -- (policy);
\draw[arrow] (policy) -- (allocator);
\draw[arrow] (allocator) -- (decision);
\draw[arrow] (decision) -- (execute);
\draw[redarrow] (execute) -- (guards);
\draw[arrow] (guards) -- (trace);
\draw[arrow] (trace) -- (decisionlayer);
\draw[feedback] (trace.north) -- (allocator.south);
\end{tikzpicture}%
}
\caption{Governed allocation loop. HAAS Studio treats policy rules, adaptive
allocation, execution effects, guard interventions, and decision-support outputs
as one auditable cycle rather than as separate post-hoc analyses.\label{fig:governed-loop-diagram}}
\end{figure}
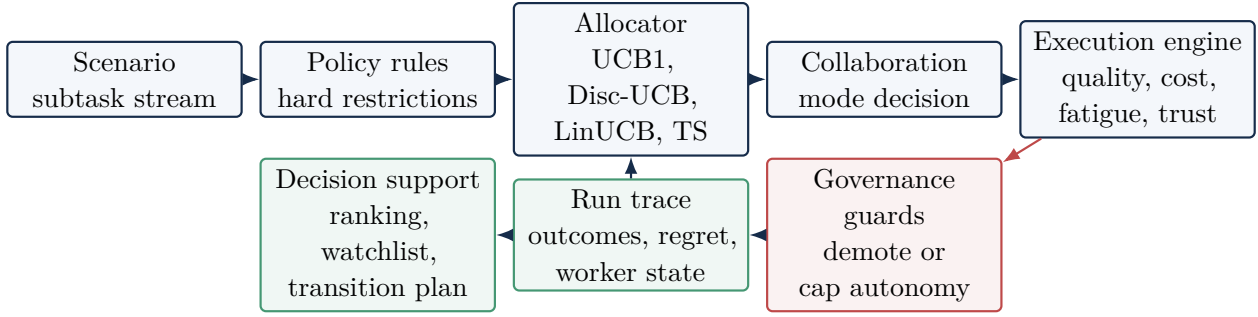

This order matters. HAAS Studio is designed so that governance is not a
post-processing patch applied after optimization. It is part of the allocation
loop itself. Figure~\ref{fig:governed-loop-diagram} summarizes the same
control flow as a closed decision cycle.

\subsection{Domains and scenarios}

The current documentation describes three implemented domains, each with its own
task catalog, worker profiles, scenarios, and KPI vocabulary. Table
\ref{tab:domains} summarizes them.

\begin{table}[H]
\centering
\small
\caption{Domain packs currently documented in HAAS Studio.}
\label{tab:domains}
\begin{tabularx}{\linewidth}{@{}>{\bfseries}lYY@{}}
\toprule
Domain & Example scenarios & Representative KPIs \\
\midrule
Software engineering &
  Standard Sprint, High Complexity, Maintenance, Deadline Crunch &
  \ui{lead\_time\_h}, \ui{defect\_escape\_rate}, \ui{cost\_per\_feature} \\
Manufacturing &
  Standard Production, Quality Crisis, Scheduled Stop, New Product Ramp-Up &
  \ui{scrap\_rate}, \ui{cost\_per\_batch}, \ui{oee} \\
Healthcare &
  Standard Screening, High-Volume Day, Complex Cases Session, Alarm Fatigue &
  \ui{diagnostic\_accuracy}, \ui{false\_positive\_rate}, \ui{alert\_fatigue\_index} \\
\bottomrule
\end{tabularx}
\end{table}

This separation is not cosmetic. It allows the same simulation core to be reused
while changing task pools, business constraints, and interpretive language. The
project documentation repeatedly emphasizes that new domains can be added by
extending the catalog layer rather than rewriting the engine.

\subsection{Human-AI co-evolution (layers L1--L6)}
\label{subsec:coevolution}

Sustained collaboration between a human and an AI system is not a static
process: both parties change. Humans learn, fatigue, and gain or lose trust;
the AI accumulates quality signal and adjusts its allocation posture. HAAS
Studio models this joint dynamic through a six-layer co-evolution
system~\cite{cummings2014,endsley1996}:

\begin{itemize}
  \item \textbf{L1 -- Observability and attribution.} After each subtask the
        engine computes an attribution confidence $\chi_t \in [0,1]$ (how much
        the outcome can be credited to the assigned agent) and a synergy score
        $\kappa_t$ (whether the joint result exceeded each agent's individual
        prediction).
  \item \textbf{L2 -- Synergy metrics.} The $\kappa_t$ history builds a
        collaborative-effectiveness indicator per task type, identifying
        human-AI pairs that generate value beyond their individual contributions.
  \item \textbf{L3 -- Internal worker model.} The bandit allocator maintains a
        per-cognitive-dimension skill model of the worker, built from observed
        outcomes. Allocation decisions thus become informed about real human
        capability, not just aggregate reward.
  \item \textbf{L4 -- RLHF propagation and AI learning phases.} Human
        corrections on AI output propagate as a lightweight RLHF
        signal~\cite{christiano2017preferences,ouyang2022rlhf}. The
        system tracks three phases (exploration, convergence, saturation) and
        exposes a \ui{rlhf\_health\_index} that indicates whether the AI is
        still improving or has plateaued.
  \item \textbf{L5 -- Comparative advantage ledger.} A per-dimension register
        $\Delta^{\mathrm{adv}}_d$ tracks where the human consistently outperforms
        the AI and vice versa, steering allocation toward modes that exploit each
        agent's strengths.
  \item \textbf{L6 -- Co-evolution guards.} The four guards described in
        Section~\ref{sec:scope} act as safety brakes against lock-in: the
        fatigue, deskilling, trust, and RLHF guards prevent the system from
        drifting into irreversible over-delegation.
\end{itemize}

\begin{figure}[H]
\centering
\resizebox{\linewidth}{!}{%
\begin{tikzpicture}[
  font=\small,
  layer/.style={draw=haasnavy, rounded corners=2pt, thick, align=left,
    text width=12.8cm, minimum height=0.82cm, fill=haasblue!5},
  guardlayer/.style={draw=haasred, rounded corners=2pt, thick, align=left,
    text width=12.8cm, minimum height=0.82cm, fill=haasred!7},
  arrow/.style={-{Latex[length=2.1mm]}, thick, draw=haasnavy}
]
\node[layer] (l1) {\textbf{L1 -- Observability and attribution:}
  attribution confidence $\chi_t$ and outcome origin are computed after each subtask.};
\node[layer, below=0.15cm of l1] (l2) {\textbf{L2 -- Synergy metrics:}
  $\kappa_t$ records whether collaboration exceeds individual predicted value.};
\node[layer, below=0.15cm of l2] (l3) {\textbf{L3 -- Internal worker model:}
  per-dimension human capability is updated from observed outcomes.};
\node[layer, below=0.15cm of l3] (l4) {\textbf{L4 -- RLHF propagation:}
  human corrections shape AI learning phase and \ui{rlhf\_health\_index}.};
\node[layer, below=0.15cm of l4] (l5) {\textbf{L5 -- Comparative advantage ledger:}
  $\Delta^{\mathrm{adv}}_d$ tracks human and AI advantage by cognitive dimension.};
\node[guardlayer, below=0.15cm of l5] (l6) {\textbf{L6 -- Co-evolution guards:}
  fatigue, deskilling, trust, and RLHF guards prevent harmful lock-in.};
\draw[arrow] (l1.east) -- ++(0.55,0) |- (l2.east);
\draw[arrow] (l2.east) -- ++(0.55,0) |- (l3.east);
\draw[arrow] (l3.east) -- ++(0.55,0) |- (l4.east);
\draw[arrow] (l4.east) -- ++(0.55,0) |- (l5.east);
\draw[arrow] (l5.east) -- ++(0.55,0) |- (l6.east);
\draw[arrow] (l6.west) -- ++(-0.65,0) |- node[pos=0.25, left, font=\scriptsize, align=center]
  {mode caps\\and recovery} (l3.west);
\end{tikzpicture}%
}
\caption{Six-layer co-evolution model. The lower guard layer can feed back into
future allocation by capping autonomy or triggering recovery actions when the
observed human-AI trajectory becomes unsafe.\label{fig:coevolution-layers}}
\end{figure}
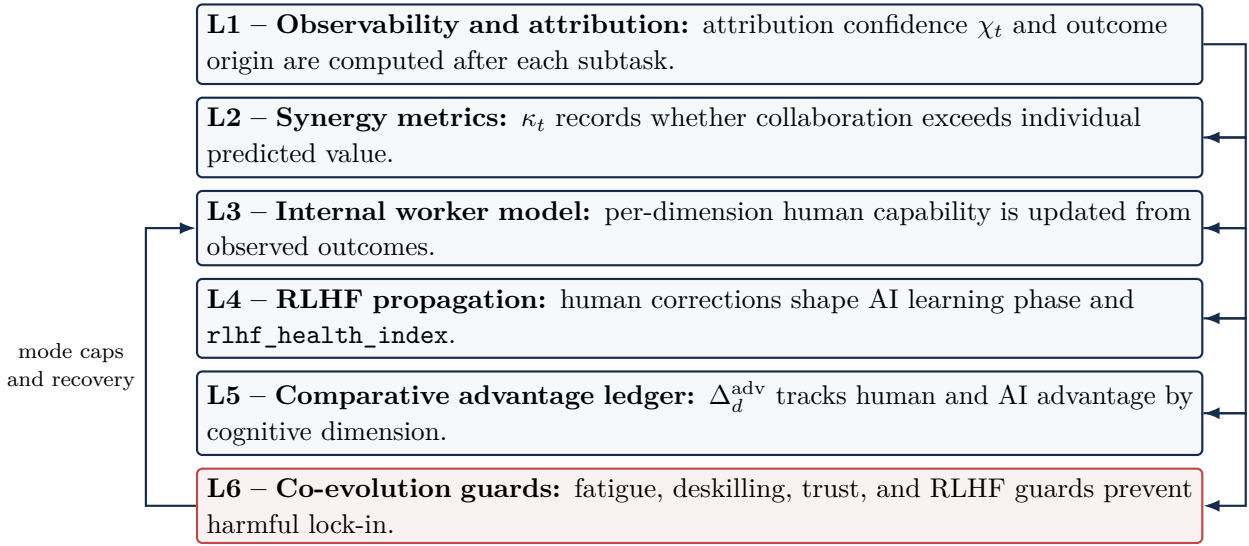

The co-evolution system encodes four design assumptions that matter for
interpreting the tool's outputs: (C1) the collaboration mode chosen today
conditions tomorrow's available capabilities; (C2) skill transfer is
dimension-specific, not generalized; (C3) allocation can improve when the
worker model (L3) is taken into account; and (C4) governance guards are needed
to reduce the risk of lock-in under repeated delegation. These assumptions are
operationalized directly in the simulation logic and in the benchmark surfaces
described throughout the paper. Figure~\ref{fig:coevolution-layers} shows
the six-layer structure.

\subsection{Deskilling and human sustainability}
\label{subsec:deskilling}

Deskilling---the progressive loss of human skills caused by excessive
delegation to AI---is one of the most documented sociotechnical risks of
automation~\cite{parasuraman2000}. HAAS Studio models and monitors it at three
levels of granularity.

\paragraph{Exposure metric.}
For each worker and cycle the engine computes
\ui{deskilling\_exposure\_recent}: the fraction of recent cycles (default
window: five sprints) in which the AI received a high share of work. Crucially,
this is disaggregated by cognitive dimension. A worker may be well-protected in
creativity but overexposed in repetitiveness if the AI systematically absorbs
routine tasks.

The per-subtask exposure signal is:
\[
  D_s = h_s \cdot \alpha_s \cdot \max\!\bigl(0,\,1 - \lambda_s\bigr)
\]
where $h_s\in[0,1]$ is the human affinity of subtask $s$ (how cognitively
suitable the task is for humans), $\alpha_s$ is the AI share, and $\lambda_s$
is the human share. The product captures the intuition that deskilling risk is
highest when a task a human should perform is instead delegated to the AI.
$D_s$ lies in $[0,1]$; typical values in software-maintenance runs with
mixed collaboration modes range from 0.50 to 0.70, reflecting the
interplay between high human affinity ($h_s \approx 0.8$) and partial AI
delegation ($\alpha_s \approx 0.6$--$0.9$).
The deployment-planning module uses $D_s$ directly to evaluate the
\emph{responsible screen}: a strategy is flagged as responsible only when
its mean $D_s$ across assigned subtasks does not exceed a configured
ceiling $\theta_{\text{resp}}$ (default~0.60).
This ceiling is set on the $[0,1]$ scale of Eq.~(5) and is therefore
independent of the per-sprint penalty rate $\rho$ used by the
sliding-window guard.

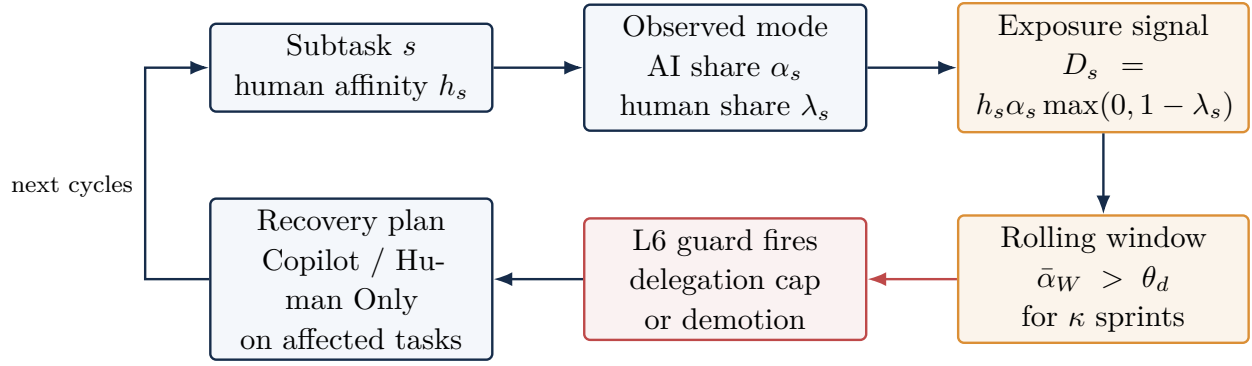
\begin{figure}[H]
\centering
\resizebox{\linewidth}{!}{%
\begin{tikzpicture}[
  font=\small,
  node distance=0.95cm and 1.05cm,
  box/.style={draw=haasnavy, rounded corners=2pt, thick, align=center,
    text width=3.05cm, minimum height=1.0cm, fill=haasblue!6},
  calc/.style={draw=haasorange, rounded corners=2pt, thick, align=center,
    text width=3.15cm, minimum height=1.0cm, fill=haasorange!10},
  action/.style={draw=haasred, rounded corners=2pt, thick, align=center,
    text width=3.05cm, minimum height=1.0cm, fill=haasred!7},
  arrow/.style={-{Latex[length=2.2mm]}, thick, draw=haasnavy},
  redarrow/.style={-{Latex[length=2.2mm]}, thick, draw=haasred}
]
\node[box] (subtask) {Subtask $s$\\human affinity $h_s$};
\node[box, right=of subtask] (share) {Observed mode\\AI share $\alpha_s$\\human share $\lambda_s$};
\node[calc, right=of share] (exposure) {Exposure signal\\$D_s=h_s\alpha_s\max(0,1-\lambda_s)$};
\node[calc, below=of exposure] (window) {Rolling window\\$\bar\alpha_W>\theta_d$\\for $\kappa$ sprints};
\node[action, left=of window] (guard) {L6 guard fires\\delegation cap\\or demotion};
\node[box, left=of guard] (recovery) {Recovery plan\\Copilot / Human Only\\on affected tasks};
\draw[arrow] (subtask) -- (share);
\draw[arrow] (share) -- (exposure);
\draw[arrow] (exposure) -- (window);
\draw[redarrow] (window) -- (guard);
\draw[arrow] (guard) -- (recovery);
\draw[arrow] (recovery.west) -- ++(-0.75,0) |- node[pos=0.22, left, font=\scriptsize, align=center]
  {next cycles} (subtask.west);
\end{tikzpicture}%
}
\caption{Deskilling guard logic. HAAS Studio converts repeated AI-heavy
allocation on human-suitable tasks into a rolling exposure signal; when the
threshold persists, the guard changes future allocation rather than merely
reporting risk.\label{fig:deskilling-guard-diagram}}
\end{figure}

The sliding-window guard monitors the $W$-sprint rolling average of the AI
ratio $\bar\alpha_W$. When $\bar\alpha_W > \theta_d = 0.80$ for at least
$\kappa = 3$ consecutive sprints, the L6 guard activates and applies a
per-sprint skill penalty:
\[
  \Delta\,\text{skill}_{t} = -\rho\,(\bar\alpha_W - \theta_d), \quad \rho = 0.003
\]
This penalty is applied per-dimension and per task type, making deskilling
traceable and recoverable through targeted mode reassignments rather than
through global rollbacks. The L6 guard applies a delegation cap when exposure exceeds the
configured threshold. Figure~\ref{fig:deskilling-guard-diagram} shows how this
calculation becomes a guard intervention rather than a descriptive warning.

These assets transform deskilling from an anecdotal concern into a measurable,
governable parameter of the deployment decision.

\section{User Workflow}
\label{sec:workflow}

\sectionrule

One of the strengths of HAAS Studio is that it does not expose the simulator as
raw parameters only. It organizes interaction as a workflow from configuration
to interpretation. The landing view organises the tool's interaction model
around a single starting surface: Figure~\ref{fig:overview} shows the workspace
entry point with its recommended workflow navigation, routing users to the setup
wizard and the analysis screens.

\screenshot[0.97\linewidth]{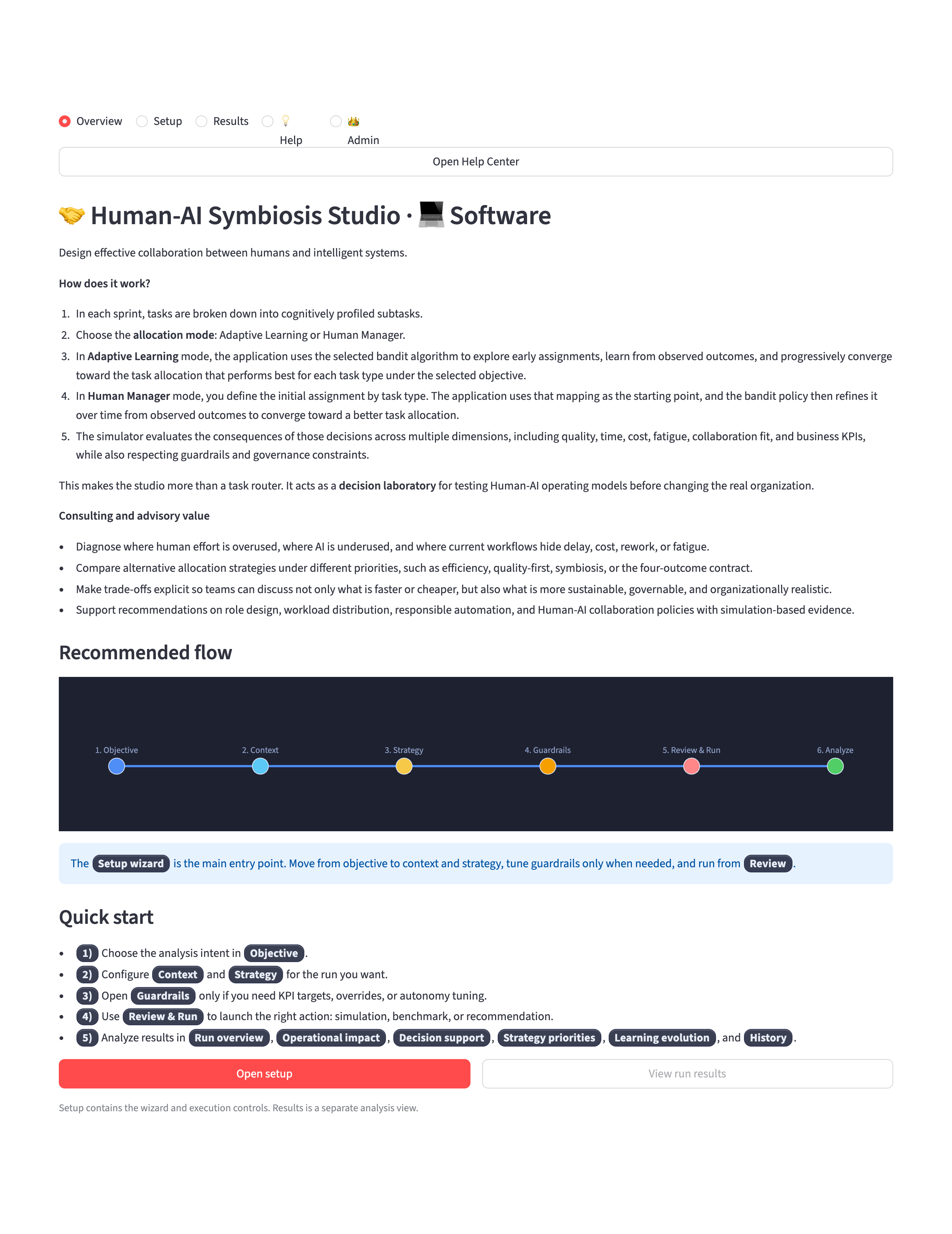}{%
Overview workspace of HAAS Studio. The landing screen introduces the tool and
its recommended workflow.\label{fig:overview}}

\subsection{Setup wizard}

Configuration is performed in a five-step wizard. The steps reflect the main
questions a user must answer before running a scenario. Figure~\ref{fig:setup-a}
shows the wizard with its five-step progression visible in the top navigation
bar, illustrating how the tool structures configuration as an ordered evidence
protocol rather than a flat parameter form.

\screenshot[0.97\linewidth]{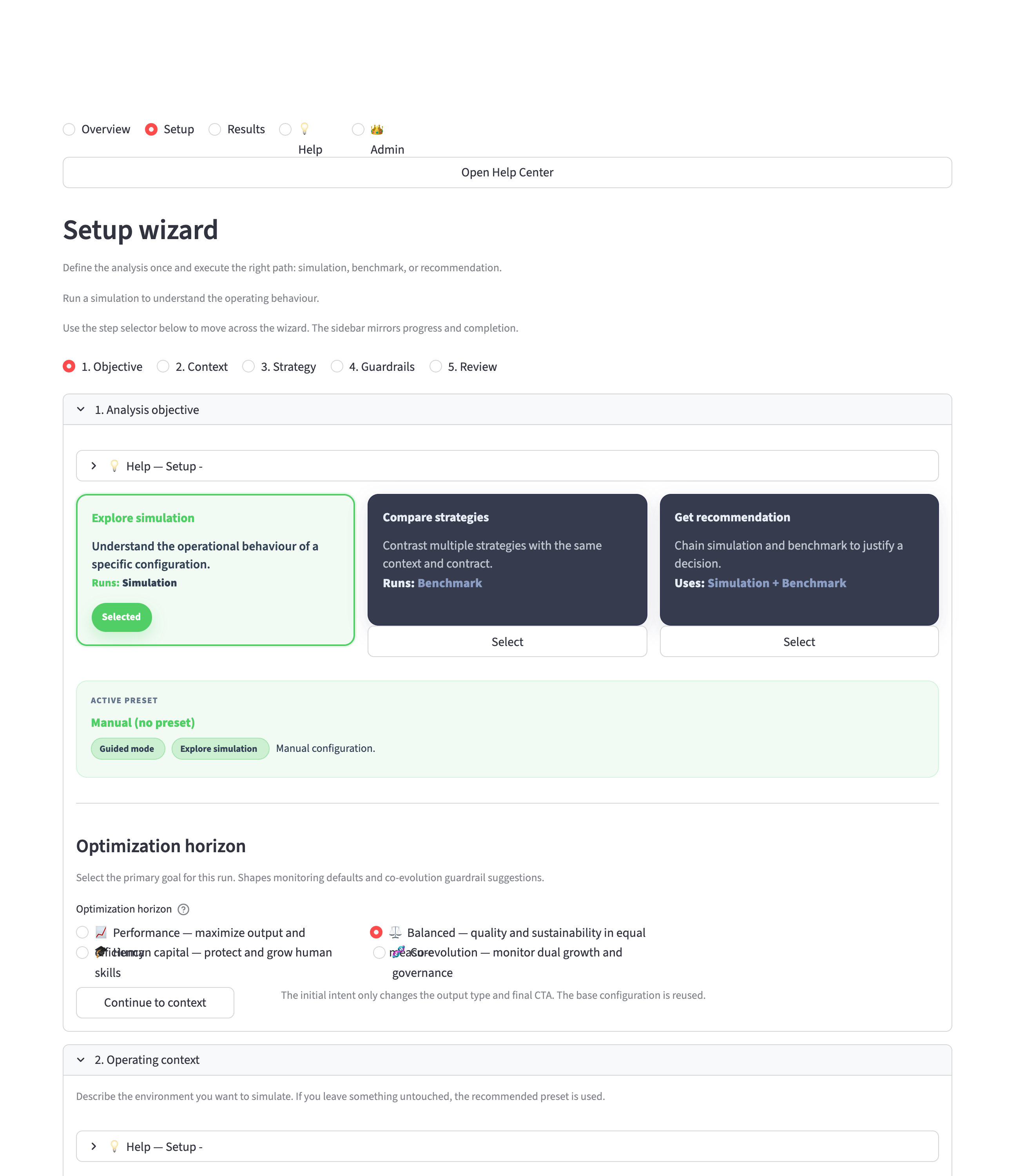}{%
Setup wizard of HAAS Studio --- the five-step progression is visible in the
navigation bar.\label{fig:setup-a}}

\begin{table}[H]
\centering
\small
\caption{Main steps of the setup workflow.}
\label{tab:wizard}
\begin{tabularx}{\linewidth}{@{}>{\bfseries}lYY@{}}
\toprule
Step & What is configured & Why it matters \\
\midrule
Objective & Explore, compare, or request recommendation; operating preset. &
  Controls the intended evidence path and execution breadth. \\
Context & Domain, scenario, backend, human profile, costs, cycles. &
  Determines the task pool and evaluation setting. \\
Strategy & Allocator, reward profile, exploration, company profile. &
  Defines how allocation learns and what it optimizes. \\
Guardrails & KPI targets, quality floor, fatigue cap, rules, restrictions. &
  Defines what counts as feasible and acceptable. \\
Review & Final configuration before run or benchmark. &
  Makes the experiment auditable and reproducible. \\
\bottomrule
\end{tabularx}
\end{table}

\subsection{Recommended defaults for new users}
\label{subsec:defaults}

Users who are learning the tool for the first time should avoid configuring the
full space from scratch. A reliable starting configuration is:
\begin{itemize}
  \item \textbf{Execution mode}: \ui{Simulation + Benchmark}
  \item \textbf{Domain}: \ui{Software}
  \item \textbf{Scenario}: \ui{Maintenance} or \ui{Standard Sprint}
  \item \textbf{Worker profile}: \ui{Mid} or \ui{Senior}
  \item \textbf{Allocator}: \ui{UCB1}
  \item \textbf{Objective}: a balanced or quality-oriented company profile
  \item \textbf{Cycles}: 10--12
  \item \textbf{Contract}: one quality floor, one fatigue cap, and one cost
        target
\end{itemize}

This baseline works well because it exposes the main decision surfaces without
requiring specialized domain interpretation. It is also the shortest path to
seeing both the detailed execution trace and the benchmark-based recommendation
layer.

\begin{infobox}
\textbf{Safe default.} If you are unsure how strict the contract should be,
start with a moderate quality floor, a visible fatigue cap, and a realistic
cost target. An extremely loose contract produces weak recommendations; an
extremely strict one often makes every strategy fail, which is informative but
not ideal for a first run.
\end{infobox}

\paragraph{Operating context.}
The \emph{Context} step is where the practitioner specifies the simulation
environment: application domain, scenario, AI backend, developer profile,
labour and AI unit costs, and the number of cycles to run.
Figure~\ref{fig:setup-b} shows this step with all its configuration fields
visible---domain selector, scenario picker, backend choice, profile assignment,
cost inputs, and cycle count---demonstrating how a single screen consolidates
all variables that define the evaluation setting.
An embedded \ui{Cycle Guidance} panel translates cycle count into practical
expectations (3--5 for a quick check, 6--10 for a learning signal, 12--20 for
a stability view). An \ui{Active Company Profile} card previews the reward
weights and guardrail defaults that the profile will inject into the next
configuration step, helping users align context and strategy before proceeding.

\screenshot[0.97\linewidth]{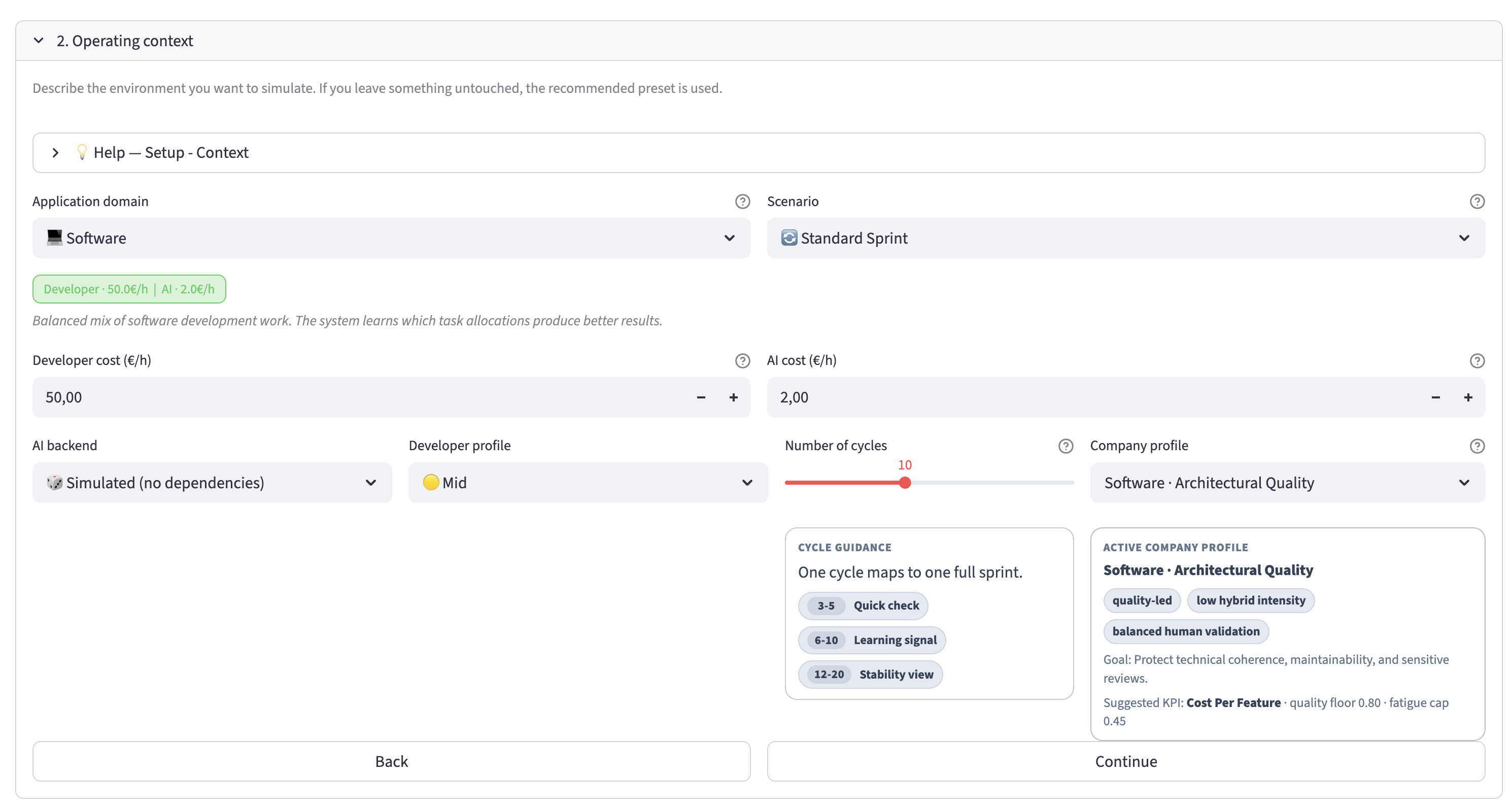}{%
Setup wizard --- Operating Context step. The practitioner selects domain,
scenario, backend, developer and company profile, costs, and cycle
count.\label{fig:setup-b}}

\paragraph{Strategy and allocator.}
The \emph{Strategy} step defines how allocation learns and what it optimizes.
Figure~\ref{fig:setup-c} shows the step with its allocation-policy selector,
optimization-objective picker, exploration-cycle guidance panel, and
adaptive-algorithm selector all visible, illustrating how each choice maps
to a distinct learning and governance posture. The user selects the initial allocation policy
(adaptive learning or manager-defined), the optimization objective (Efficiency,
Quality First, Symbiosis, or Four Outcome), and the number of initial
exploration cycles before the bandit takes over. The step also exposes the
\textbf{adaptive learning algorithm} selector: UCB1, Discounted-UCB, LinUCB,
or Thompson Sampling. Finally, it includes a \textbf{company profile} selector
--- pre-configured bundles of reward weights, guardrail thresholds, and AI
share caps that encode a specific organizational value stance. The current
release ships 16 profiles. In this paper we refer to them using English
display labels, including cross-domain profiles such as
\ui{Balanced\_Wellbeing} and \ui{Scalable\_Automation}, and domain-specific
profiles such as \ui{Software\_Quality\_Architecture} and
\ui{Manufacturing\_Safety\_Quality}. Selecting a profile automatically
populates the reward and guardrail configuration, giving practitioners a
starting point grounded in recognizable operational postures rather than raw
numeric weights.

\screenshot[0.97\linewidth]{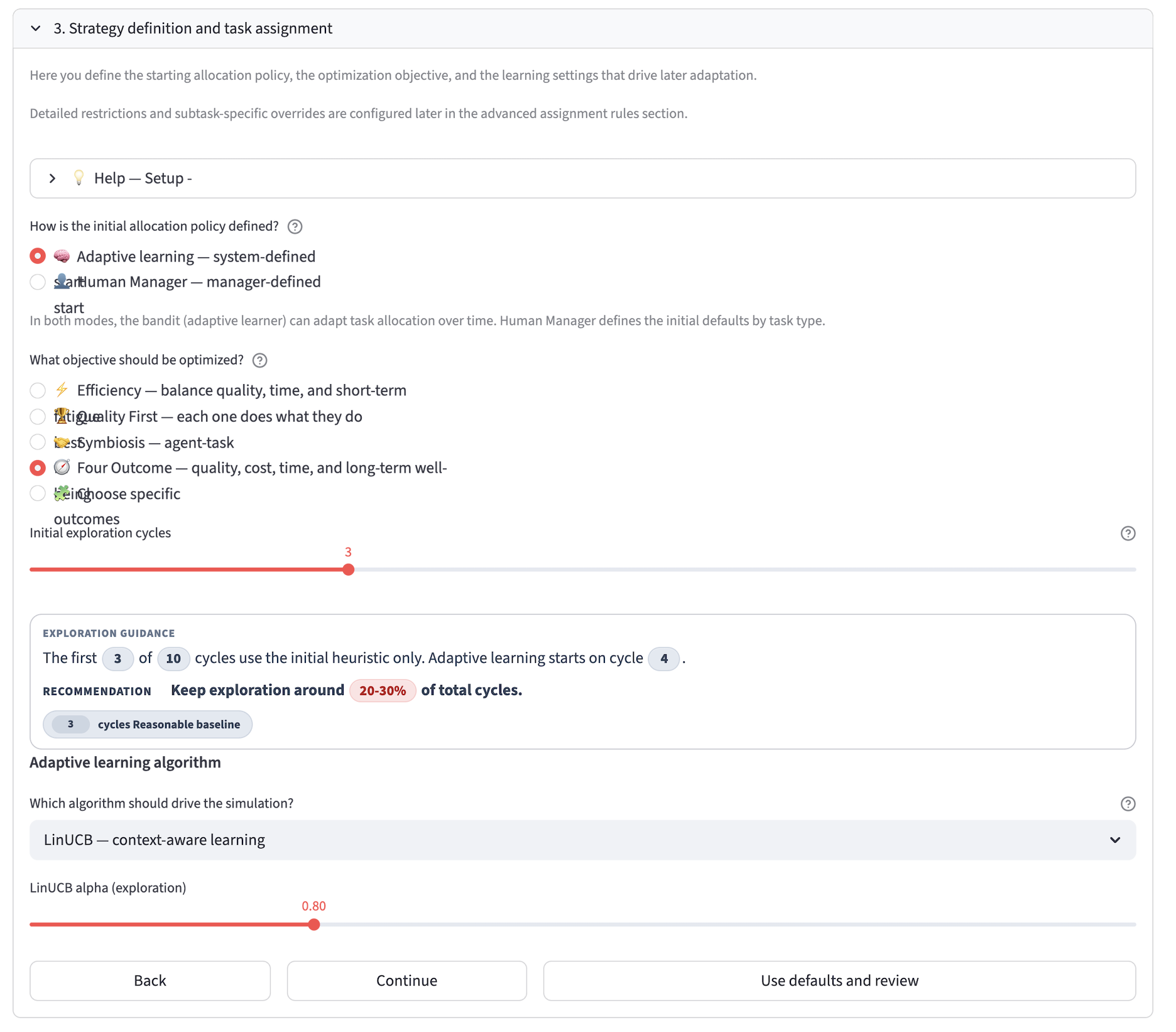}{%
Setup wizard --- Strategy step, showing allocation policy, optimization
objective, exploration cycle guidance, and allocator
selection.\label{fig:setup-c}}

The source manual is useful here because it treats the wizard not as a generic
form, but as an evidence protocol. The recommended order of configuration is
part of the methodology: define the operational question first, then the
scenario, then the learning strategy, and only then the contract and guardrails.

\subsection{How to use the wizard well}

Each setup step answers a different operational question:
\begin{itemize}
  \item \textbf{Objective}: what kind of evidence do I need?
  \item \textbf{Context}: what environment am I simulating?
  \item \textbf{Strategy}: how should the tool learn, and what should it value?
  \item \textbf{Guardrails}: what outcomes are acceptable?
  \item \textbf{Review}: is the run auditable before launch?
\end{itemize}

Common setup mistakes are predictable:
\begin{itemize}
  \item running \ui{Simulation} and then expecting benchmark-only views such as
        \ui{Decision support};
  \item leaving the contract implicit and later interpreting the ranking as a
        deployment recommendation;
  \item choosing an aggressive automation posture before seeing a balanced
        baseline;
  \item changing scenario, contract, and profile at the same time and then
        trying to attribute the outcome difference to only one factor.
\end{itemize}

\subsection{Results surfaces}

HAAS Studio offers three execution modes, each producing a different set of
analysis surfaces. Understanding which mode was used is essential for
interpreting what the workspace displays.

\begin{description}
  \item[\ui{Simulation}] Runs a single strategy for the configured number of
    cycles. Produces a detailed operational trace for that strategy.
    Available surfaces: \ui{Run overview}, \ui{Operational impact},
    \ui{Co-evolution}, \ui{Learning evolution}, \ui{History},
    \ui{Live Twin}, \ui{Planning}.
  \item[\ui{Benchmark}] Runs all active allocators across multiple random seeds.
    Produces comparative evidence across strategies under uncertainty.
    Available surfaces: \ui{Decision support}, \ui{Strategy priorities},
    \ui{History}.
  \item[\ui{Simulation + Benchmark}] Combines both: a single-strategy trace
    \emph{and} a multi-seed comparative run. All surfaces are available.
\end{description}

After execution, HAAS Studio routes the user to the analysis workspace. The
tabs that appear depend on the execution mode; surfaces requiring benchmark
data (\ui{Decision support}, \ui{Strategy priorities}) are only populated when
at least one benchmark has been run. Figure~\ref{fig:results} shows the
results workspace after a \ui{Simulation + Benchmark} run, with all nine
analysis tabs active simultaneously, illustrating the full scope of evidence
surfaces available when both execution modes are combined.

\screenshot[0.46\linewidth]{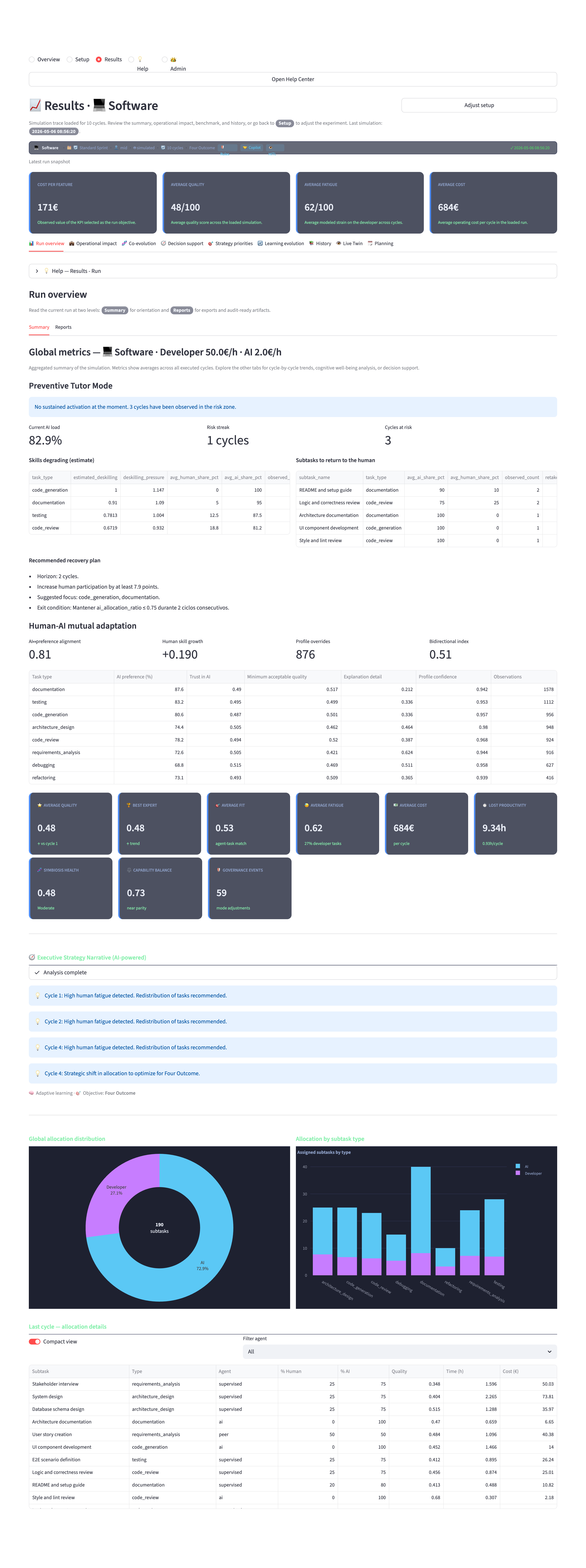}{%
Results workspace of HAAS Studio after a Simulation + Benchmark run, showing
all analysis surfaces active.\label{fig:results}}

\begin{table}[H]
\centering
\small
\caption{Analysis surfaces of HAAS Studio and their execution-mode availability.
  \textbullet{} = always available; $\circ$ = requires benchmark data;
  $\star$ = requires persistent worker (\ui{Live Twin} enabled).}
\label{tab:tabs}
\begin{tabularx}{\linewidth}{@{}>{\bfseries}lYccc@{}}
\toprule
Tab & Main content & Sim. & Bench. & Sim.+B. \\
\midrule
Run overview & Summary metrics, allocation views, last-cycle detail, reports/downloads. &
  \textbullet & & \textbullet \\
Operational impact & KPI contract, trends, well-being, corrective actions, subtask detail. &
  \textbullet & & \textbullet \\
Co-evolution & Skill trajectories, capability ledger, governance guards. &
  \textbullet & & \textbullet \\
Decision support & Ranking, feasible set, Pareto frontier, deployment view. &
  & $\circ$ & $\circ$ \\
Strategy priorities & Preference-sensitive ranking, strategy profiles. &
  & $\circ$ & $\circ$ \\
Learning evolution & Allocation evolution, learning status, cross-domain comparison. &
  \textbullet & & \textbullet \\
History & Saved runs and longitudinal comparison. &
  \textbullet & \textbullet & \textbullet \\
Live Twin & Persistent worker state, alerts, drift monitor. &
  $\star$ & & $\star$ \\
Planning & Multi-horizon projections and fatigue/trust forecasts. &
  $\star$ & & $\star$ \\
\bottomrule
\end{tabularx}
\end{table}

The tool encourages a specific reading order within each mode. For
\ui{Simulation} runs, users should first verify contract compliance in
\ui{Operational impact}, then inspect temporal trends and well-being in
\ui{Co-evolution} and \ui{Learning evolution} before drawing conclusions. For
\ui{Benchmark} runs, the recommended entry point is \ui{Decision support},
which aggregates multi-seed results into a ranked, filtered recommendation.
This ordering prevents premature conclusions from single-run metrics or from
efficiency rankings that ignore governance constraints.

\paragraph{Run overview.}
The first tab populated after any simulation run. It presents the top-level
summary of the run through aggregate KPIs (quality, cost, human/AI split,
symbiosis health), allocation views, and last-cycle detail. The tab is split into \ui{Summary} and \ui{Reports}: the first supports
quick reading of the run, while the second groups export and reporting
artifacts. It answers the question \emph{what did this run look like at a
glance?} before the user moves to the contract and trend views.

The \ui{Summary} view is already illustrated in the general results workspace
(Figure~\ref{fig:results} shows all analysis tabs active after a
\ui{Simulation + Benchmark} run); Figure~\ref{fig:r1a} focuses instead on the
\ui{Reports} subtab, which complements the summary view with export and
reporting artifacts, showing what is available for download once a run
is complete.

\screenshot[0.60\linewidth]{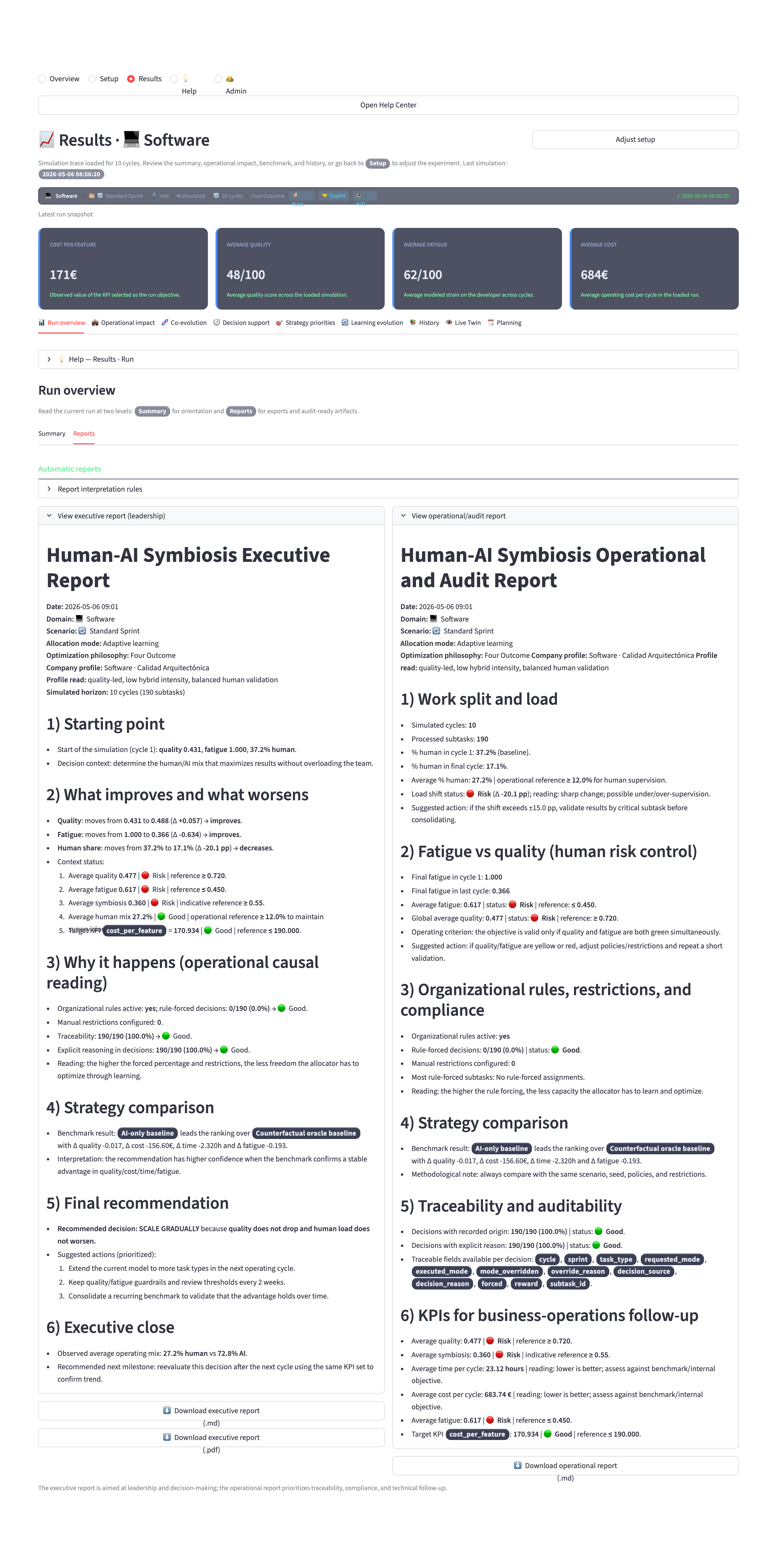}{%
\ui{Run overview} tab --- \ui{Reports} subtab after a completed simulation run,
showing export and reporting artifacts for the current run.\label{fig:r1a}}

\paragraph{Operational impact.}
The core diagnostic tab for simulation runs. It is organized into four
subtabs: \ui{KPI contract}, \ui{Trends and trade-offs}, \ui{Well-being}, and
\ui{Corrective actions}. In this paper, the tab is documented through five
captures: one for \ui{KPI contract}, one for \ui{Trends and trade-offs}, one
for \ui{Well-being}, and two consecutive captures for \ui{Corrective actions},
which is visually longer than a single page-sized screenshot. Together these
views show the contract status, cycle-by-cycle trends for quality, fit,
allocation, fatigue, cost, and counterfactual regret, the human-load and
monotony views, and the corrective recommendation workflow used when the
current run requires operational adjustment. The \ui{KPI contract} subtab also
includes a subtask-detail view for inspecting which task types drove the
observed human--AI split. The five corresponding views are documented
sequentially: Figure~\ref{fig:r1b} shows the contract compliance panel with
governance events log; Figure~\ref{fig:r1b-trends} shows cycle-by-cycle
quality, fit, allocation, fatigue, cost, and regret trajectories, revealing
when the run drifted; Figure~\ref{fig:r1b-wellbeing} shows fatigue, monotony,
repetitive exposure, and human-load detail, supporting well-being diagnosis;
Figure~\ref{fig:r1b-corrective-1} shows the redistribution recommendation
and initial-vs-recommended configuration comparison; and
Figure~\ref{fig:r1b-corrective-2} shows the cognitive-weight sensitivity
ranking and diagnostic report that guide the corrective action workflow.

\screenshot[0.97\linewidth]{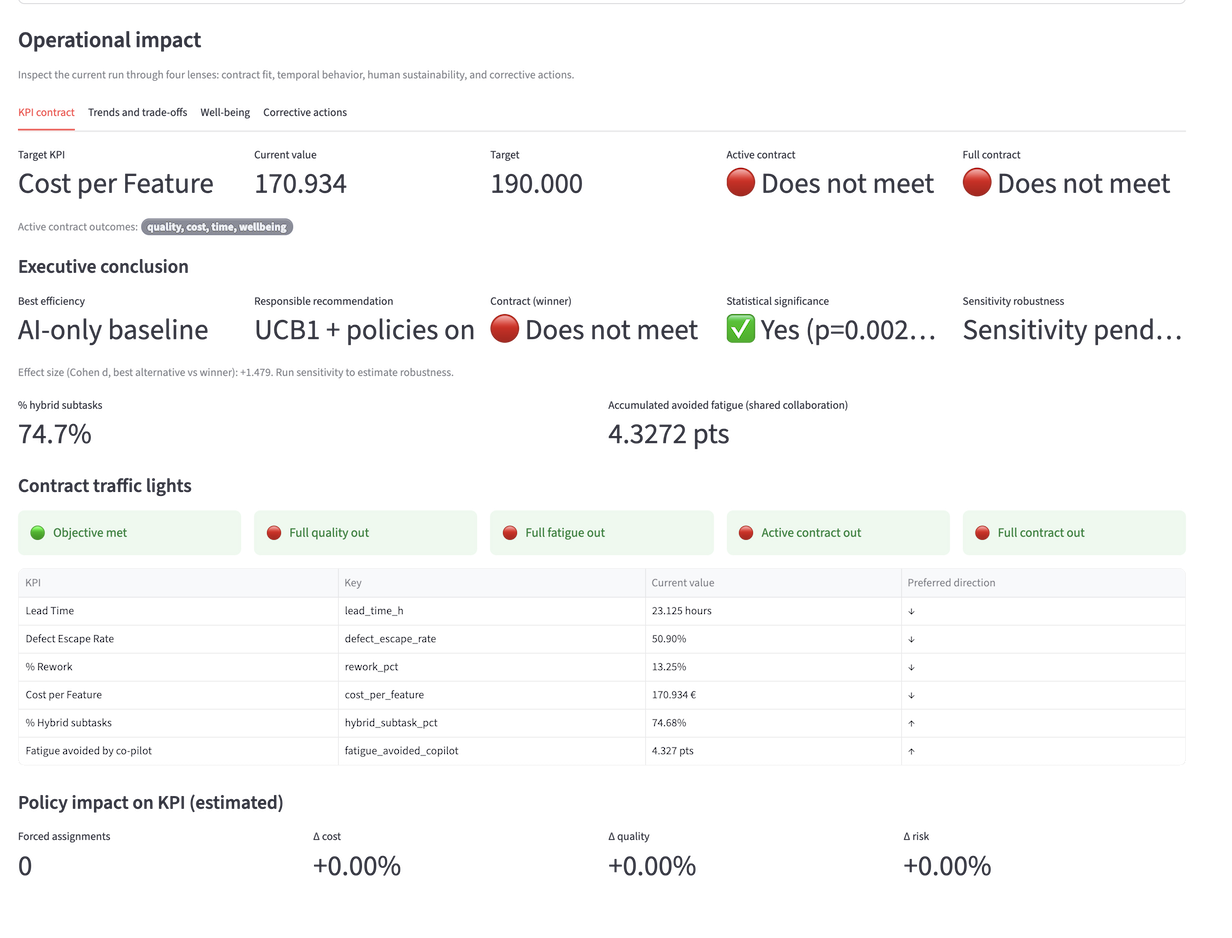}{%
\ui{Operational impact} tab --- contract compliance panel and subtask
distribution by task type, with the governance events log
visible.\label{fig:r1b}}

\screenshot[0.88\linewidth]{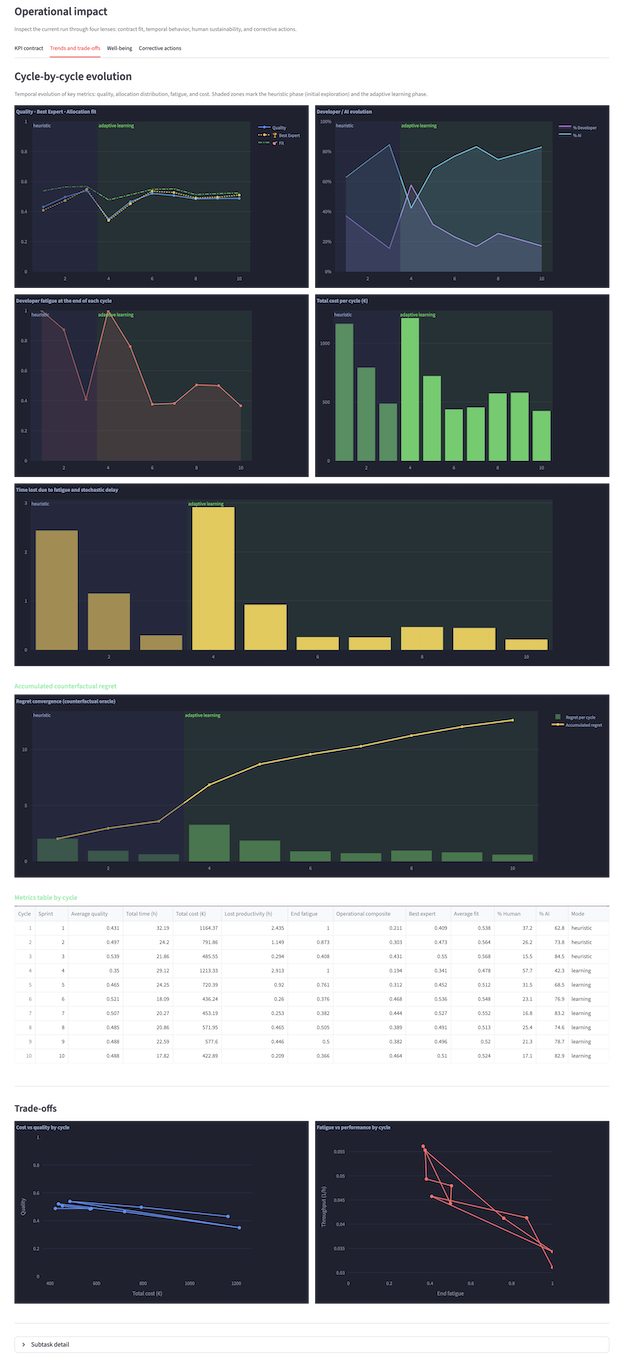}{%
\ui{Operational impact} tab --- \ui{Trends and trade-offs} subtab, showing
cycle-by-cycle quality, fit, allocation, fatigue, cost, and regret
trajectories.\label{fig:r1b-trends}}

\screenshot[0.97\linewidth]{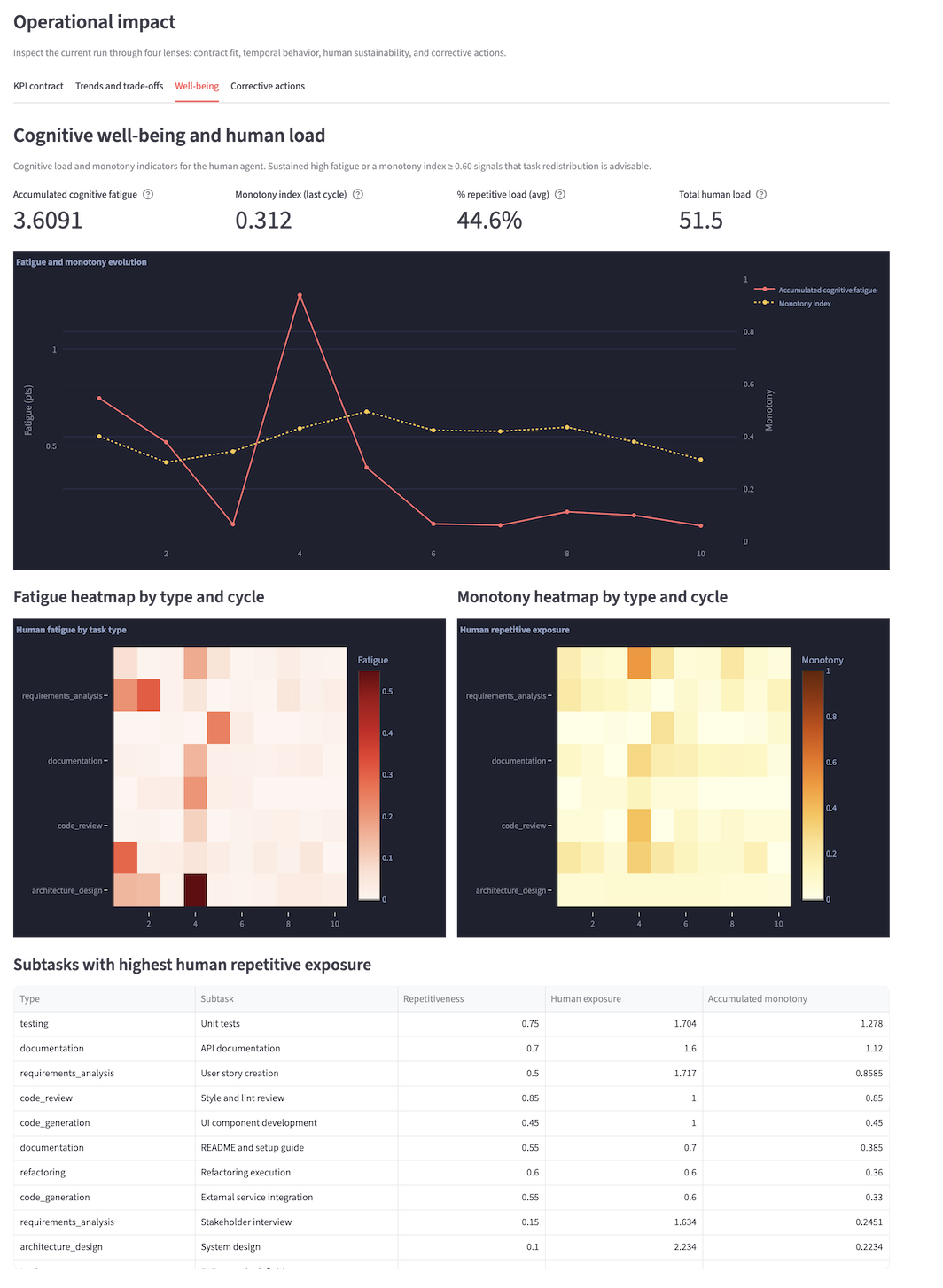}{%
\ui{Operational impact} tab --- \ui{Well-being} subtab, showing fatigue,
monotony, repetitive exposure, and the human-load detail table for the current
run.\label{fig:r1b-wellbeing}}

\screenshot[0.88\linewidth]{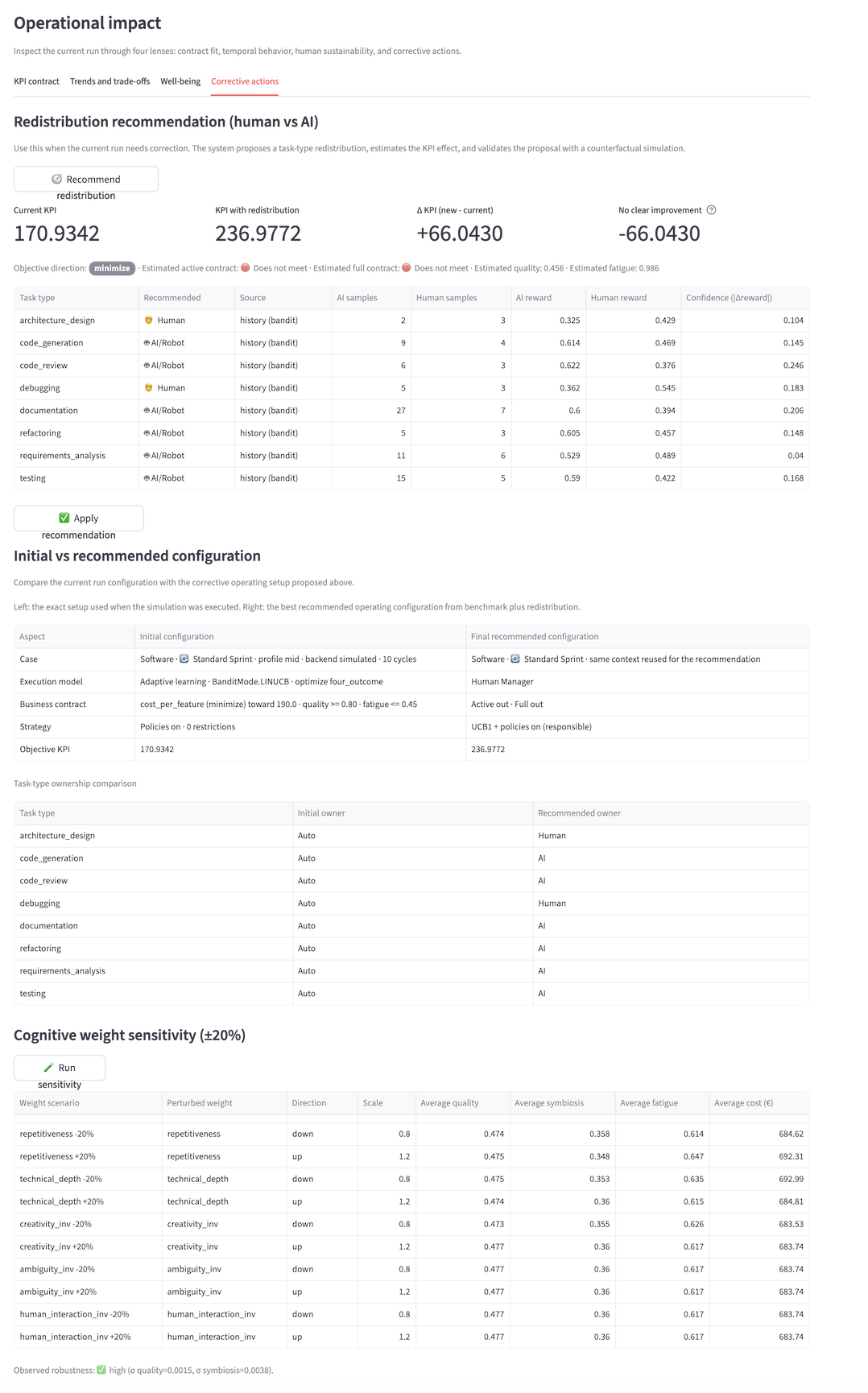}{%
\ui{Operational impact} tab --- \ui{Corrective actions} subtab (part~1),
showing the redistribution recommendation, the initial-vs-recommended
configuration comparison, and the task-ownership changes proposed for the
current run.\label{fig:r1b-corrective-1}}

\screenshot[0.97\linewidth]{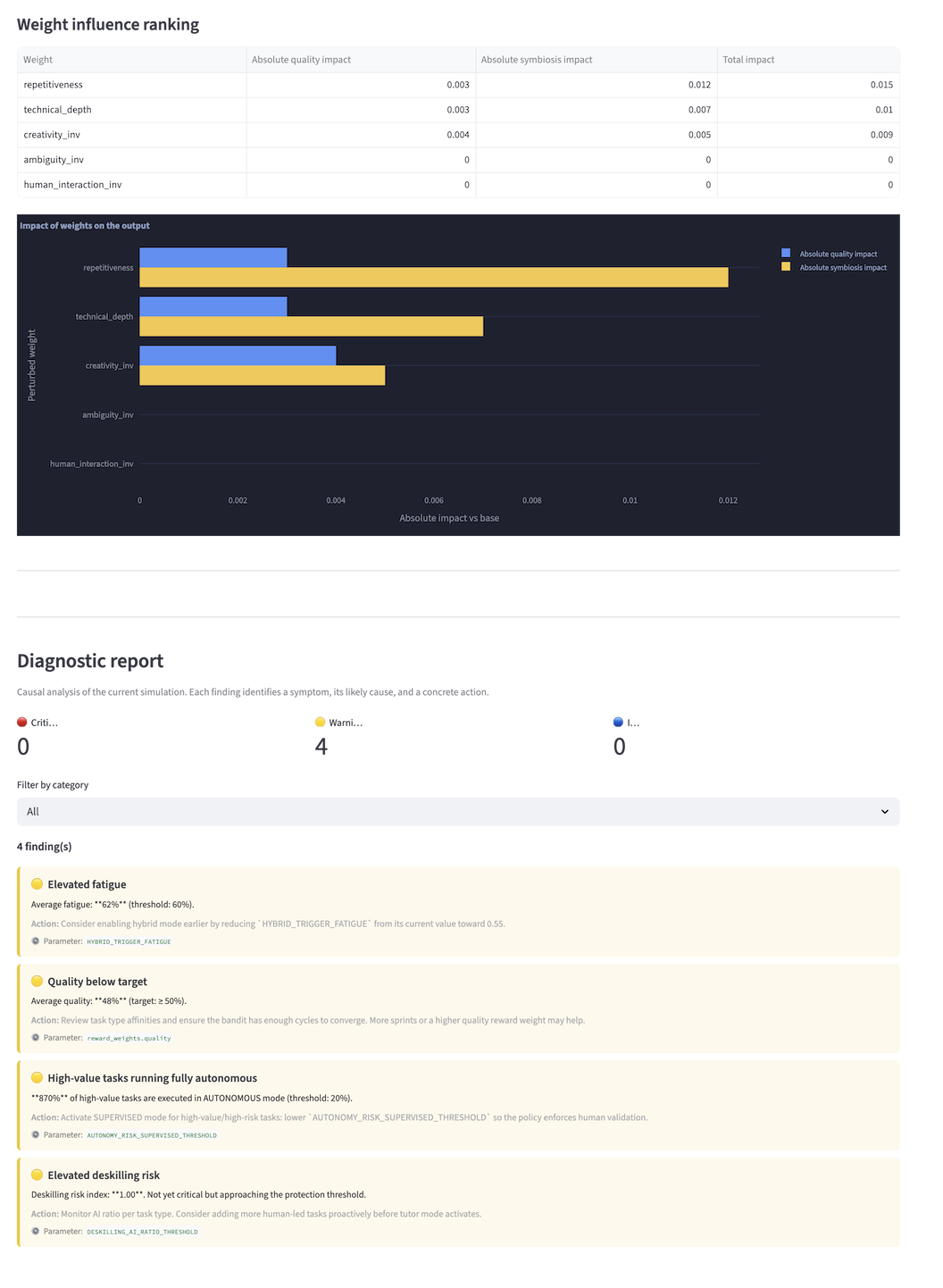}{%
\ui{Operational impact} tab --- \ui{Corrective actions} subtab (part~2),
showing the cognitive-weight sensitivity ranking and the diagnostic report
generated for the current run.\label{fig:r1b-corrective-2}}

\paragraph{Co-evolution.}
The co-evolution model is surfaced to the user through four analytical
panels. \textbf{Contribution attribution} charts the proportion of
human-led, AI-led, and collaborative outcomes across cycles, with the
synergy score~$\kappa_t$ overlaid. \textbf{Human skill dynamics} shows
per-dimension skill trajectories as a heatmap, making deskilling trends
visible at the cognitive level rather than in aggregate. \textbf{Capability
ledger} displays the $\Delta^{\mathrm{adv}}_d$ register per cognitive
dimension and flags inversion alerts when the AI has gained comparative
advantage in a dimension the human previously dominated.
\textbf{Recovery planning} translates active inversion alerts into
corrective-action suggestions, such as reassigning specific task types
back to \ui{Copilot} or \ui{Human Only} for a fixed number of cycles.

\paragraph{Learning evolution.}
The tab that makes the allocator's adaptation pattern visible. It is organized
into three subtabs: \ui{Allocation evolution}, \ui{Learning status}, and
\ui{Cross-domain comparison}. In this paper, the tab should be documented
through separate captures of these three subtabs rather than through a single
condensed view. The first compares initial or heuristic task-type allocations
with the currently learned ones. The second summarizes learned preferences,
reward convergence, changes relative to the initial heuristic, and final human
skills for the current run. The third runs the same setup across the three
domains to show how allocator behavior changes with context. Together they answer the question \emph{how did the allocator's behavior
evolve, and what kind of preference structure did it learn?}
Figure~\ref{fig:r1c-learn} shows the allocation evolution subtab for the
active allocator, making learned task-type preferences directly visible.
Figure~\ref{fig:r1c-status} shows the learning status subtab with convergence
indicators and the current allocation posture, confirming whether the bandit
has stabilised. Figure~\ref{fig:r1c-cross} shows the cross-domain comparison
subtab with the same setup running across software, manufacturing, and
healthcare, illustrating how context changes the allocator's learned
behaviour.

\screenshot[0.97\linewidth]{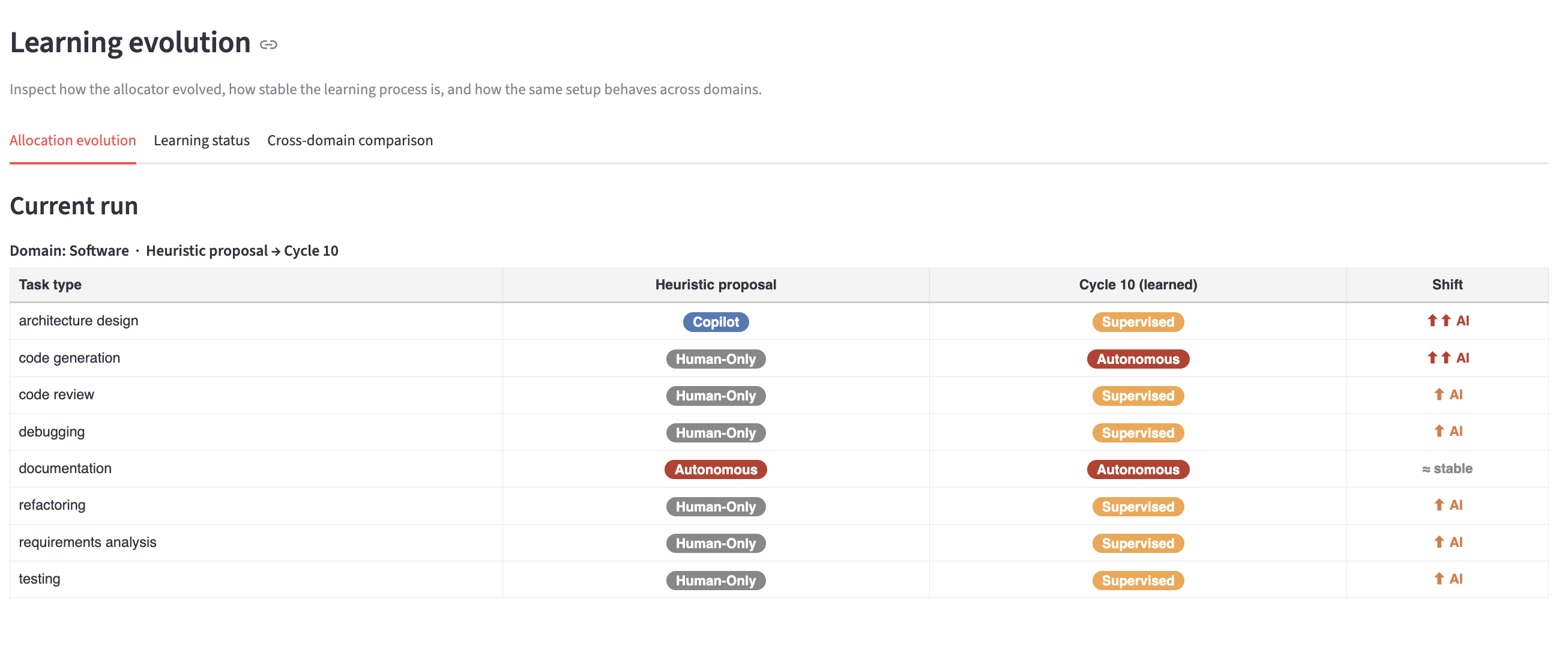}{%
\ui{Learning evolution} tab --- \ui{Allocation evolution} subtab for the
active allocator.\label{fig:r1c-learn}}

\screenshot[0.88\linewidth]{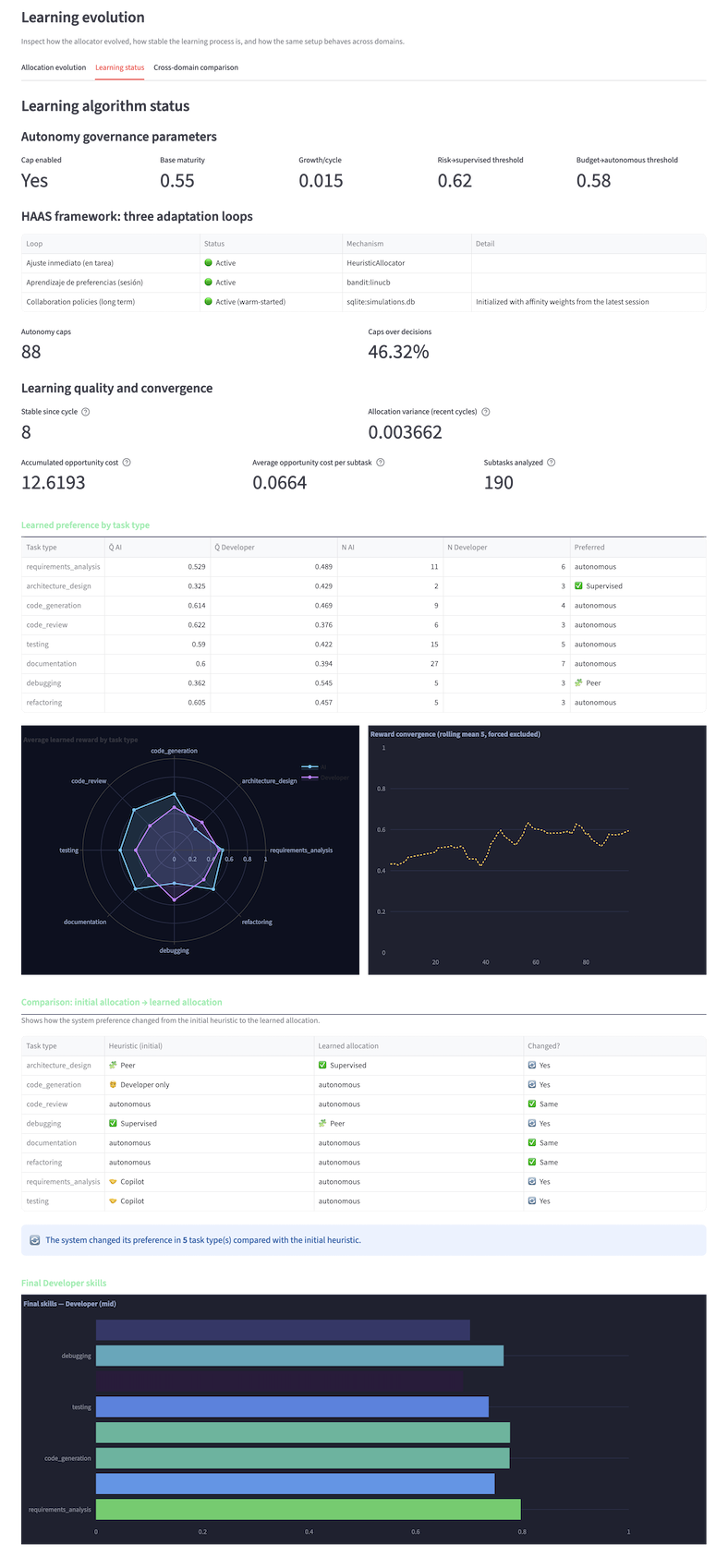}{%
\ui{Learning evolution} tab --- \ui{Learning status} subtab, summarizing the
allocator's learned preferences, convergence indicators, and current
posture.\label{fig:r1c-status}}

\screenshot[0.97\linewidth]{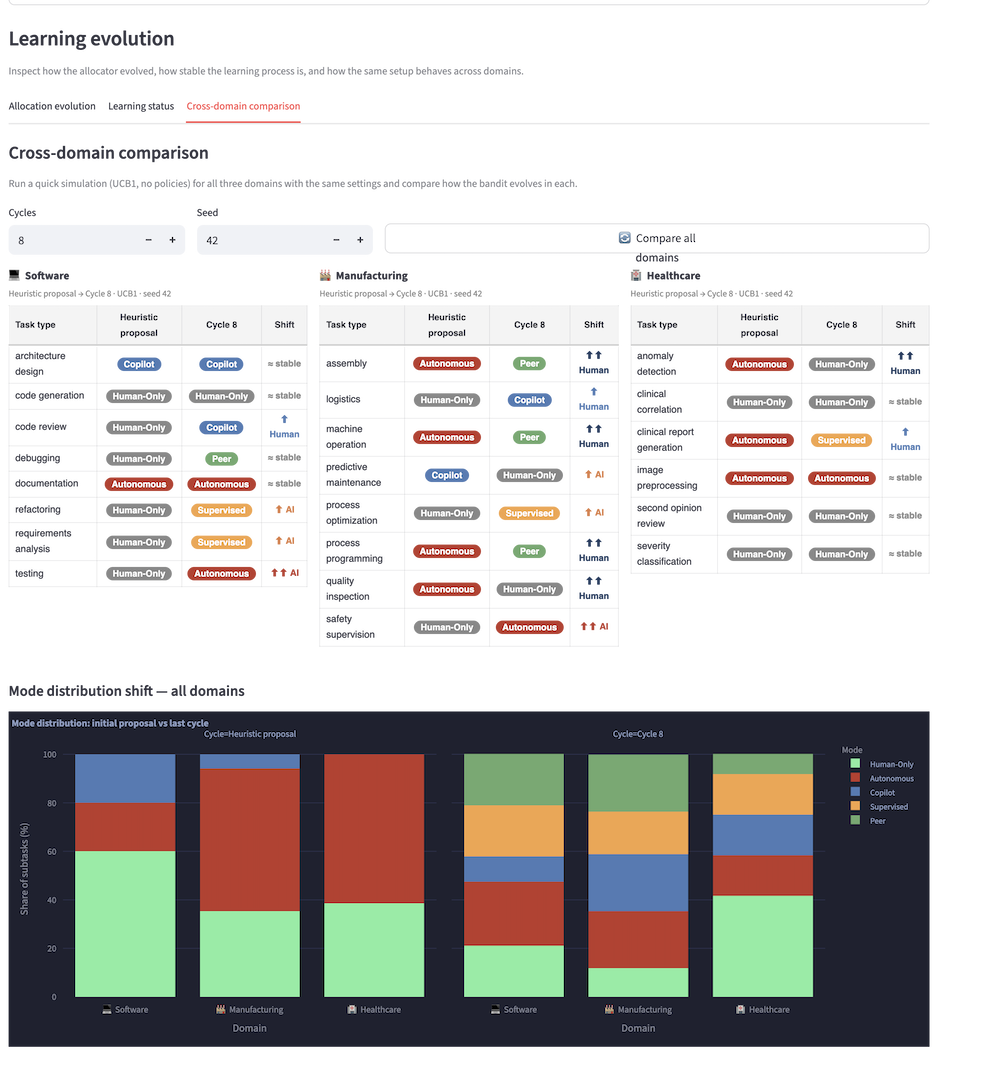}{%
\ui{Learning evolution} tab --- \ui{Cross-domain comparison} subtab, showing
how the same setup evolves across software, manufacturing, and
healthcare.\label{fig:r1c-cross}}

\paragraph{History.}
A persistent record of all saved runs and benchmarks. In the paper, this tab
should be described explicitly as a two-part workspace: a saved-run register
and a longitudinal comparison view. The register lets users reopen prior runs and benchmarks (Figure~\ref{fig:r1d-register}
shows the saved-run and saved-benchmark register with all prior executions
listed and accessible, confirming that simulation history is preserved across
sessions), while the comparison view loads two or more saved runs side by side
(Figure~\ref{fig:r1d-comparison} shows this longitudinal comparison view
with KPI evolution visible across runs, enabling direct inspection of how
deployment parameters and outcome metrics changed between configurations). It is the recommended entry point for
stress-testing a deployment recommendation by comparing variants across saved
configurations.

\screenshot[0.88\linewidth]{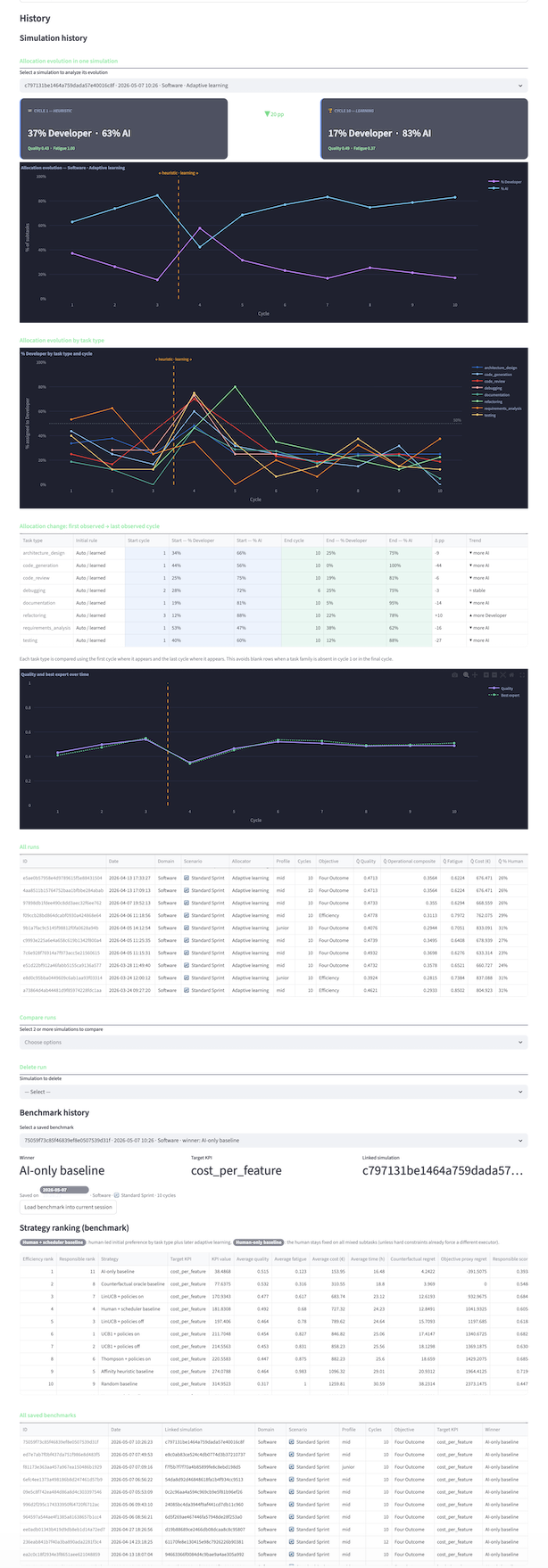}{%
\ui{History} tab --- saved-run and saved-benchmark register, showing the
persistent workspace used to reopen prior executions.\label{fig:r1d-register}}

\screenshot[0.97\linewidth]{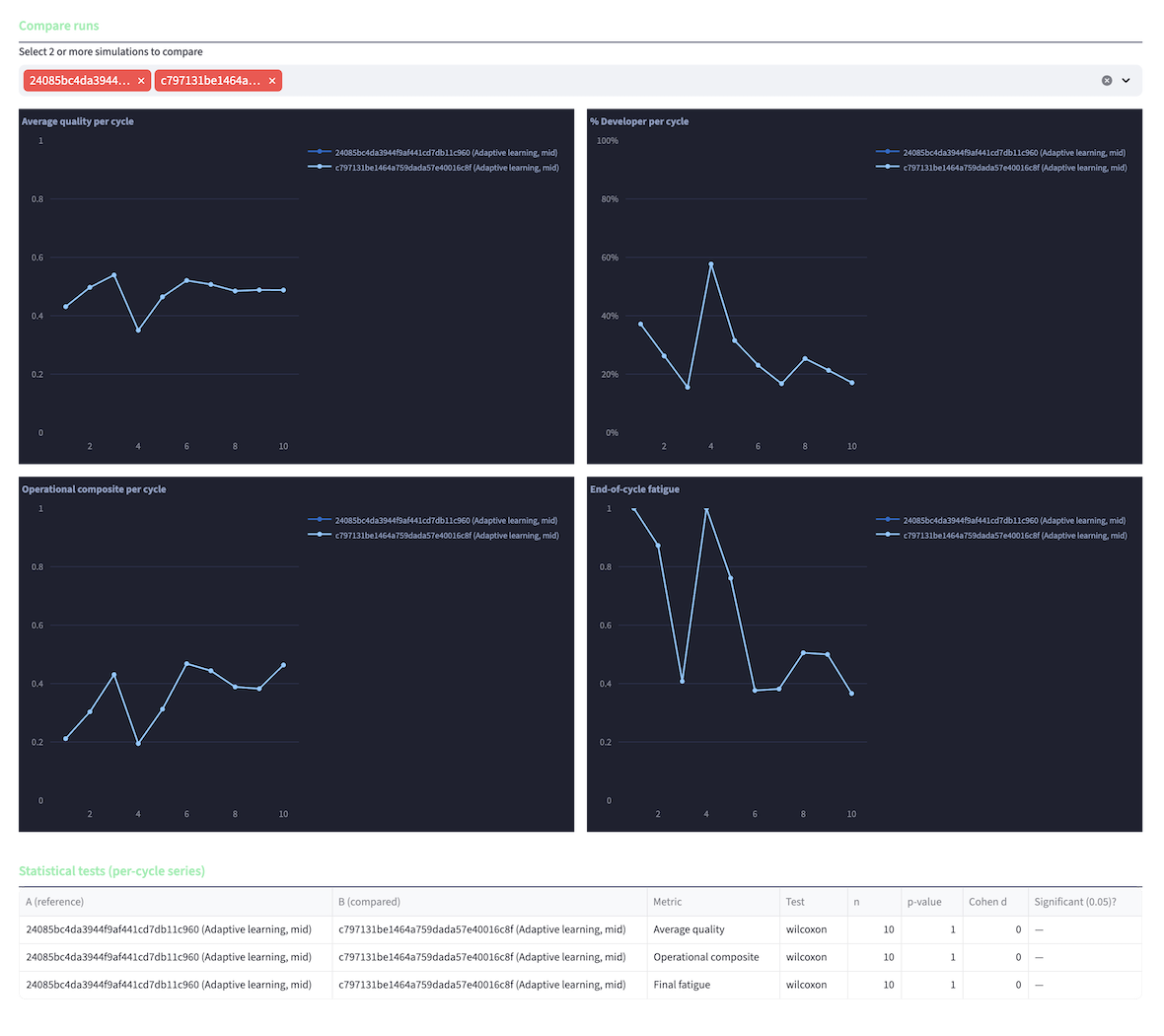}{%
\ui{History} tab --- longitudinal comparison view, used to inspect two or more
saved runs side by side.\label{fig:r1d-comparison}}

\subsection{The \ui{Decision support} tab}

The \ui{Decision support} tab is the main comparative workspace for
benchmark-based decisions. It is organized into four subtabs:
\ui{Benchmark ranking}, \ui{Deployment decision}, \ui{Deployment planning},
and \ui{Advanced comparison}. Together they help the user move from raw
benchmark results to an operational deployment choice.

\ui{Benchmark ranking} presents the compared strategies in ranked form and
makes visible which options remain acceptable under the current benchmark.
\ui{Deployment decision} summarizes the deployability signal derived from the
benchmark and separates strong benchmark performers from strategies that remain
suitable under the active contract. \ui{Deployment planning} turns the
benchmark into an actionable operating view through the viable set, Pareto
frontier, recommended operating models, risk watchlist, and transition plan.
\ui{Advanced comparison} supports deeper side-by-side inspection when the user
needs to study trade-offs beyond the headline recommendation.

Internally, the recommendation logic combines feasible-set filtering, Pareto
analysis, and multi-criteria comparison~\cite{topsis1981}. From the user
perspective, what matters is the separation between \term{best benchmark
result} and \term{deployable operating choice}.

Three design decisions shape the Pareto frontier shown in
\ui{Deployment planning} and are worth making explicit.
\textbf{(i) Human-participation pre-filter.} Before computing dominance,
the engine removes any strategy whose mean human participation falls below
a configurable floor (default~15\,\%). Fully autonomous baselines therefore
never appear on the frontier regardless of their efficiency score; the
comparison is always among strategies that preserve a minimum level of
human involvement, in line with the governance contract.
\textbf{(ii) Lexicographic treatment of responsible strategies.}
The responsible screen is not an objective to be optimised alongside
quality and fatigue: it is a first-order constraint. Any strategy that
passes the responsible screen is always shown on the frontier even if it
is dominated on quality or fatigue by a non-responsible strategy. This
prevents a purely efficient but governance-unsafe option from hiding the
responsible alternative from the decision-maker.
\textbf{(iii) Minimum-cardinality padding.} When the surviving set
contains fewer than three Pareto-optimal strategies (e.g.\ under a very
tight quality contract), the visualisation is padded to at least three
entries by appending the highest-scoring non-dominated candidates from the
full strategy pool. Padded entries are included for legibility only; they
are not formal members of the Pareto set and should not be interpreted as
contract-admissible recommendations.

\paragraph{Benchmark ranking.}
This subtab is the direct ranking table for the current benchmark.
Figure~\ref{fig:decision-ranking} shows the full table with strategies ranked
by efficiency alongside responsibility, KPI value, cost, time, fatigue, and
deployability columns, illustrating how a single view separates raw performance
order from governance-filtered acceptability.
\begin{itemize}
  \item \textbf{How to read it.} First identify which rows remain deployable
        under the active benchmark contract. Only then compare the surviving
        strategies by efficiency rank and responsible rank.
  \item \textbf{What the example shows.} \ui{Counterfactual oracle baseline}
        appears near the top on raw efficiency (rank~2, KPI value 77.6375,
        fatigue 0.316, cost 310.55~EUR) but does not receive the deployability
        checks. By contrast, \ui{UCB1 + policies on} is only efficiency rank~6
        yet becomes responsible rank~1 and retains all three green deployment
        checks.
  \item \textbf{Interpretation.} If a strategy looks strong on raw KPI or time
        but weak on the responsible screen, the figure should be read as a
        warning that benchmark performance alone is not sufficient for
        deployment. The table is therefore not asking only ``who won?'' but
        also ``who won without creating an unacceptable operational or
        governance profile?''
\end{itemize}

\screenshot[0.97\linewidth]{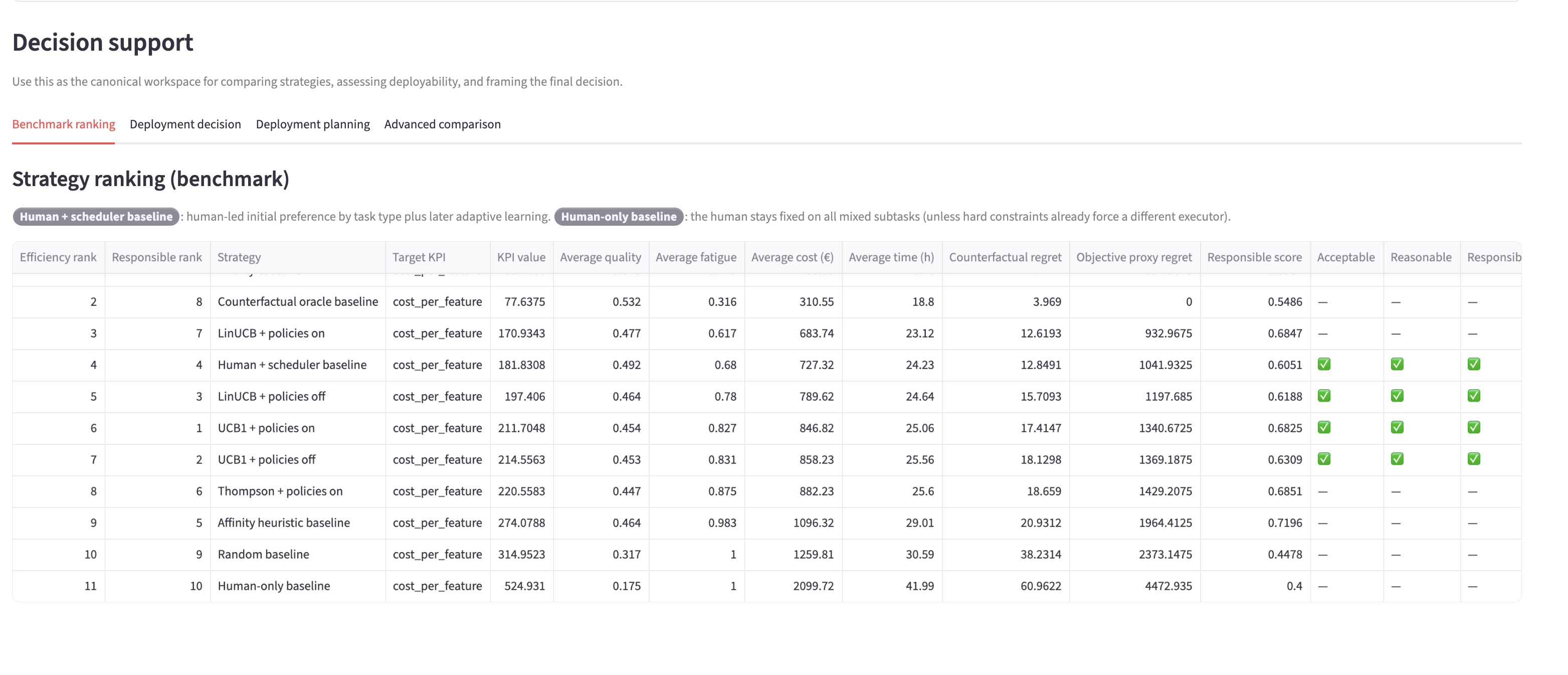}{%
\ui{Decision support} --- \ui{Benchmark ranking} subtab, showing the benchmark
table with efficiency rank, responsible rank, KPI value, cost, time, fatigue,
and deployability-oriented columns.\label{fig:decision-ranking}}

\paragraph{Deployment decision.}
This subtab is the clearest place to understand what the tool is actually
recommending. Figure~\ref{fig:decision-deployment} shows the staged
deployment-decision view---benchmark pool, acceptance stage, recommended
configuration, and deployment watchpoints---demonstrating how the tool
translates a benchmark result into an explicit go/no-go deployment signal.
\begin{itemize}
  \item \textbf{How to read it.} Read the figure from top to bottom as a
        narrowing funnel: benchmark pool, acceptance stage, recommended
        configuration, then deployment watchpoints.
  \item \textbf{What the example shows.} The benchmark starts from
        11 evaluated strategies, only 4 pass the acceptability screen, and the
        responsible recommendation becomes \ui{UCB1 + policies on}.
  \item \textbf{Why the result is not ``fully clear''.} The panel still labels
        the outcome \textbf{Conditional} and reports that no strategy passes
        the full contract. The reason is visible in the metrics: governance
        violations remain at 0.000, deskilling risk at 0.000\%, and
        responsible score at 68.2\%, but modeled fatigue is still relatively
        high at 0.827.
  \item \textbf{Interpretation.} A recommendation here is not merely the
        top-scoring strategy; it is the strategy that remains defensible once
        the contract, constraints, and visible risks are applied. The yellow
        note makes that fragility explicit: if priorities shift toward cost and
        time or toward human sustainability, the preferred option changes to
        \ui{Human + scheduler baseline}.
\end{itemize}

\screenshot[0.68\linewidth]{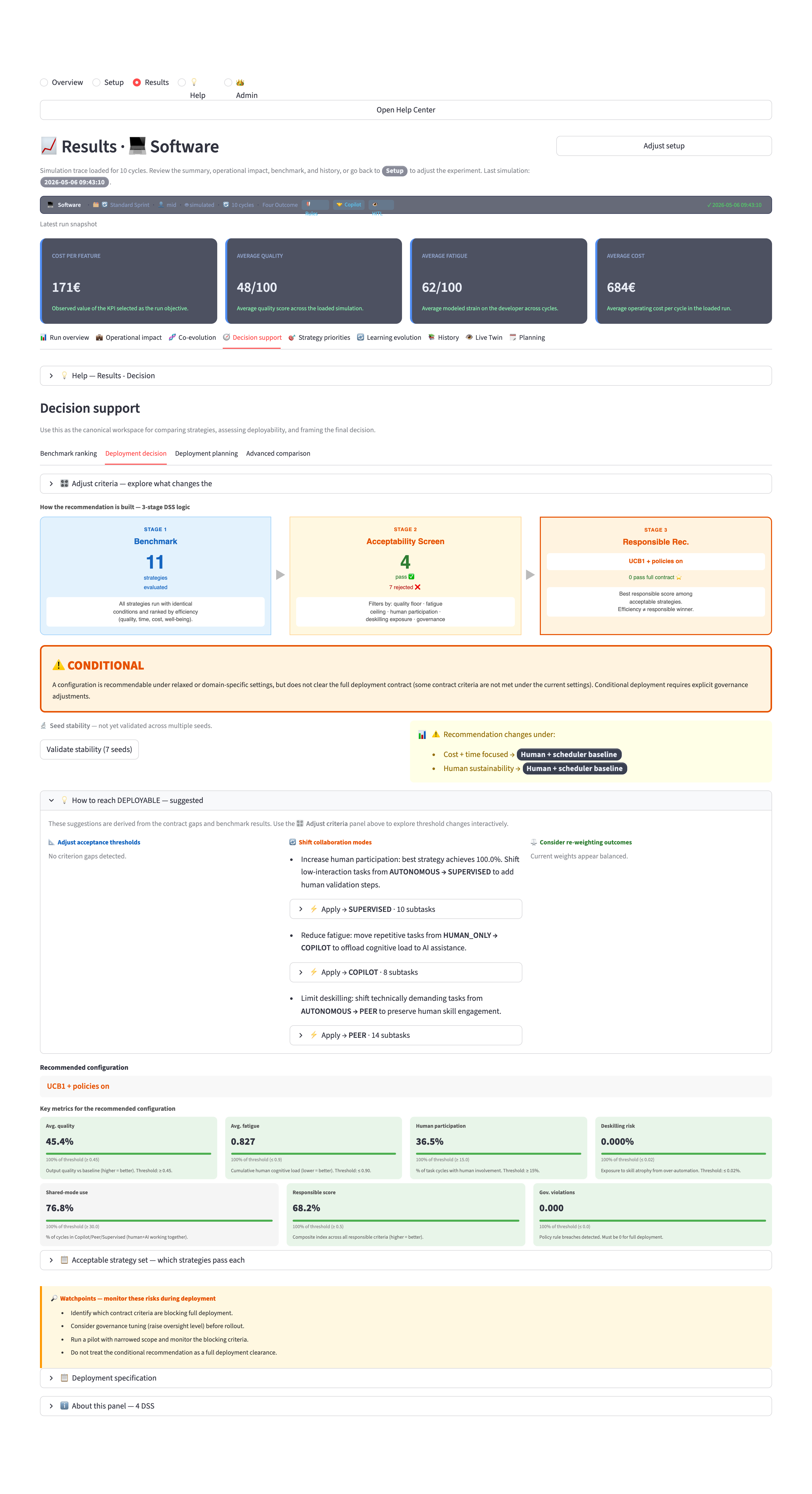}{%
\ui{Decision support} --- \ui{Deployment decision} subtab, summarizing the
benchmark pool, acceptance stage, recommended configuration, and deployment
watchpoints.\label{fig:decision-deployment}}

\paragraph{Deployment planning.}
This subtab translates the benchmark into an operational plan. In the current
UI, this is where the feasible set, Pareto frontier table, recommended
operating models, risk watchlist, observed mode overrides, and task-level
transition plan are consolidated. Figure~\ref{fig:decision-planning} shows
this subtab with the viable set, Pareto frontier section, recommended
operating models, risk watchlist, and mode-transition plan all visible in a
single scrollable workspace, confirming that the Pareto view and the transition
plan belong together here rather than in a separate standalone tab.
\begin{itemize}
  \item \textbf{How to read it.} Move through the subtab in sequence:
        viable set, Pareto frontier, recommended operating models, risk
        watchlist, and finally the transition plan.
  \item \textbf{What the example shows first.} The \ui{Viable strategies}
        counter is 0, so there is no fully admissible strategy under the
        current contract, yet the Pareto frontier still contains
        non-dominated options drawn from strategies that satisfy the minimum
        human-participation floor (see the design-decision note above).
  \item \textbf{What the frontier means here.} The frontier entries make
        the trade-off concrete: strategies with lower fatigue and cost appear
        at the left of the scatter plot, while those with higher quality
        appear higher on the vertical axis. At least one diamond marker
        identifies any strategy that also passes the responsible screen,
        providing a governance-safe reference point even when no strategy
        is fully contract-admissible.
  \item \textbf{How the plan becomes operational.} The transition plan converts
        those differences into task-level action. For example, the figure
        recommends moving \ui{architecture\_design} from \ui{peer} to
        \ui{autonomous} because the best counterfactual score (0.591) exceeds
        the best observed supervised outcome (0.557), while
        \ui{documentation} remains \ui{autonomous} because observed and
        counterfactual evidence align.
  \item \textbf{Interpretation.} This subtab should be read less as a static
        report and more as the bridge between analytical selection and
        operational rollout. The watchlist clarifies the governance meaning of
        the result: the plan is useful, but it is still a frontier-based
        guidance layer rather than confirmed contract clearance.
\end{itemize}

\screenshot[0.88\linewidth]{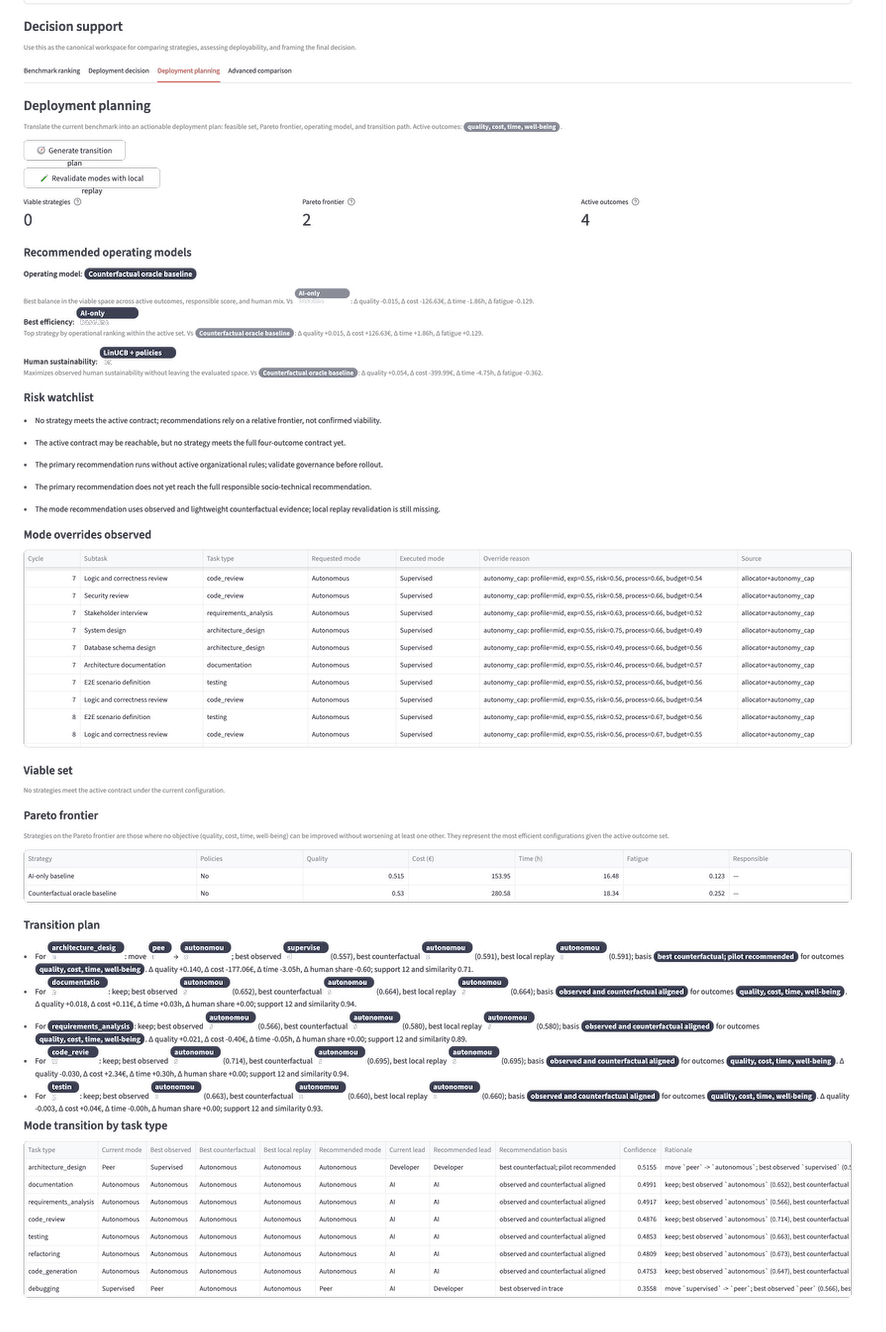}{%
\ui{Decision support} --- \ui{Deployment planning} subtab, showing the viable
set, Pareto frontier section, recommended operating models, risk watchlist, and
mode-transition plan.\label{fig:decision-planning}}

\paragraph{Advanced comparison.}
This subtab is a deeper comparison workspace. It is useful when the user needs
to inspect benchmark families and strategy trade-offs beyond the headline
recommendation and the deployment plan. Figure~\ref{fig:decision-advanced}
shows the unified strategy comparison view with all strategies laid out for
side-by-side inspection, illustrating how the subtab supports detailed
analysis that would otherwise require manual cross-tab navigation.
\begin{itemize}
  \item \textbf{How to read it.} Read this figure comparatively rather than
        sequentially: scan across strategies to detect where one family
        consistently dominates, where trade-offs cross over, and where two
        candidates are effectively tied on outcomes but differ in risk exposure
        or governance fit.
  \item \textbf{What the example shows.} The \ui{AI-only baseline} is rank~1 on
        raw efficiency, with KPI value 38.4868, cost 153.95~EUR,
        time 16.48~hours, and fatigue 0.123, but it has responsible rank~11 and
        receives no deployability checks.
  \item \textbf{What changes under responsible reading.} By contrast,
        \ui{LinUCB + policies on} is only rank~3 on efficiency, yet it reaches
        responsible rank~2 with responsible score 0.6782 and all three green
        deployability checks. \ui{UCB1 + policies on} shows the same pattern
        even more strongly: rank~6 by efficiency, but responsible rank~1 with
        responsible score 0.6836.
  \item \textbf{Interpretation.} When the headline recommendation feels too
        close to call, this view helps the user justify why one benchmark
        family is more robust, more balanced, or simply easier to defend than
        another. It is precisely the place where a bandit family can be shown
        to be operationally preferable even though a baseline remains faster or
        cheaper on the surface.
\end{itemize}

\screenshot[0.97\linewidth]{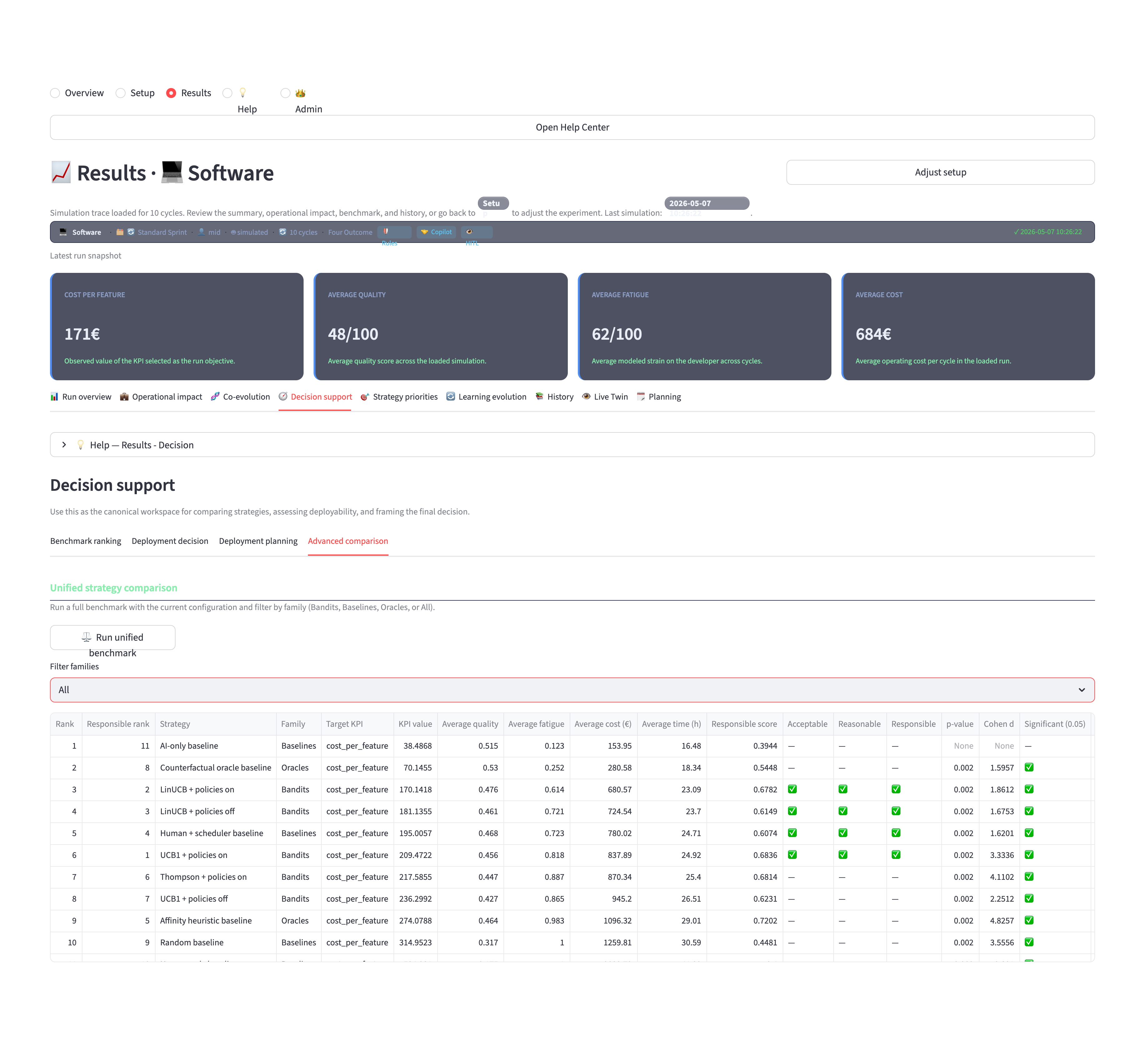}{%
\ui{Decision support} --- \ui{Advanced comparison} subtab, showing the unified
strategy comparison view for deeper side-by-side inspection.\label{fig:decision-advanced}}

\clearpage

\subsection{The \ui{Strategy priorities} tab}

Whereas \ui{Decision support} filters and ranks strategies by their
compliance with the active contract, \ui{Strategy priorities} addresses a
complementary question: how does the ranking change when the user's own
value preferences are made explicit? The tab provides a set of interactive
preference sliders that let the user encode a priority order over the four
outcome dimensions (quality, cost, time, well-being) as numerical weights.
The ranking is recomputed in real time, making visible how sensitive the
top recommendation is to the organisation's actual priorities.

The key analytical use of this tab is identifying \textit{robustly good}
strategies: those that rank in the top tier under all reasonable weight
configurations. A strategy that only dominates under a narrow weight
combination should be treated as locally optimal rather than generally
deployable.

The tab also includes a secondary \ui{Strategy profiles} subtab, but this is
best understood as a visual profile view derived from the same priority
settings rather than as a separate decision layer.

For documentation purposes, the main reference view is the ranking subtab.
Figure~\ref{fig:strategy-priorities} shows the preference sliders over the
four outcome dimensions alongside the recomputed strategy ranking. The figure
makes explicit how the top recommendation shifts as organisational priorities
change. Recipe~9 in Appendix~\ref{app:analytical-recipes} walks through a
structured sensitivity analysis using this tab.
\begin{itemize}
  \item \textbf{What the sliders encode in the example.} The figure visibly
        privileges quality (60), fatigue (50), cost (40), time (30), and
        cumulative regret (25), while leaving human participation, shared mode,
        reward, and counterfactual regret at zero.
  \item \textbf{What the example ranking becomes.} Under that weighting, the
        best strategy becomes \ui{AI-only baseline} with a weighted score of
        0.986, followed by \ui{Counterfactual oracle baseline} at 0.907, with a
        substantial drop to \ui{LinUCB + policies on} at 0.693.
  \item \textbf{Why that strategy wins.} The summary line makes the reason
        explicit: the winning baseline is best of the 11 on time
        (16.480~h), cost (153.950~EUR), fatigue (0.123), and cumulative regret
        (-391.508), while still remaining second on quality (0.515).
  \item \textbf{Interpretation.} The tab should not be read as a replacement
        for deployability screening, but as a demonstration that once the user
        emphasizes pure performance and de-emphasizes governance-relevant human
        factors, the ranking can move sharply toward highly automated options.
        The bar chart and normalized profile view reinforce that point by
        showing a wide gap between the top two strategies and the rest of the
        field.
\end{itemize}

\screenshot[0.88\linewidth]{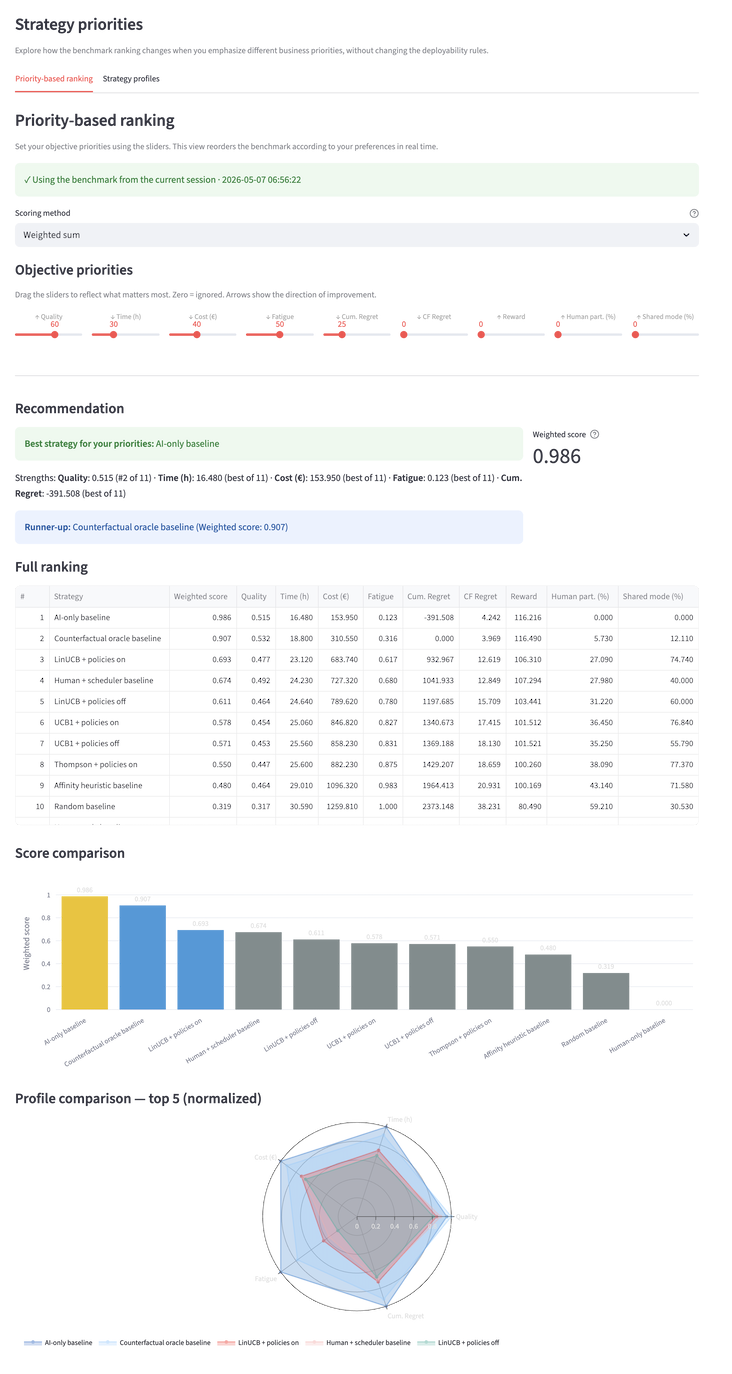}{%
\ui{Strategy priorities} tab showing the preference sliders over the four
outcome dimensions and the recomputed strategy ranking for the current weight
configuration.\label{fig:strategy-priorities}}

\subsection{Persistent worker analysis}

Two advanced modules extend the standard run-and-benchmark workflow:
\begin{itemize}
  \item \textbf{Live Twin} stores worker state across sessions, including
        fatigue, skill evolution, trust, and learning traces;
  \item \textbf{Planning} performs forward-looking simulations from that state to
        estimate fatigue, trust, and risk over a configurable horizon.
\end{itemize}

These modules are important because they transform the tool from a single-run
simulator into a longitudinal decision-support environment. The manual uses them
especially in advanced cases such as \emph{New Product Ramp-Up}, where the
question is not only what works now, but whether the human can sustain the next
set of cycles without critical fatigue or skill erosion. In practical terms,
these views are populated by repeated \ui{Run twin session} executions with the
same persistent worker ID inside the \ui{Live Twin} workflow. Standard
\ui{Simulation} runs and benchmark runs alone are not sufficient: the tabs
require accumulated twin state across sessions before the \ui{Live Twin}
tab (Figure~\ref{fig:live-twin} shows the persistent worker state after
several sessions, including current fatigue, skill level, trust score, active
alerts, and the longitudinal evolution trace, demonstrating how repeated
executions build a multi-session worker history) and the \ui{Planning} tab
(Figure~\ref{fig:planning} shows the multi-horizon fatigue and trust
projections with risk threshold bands, illustrating the forward-looking
capability unlocked by accumulated twin state) become informative. The figures
make the distinction between \emph{state}, \emph{recent history}, and
\emph{forward risk} concrete.

\paragraph{Reading the \ui{Live Twin} view.}
\begin{itemize}
  \item \textbf{What the saved state says.} In Figure~\ref{fig:live-twin}, the
        saved worker state for the \ui{tecnico} profile shows trust already at
        1.000 and current fatigue at 0.000 by sprint~12, which might look fully
        stable if read in isolation.
  \item \textbf{What the recent session adds.} The recent four-sprint session
        immediately adds nuance: average quality is only 0.454, average fatigue
        is 0.397, and fatigue rose from 0.176 to 0.538 before easing to 0.479.
  \item \textbf{What the alerts mean.} There are no critical or warning alerts,
        but there are 3 informational alerts and the drift monitor still
        recommends recalibration, with a fatigue drift magnitude of 0.045 and a
        quality drift magnitude of 0.094.
  \item \textbf{Interpretation.} The twin view separates ``the worker is safe
        right now'' from ``the model is still seeing behaviour worth
        rechecking.''
\end{itemize}

\paragraph{Reading the \ui{Planning} view.}
\begin{itemize}
  \item \textbf{What the horizon setup is.} In Figure~\ref{fig:planning},
        6 future sprints are projected with 3 Monte Carlo samples.
  \item \textbf{What the summary badges say.} Fatigue stays safe, trust stays
        healthy, but AI share peaks at 91\%, creating a deskilling risk signal.
  \item \textbf{What the table shows numerically.} Projected fatigue remains
        below the warning threshold of 0.70, peaking at 0.679 in sprint~15,
        while projected trust remains fixed at 1.000 and projected quality
        rises from 0.435 to 0.520 by sprint~18. Yet sprint~17 projects an
        AI share of 90.8\%.
  \item \textbf{Interpretation.} The message is not ``future risk is low'' in a
        general sense, but rather ``physiological and trust risk look
        controlled while capability-exposure risk may accumulate through
        excessive automation.'' Together, the two figures show why persistent
        worker analysis is a longitudinal governance layer rather than only a
        worker-status dashboard.
\end{itemize}

\screenshot[0.88\linewidth]{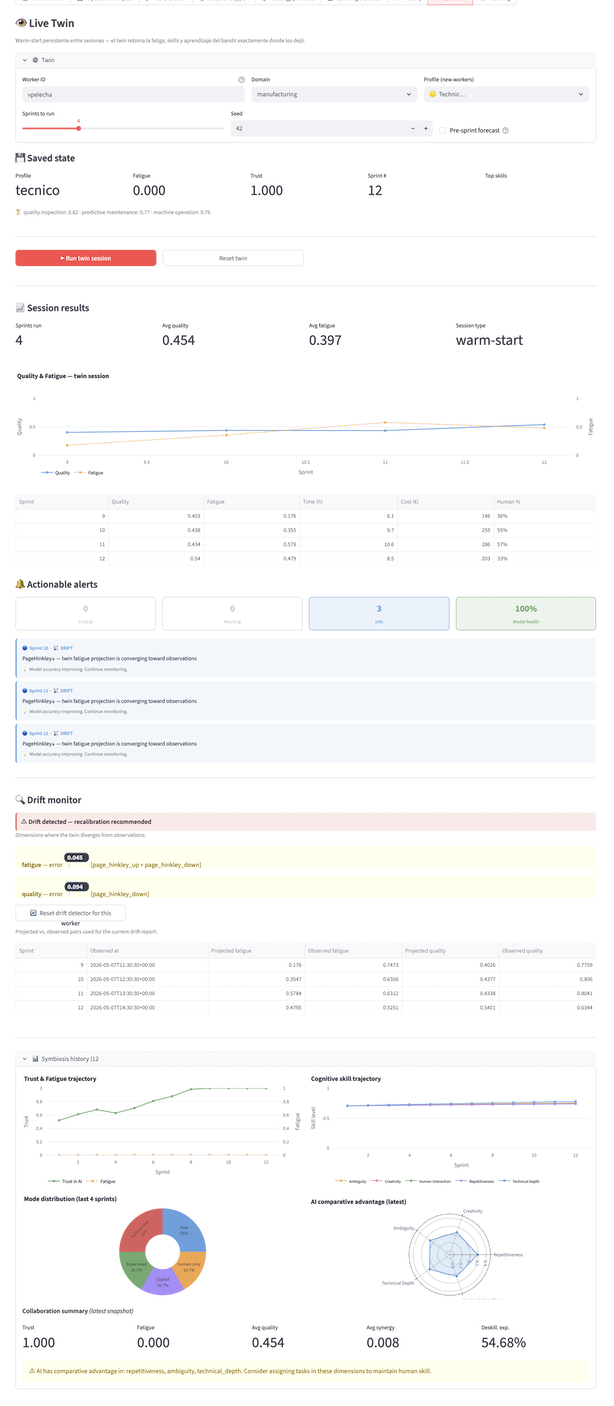}{%
\ui{Live Twin} tab for a persistent worker after several accumulated sessions,
showing current fatigue, skill level, trust score, active alerts, and the
longitudinal evolution trace.\label{fig:live-twin}}

\screenshot[0.97\linewidth]{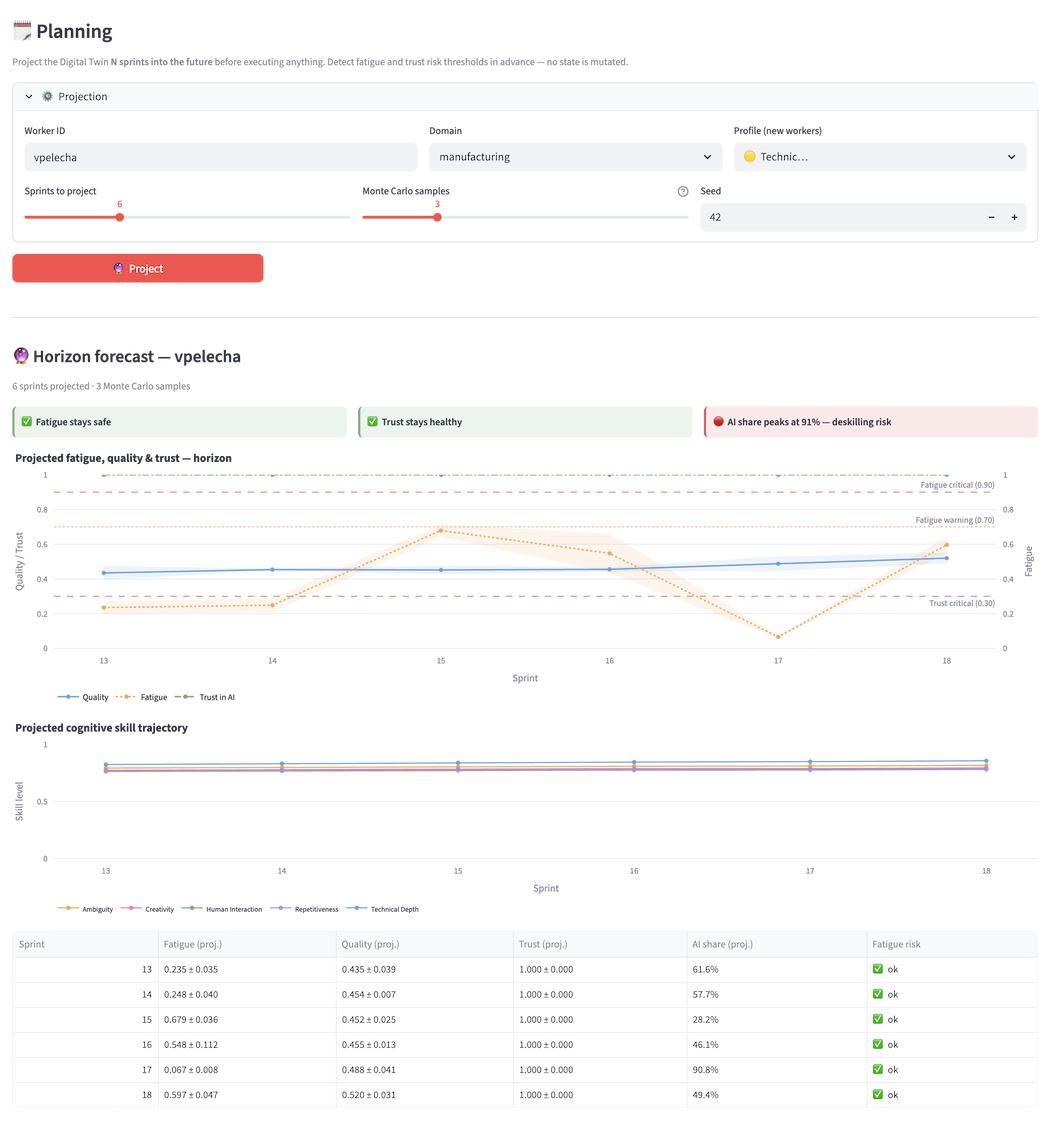}{%
\ui{Planning} tab showing multi-horizon projections of fatigue and trust from
the current worker state, with risk threshold bands
indicated.\label{fig:planning}}

\subsection{How to read the workspace without getting lost}
\label{subsec:reading-results}

The workspace is easiest to interpret if each tab is treated as answering one
primary question:
\begin{itemize}
  \item \textbf{\ui{Run overview}}: what happened overall in this run?
  \item \textbf{\ui{Operational impact}}: did the run satisfy the contract, and
        where did it drift?
  \item \textbf{\ui{Co-evolution}}: is the human--AI relationship improving,
        degrading, or locking into a risky pattern?
  \item \textbf{\ui{Learning evolution}}: how did the allocator adapt?
  \item \textbf{\ui{Decision support}}: which strategies remain deployable?
  \item \textbf{\ui{Strategy priorities}}: does the recommendation remain stable
        if organizational priorities change?
  \item \textbf{\ui{History}}: is the result robust across saved variants?
  \item \textbf{\ui{Live Twin}} / \textbf{\ui{Planning}}: is the worker state
        sustainable across sessions and future horizons?
\end{itemize}

The recommended reading sequence depends on execution mode:
\begin{enumerate}
  \item For \ui{Simulation}: \ui{Run overview} $\rightarrow$
        \ui{Operational impact} $\rightarrow$ \ui{Learning evolution}
        $\rightarrow$ \ui{History}.
  \item For \ui{Benchmark}: \ui{Decision support} $\rightarrow$
        \ui{Strategy priorities} $\rightarrow$ \ui{History}.
  \item For \ui{Simulation + Benchmark}: \ui{Run overview} $\rightarrow$
        \ui{Operational impact} $\rightarrow$ \ui{Decision support}
        $\rightarrow$ \ui{History}, using \ui{Co-evolution} and
        \ui{Learning evolution} to explain why the recommendation looks the way
        it does.
\end{enumerate}

\clearpage

\subsection{Task-oriented recipes}
\label{sec:recipes}

The subsections above describe the tool's surfaces; this subsection describes
how to \emph{operate} them. Each recipe names a concrete question about running,
configuring, or navigating the tool, and lists the steps needed to answer it.
To keep the main text focused on the artifact rather than on exhaustive
step-by-step usage, Table~\ref{tab:recipes} summarizes four representative
workflows. Advanced operational recipes (R5--R7) and analytical recipes
(R8--R11) are collected in the appendices.

\begin{table}[H]
\centering
\small
\caption{Representative workflows retained in the main text.
Advanced operational and analytical recipes are collected in the appendices.}
\label{tab:recipes}
\begin{tabularx}{\linewidth}{@{}>{\bfseries}clY@{}}
\toprule
\# & Workflow question & Primary tabs \\
\midrule
1 & Explore a single run & Run overview, Operational impact \\
2 & Compare strategies; select deployment option &
    Decision support, History \\
3 & Monitor co-evolution; detect skill degradation &
    Co-evolution, Learning evolution (requires full simulation) \\
4 & Track a persistent worker across sessions &
    Live Twin, Planning \\
\bottomrule
\end{tabularx}
\end{table}

\paragraph{Recipe~1 --- Explore a single run and read its results.}
\textit{Question:} how does a given allocator behave on a specific domain
and scenario, and does it satisfy the operational contract?

\begin{enumerate}
  \item In the wizard, set \ui{Objective} to \ui{Explore simulation} and keep
        the \ui{Operating preset} at \ui{Standard evaluation}.
  \item In \ui{Context}, choose a domain (e.g.\ \ui{Software}), a scenario
        (e.g.\ \ui{Maintenance}), and a human profile that reflects the target
        worker. Set run length to 10~cycles.
  \item In \ui{Strategy}, select \ui{UCB1} as allocator and \ui{Quality First}
        as reward profile. Leave exploration cycles at the default (3).
  \item In \ui{Guardrails}, set a cost target, a minimum quality floor
        (e.g.\ 0.80), and a fatigue cap (e.g.\ 0.45). These define the
        operational contract against which the run will be evaluated.
  \item Run the simulation. On completion, open the \ui{Run overview} tab
        (Figure~\ref{fig:r1a} shows the \ui{Reports} subtab with export
        and reporting artifacts, which complements the aggregate KPI summary
        visible in the \ui{Summary} subtab). It is fed directly from the
        simulation trace and aggregates the cycle-level outcomes into a
        top-level snapshot organised in three layers that should be inspected
        in order:
        \begin{itemize}
          \item \term{Headline KPI cards.} Four cards summarise the run
                averages. In the reference run shown in
                Figure~\ref{fig:r1a}: \texttt{cost\_per\_feature} 152~€
                (well below the \texttt{190} target, the efficiency signal
                looks healthy), \texttt{average\_quality} 50/100 (moderate),
                \texttt{average\_fatigue} 63/100 (elevated, flag for the
                well-being check in step~6) and \texttt{average\_cost}
                607~€/cycle (operational, contextual rather than
                contractual).
          \item \ui{Summary} \term{sub-tab --- co-evolution signals.}
                Consolidates the human--AI adaptation indicators
                (Section~\ref{subsec:coevolution}):
                \term{AI--preference alignment} 0.82,
                \term{human skill growth} $+0.172$,
                \term{profile overrides} 927 (a high count signals strong
                governance pressure) and
                \term{bidirectional index} 0.50. The
                \term{Preventive Tutor Mode} card flags any sustained risk;
                in the reference run \emph{``no sustained activation,
                2~cycles in the risk zone''} is a soft warning, not yet
                actionable.
          \item \term{Allocation distribution.} The
                \term{Global allocation distribution} donut and the
                \term{Last cycle --- allocation details} table close the
                tab. Use them to confirm that the human/AI split
                (71\,\%\,/\,29\,\% in the reference run) is consistent with
                the task pool and the company profile selected in step~3.
        \end{itemize}
  \item Open \ui{Operational impact} (Figure~\ref{fig:r1b} shows the
        \ui{KPI contract} subtab with the contract compliance panel,
        subtask distribution by task type, and the governance events log
        visible below the traffic lights, illustrating how contractual
        pass/fail is decomposed at the KPI level). This is the contractual
        reading of the run, organised in four sub-tabs that should be
        inspected in order:
        \begin{itemize}
          \item \ui{KPI contract} shows the target KPI and its current value
                against the configured target. In the reference run
                \texttt{lead\_time\_h} is 20.278 against a target of 18.000,
                so both \term{Active contract} and \term{Full contract}
                indicators are red (``Does not meet''). The five
                \term{Contract traffic lights} below decompose the failure
                (objective, full quality, full fatigue, active and full
                contract). The KPI table at the bottom lists every
                contractual indicator with its preferred direction; the first
                row whose direction is wrong is the proximate cause of the
                failure (here \texttt{defect\_escape\_rate} 50.90\,\% and
                \texttt{rework\_pct} 27.78\,\% both pull lead time up).
          \item \ui{Trends and trade-offs} plots the cycle-by-cycle evolution
                of quality, fatigue and cost together with the
                \term{Accumulated counterfactual regret} curve $R_C$ (see
                Section~\ref{subsec:oracle}). Use it to localise \emph{when}
                the run drifted: a quality dip in a specific sprint, a
                fatigue spike, or a regret slope that does not flatten by
                cycle~6.
          \item \ui{Well-being} carries the per-cycle fatigue, trust and
                skill trajectories. Inspect this whenever
                \texttt{average\_fatigue} from step~5 sits above the cap
                configured in step~4.
          \item \ui{Corrective actions} hosts the governance events log and
                the suggested adjustments emitted by the policy engine ---
                the actionable handoff to step~7.
        \end{itemize}
        Cross-check the headline cards from step~5 against the contract
        signals here: the cards summarise \emph{averages} across the run,
        whereas this tab evaluates \emph{feasibility} against the operational
        contract. A run can post a healthy average and still fail the
        contract on a single binding KPI.
  \item If the contract is not met (any red traffic light in step~6),
        the question to answer is \emph{which knob to turn}, not whether
        to retry. Translate the diagnostics from \ui{Well-being} and
        \ui{Corrective actions} into one of three response patterns:
        \begin{itemize}
          \item \term{Sustainability fix.} If the \ui{Well-being} trace
                shows the fatigue cap from step~4 crossed in two or more
                consecutive cycles, the failure is human-side. Either
                tighten the fatigue guardrail (lower cap in step~4) so
                the policy engine can demote autonomy earlier, or reduce
                autonomy directly in step~3 by enforcing \ui{Copilot} on
                tasks currently allowed to run \ui{Supervised}.
          \item \term{Allocator/policy fix.} If the \term{governance
                events log} in \ui{Corrective actions} shows repeated
                firings on the same task type, and the
                \term{profile overrides} count from step~5 is high
                (927 in the reference run), the allocator is fighting
                the policy. Revisit the company profile or the reward
                weights in step~3 --- relaxing guardrails here would
                only mask the conflict.
          \item \term{Convergence fix.} If neither pattern holds and the
                failure is concentrated on a handful of cycles
                (especially the early ones), the bandit has not
                converged. Increase exploration cycles in step~3 or
                extend run length in step~2.
        \end{itemize}
        Re-run with the adjusted setup and save the new run to
        \ui{History} alongside the failing baseline. Recipe~3 compares
        runs side by side --- both runs together establish the causal
        chain between guardrail tuning and contract feasibility, which
        is what makes a deployment recommendation defensible.
\end{enumerate}

\begin{infobox}
\textbf{What to look for.} A passing run should show quality at or above the
floor across all cycles, fatigue below the cap, and a subtask distribution
consistent with the task pool. If the run fails the contract, the corrective
actions panel will suggest which guardrail or reward weight to adjust before
re-running.
\end{infobox}

\paragraph{Recipe~2 --- Compare strategies and select a deployment option.}
\textit{Question:} which allocation strategy is most defensible for
deployment, accounting for performance, governance, and risk?

\begin{enumerate}
  \item In the wizard, set \ui{Objective} to \ui{Compare strategies} and keep
        the \ui{Operating preset} at \ui{Standard evaluation}.
  \item Configure context and guardrails as in Recipe~1. Select a company
        profile in the \ui{Strategy} step (e.g.\
        \ui{Software\_Quality\_Architecture}) to anchor the reward weights and
        guardrail defaults to a recognizable operational posture.
  \item Run the benchmark. The tool will execute the configured scenario across
        multiple allocators (UCB1, Thompson Sampling, LinUCB, Discounted-UCB)
        and seeds.
  \item Open \ui{Decision support} and read the subtabs in this order:
        \begin{enumerate}[label=\alph*.]
          \item \ui{Benchmark ranking} subtab: Figure~\ref{fig:r2a} shows
                the strategy ranking table with feasible-set and
                responsible-screen columns visible alongside average quality,
                cost, time, and fatigue, illustrating the separation between
                the efficiency winner (top row by reward) and the deployment
                recommendation (the strategy that also passes the responsible
                screen).

          \item \ui{Deployment decision} subtab: Figure~\ref{fig:decision-deployment}
                shows the responsible screen with the deployability signal
                (\textsc{deploy} / \textsc{conditional} / \textsc{no-deploy}),
                the acceptable set filtered for fatigue ceiling, quality floor,
                deskilling cap, and governance level, and contract-gap bars
                showing how close the best available strategy came to each
                unmet threshold, confirming that the deployment recommendation
                is drawn only from strategies that pass \emph{both} the
                viable-set filter and this screen.

          \item \ui{Deployment planning} subtab. Read its panels in this
                internal order:
                \begin{enumerate}[label=\roman*.]
                  \item \textbf{Viable set}: Figure~\ref{fig:decision-planning}
                        shows the viable-set panel with \emph{Active contract},
                        \emph{Full contract}, and \emph{Responsible score}
                        columns for each strategy, making explicit which
                        options satisfy all guardrails and which are
                        disqualified regardless of benchmark rank.

                  \item \textbf{Pareto frontier}: Figure~\ref{fig:r2b} shows
                        the Pareto frontier scatter plot (x = avg fatigue,
                        y = avg quality; marker size $\propto$ cost; diamond =
                        passes responsible screen) together with the detail
                        table, illustrating which frontier strategies are also
                        governably deployable.

                  \item \textbf{Risk watchlist}: the same
                        Figure~\ref{fig:decision-planning} shows the risk
                        watchlist inside the \ui{Deployment planning} workspace.
                        Read this panel as the benchmark-generated list of
                        instability flags, governance failures, or deskilling
                        exposure. Any item listed here should be resolved before
                        committing to a deployment plan.

                  \item \textbf{Transition plan} and
                        \textbf{Mode transition by task type}:
                        Figure~\ref{fig:r2b} shows the mode-transition table
                        with suggested mode shifts per task type (e.g.\
                        \term{Code generation} toward \ui{Supervised};
                        \term{Architecture review} toward \ui{Copilot}). Each
                        row lists \emph{Current mode},
                        \emph{Recommended mode}, \emph{Confidence}, and
                        \emph{Rationale}. \textbf{This view requires a full
                        simulation trace} --- it is not populated by a
                        benchmark-only run. To unlock it, run
                        \ui{Explore simulation} or \ui{Get recommendation} from
                        the \ui{Setup} tab first, then return to
                        \ui{Decision support} and click
                        \ui{Generate transition plan}.
                \end{enumerate}
        \end{enumerate}
  \item Save the run to \ui{History}. To stress-test the recommendation,
        duplicate the configuration, change one variable (e.g.\ company
        profile or cost ratio), re-run, and compare in the \ui{History} tab.
\end{enumerate}

\screenshot[0.97\linewidth]{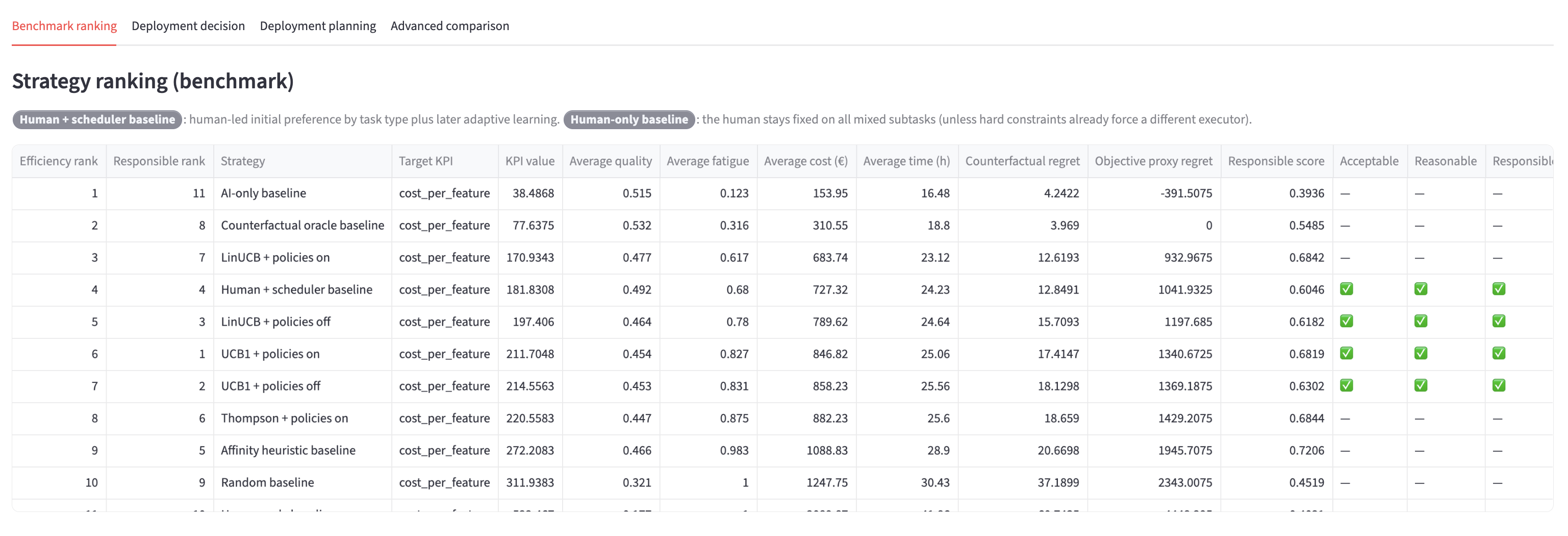}{%
Decision support tab --- strategy ranking table with the feasible-set
and responsible-screen columns visible, showing the separation between
efficiency winner and deployment
recommendation.\label{fig:r2a}}

\screenshot[0.97\linewidth]{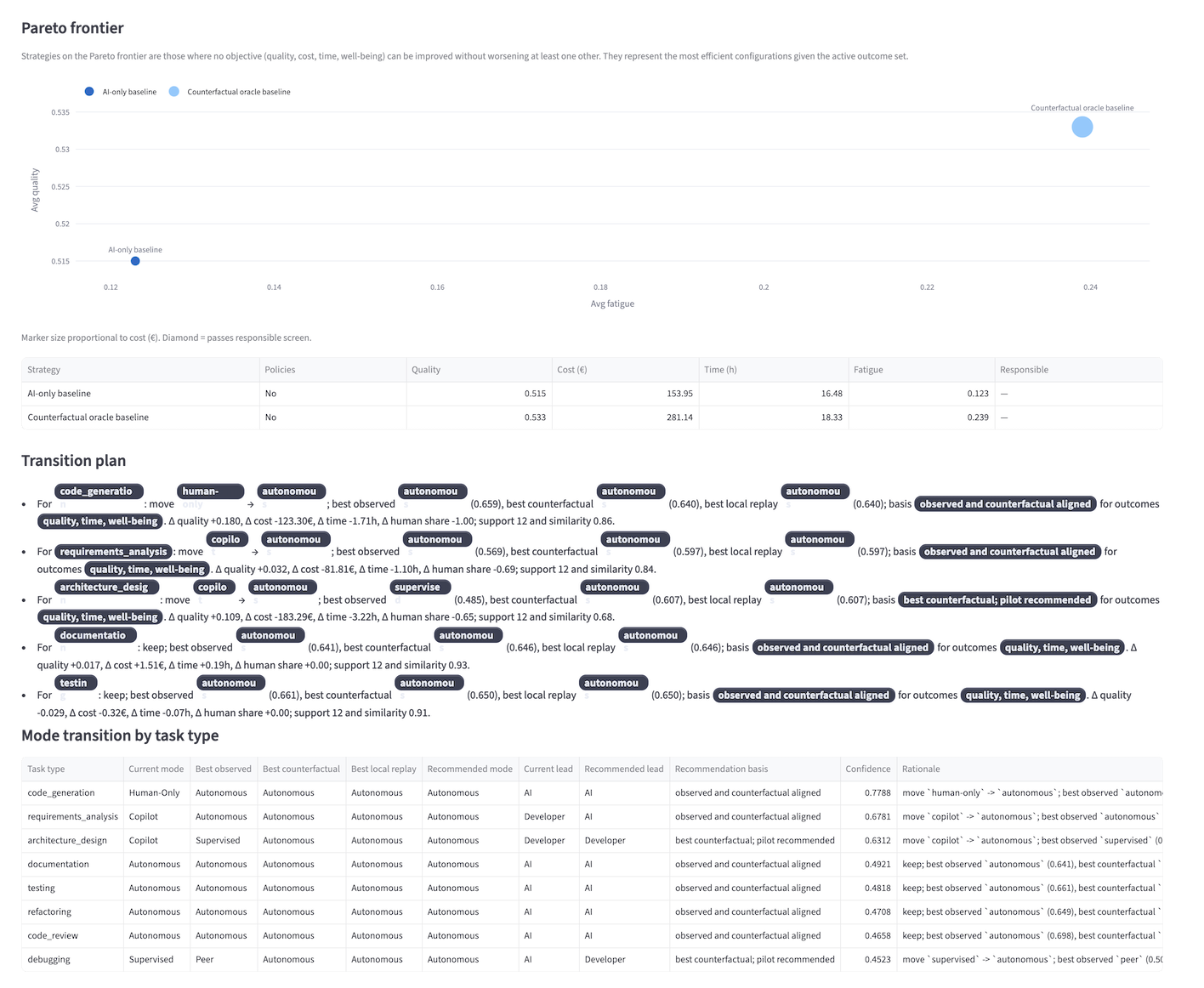}{%
\ui{Deployment planning} subtab showing (top to bottom): the Pareto
frontier scatter plot (quality vs.\ fatigue, marker size proportional to
cost, diamond = passes responsible screen), the Pareto frontier detail
table, and the \textbf{Mode transition by task type}
table.\label{fig:r2b}}

\begin{infobox}
\textbf{Key distinction.} The efficiency winner and the deployment
recommendation are often different strategies. The tool makes this separation
explicit: use the responsible screen result, not the ranking position, as
the basis for any deployment decision.
\end{infobox}

\paragraph{Recipe~3 --- Monitor co-evolution and detect skill degradation.}
\textit{Question:} is the allocation strategy preserving human skills over
the run, or is it creating deskilling exposure in critical cognitive
dimensions?

\begin{enumerate}
  \item Set \ui{Strategy} to \ui{LinUCB}, reward profile to \ui{Four Outcome},
        and co-evolution horizon to \ui{balanced}. Use a scenario with task
        diversity (e.g.\ \ui{Software / Standard Sprint}, 12~cycles, Senior
        profile).
  \item Run a full simulation: use \ui{Explore simulation} or
        \ui{Run sim + benchmark} from the \ui{Setup} wizard.
        \textbf{A benchmark-only run is not sufficient} --- the
        \ui{Co-evolution} tab requires the subtask-level trace produced by
        a simulation. Figure~\ref{fig:r3-overview} shows the kind of
        simulation-plus-results workflow needed for this recipe, ending with
        the \ui{Co-evolution} tab open on the loaded run.
  \item Open \ui{Co-evolution} and read its panels in this order:
        \begin{enumerate}[label=\alph*.]
          \item \textbf{Contribution attribution}: Figure~\ref{fig:r3a} shows
                how outcome quality is attributed across origin types.
                In the displayed run, \ui{AI-led} subtasks have the highest
                average quality (0.516 across 63 subtasks), while
                \ui{Human correction} covers the largest volume
                (107 subtasks) at a lower average quality of 0.473. The
                time-series is useful because it shows a sharp AI-attributed
                quality drop around cycle~4 (to roughly 0.21) even though the
                confidence signal $x_t$ remains high at about 0.97--0.98. That
                combination indicates a performance shock rather than a simple
                low-confidence episode.

          \item \textbf{Human skill dynamics}: start with the run-level
                task-type trajectory in Figure~\ref{fig:r3-overview}, which
                shows that several task types dip sharply around cycle~4 and
                then recover by cycles~6--8. Then switch to the cognitive
                dimension view in Figure~\ref{fig:r3b}, which is the cleaner
                governance diagnostic. In that figure, all five dimensions are
                improving rather than degrading: \term{technical depth} grows
                from 0.871 to 0.990 ($\Delta = 0.119$), \term{ambiguity} from
                0.863 to 0.939 ($\Delta = 0.076$), and even the weakest gain,
                \term{creativity}, still rises from 0.860 to 0.884
                ($\Delta = 0.025$). The correct interpretation is that this
                run shows short-term volatility by task type, but not
                deskilling in the modeled cognitive dimensions.

          \item \textbf{Capability ledger}: Figure~\ref{fig:r3c} translates
                capability change into an audit trail for inversion risk. The
                most important pattern in the example is temporal: cycle~1
                shows a large negative relative human gain (-0.4571) and one
                inversion candidate, but from cycle~5 onward the inversion
                candidate count drops to 0. By the latest cycle, the strongest
                comparative AI advantage remains in \term{repetitiveness}
                (0.1235), while \term{technical depth} is close to neutral
                (0.0056). This is the panel that tells the reader whether a
                temporary AI edge is consolidating into a structurally risky
                reversal.

          \item \textbf{Recovery planning}: Figure~\ref{fig:r3d} converts the
                ledger into action. In the example, only
                \term{repetitiveness} is flagged, with AI advantage 0.201 and
                the likely task types \ui{documentation} and \ui{testing}. The
                recommended response is explicit: \ui{Return to human ---
                high inversion risk}. The remaining dimensions all show
                \ui{No action needed}. This is the point where the user moves
                from diagnosis to intervention.
        \end{enumerate}
  \item Open \ui{Learning evolution} and inspect how task-type allocations
        changed under LinUCB relative to the initial heuristic.
        Figure~\ref{fig:r3e} shows a clear contextual reallocation pattern.
        \ui{architecture design} moves from \ui{Copilot} to
        \ui{Supervised}. \ui{code generation}, \ui{code review},
        \ui{debugging}, \ui{refactoring}, and \ui{testing} all move from
        \ui{Human Only} to \ui{Supervised}. \ui{documentation} remains
        \ui{Autonomous}, while \ui{requirements analysis} settles at
        \ui{Peer}. This is exactly the kind of task-type specialization that
        contextual allocation should reveal.

        If needed, use \ui{Cross-domain comparison}
        (Figure~\ref{fig:r1c-cross}) to test whether the same setup behaves
        differently across software, manufacturing, and healthcare.
\end{enumerate}

\screenshot[0.97\linewidth]{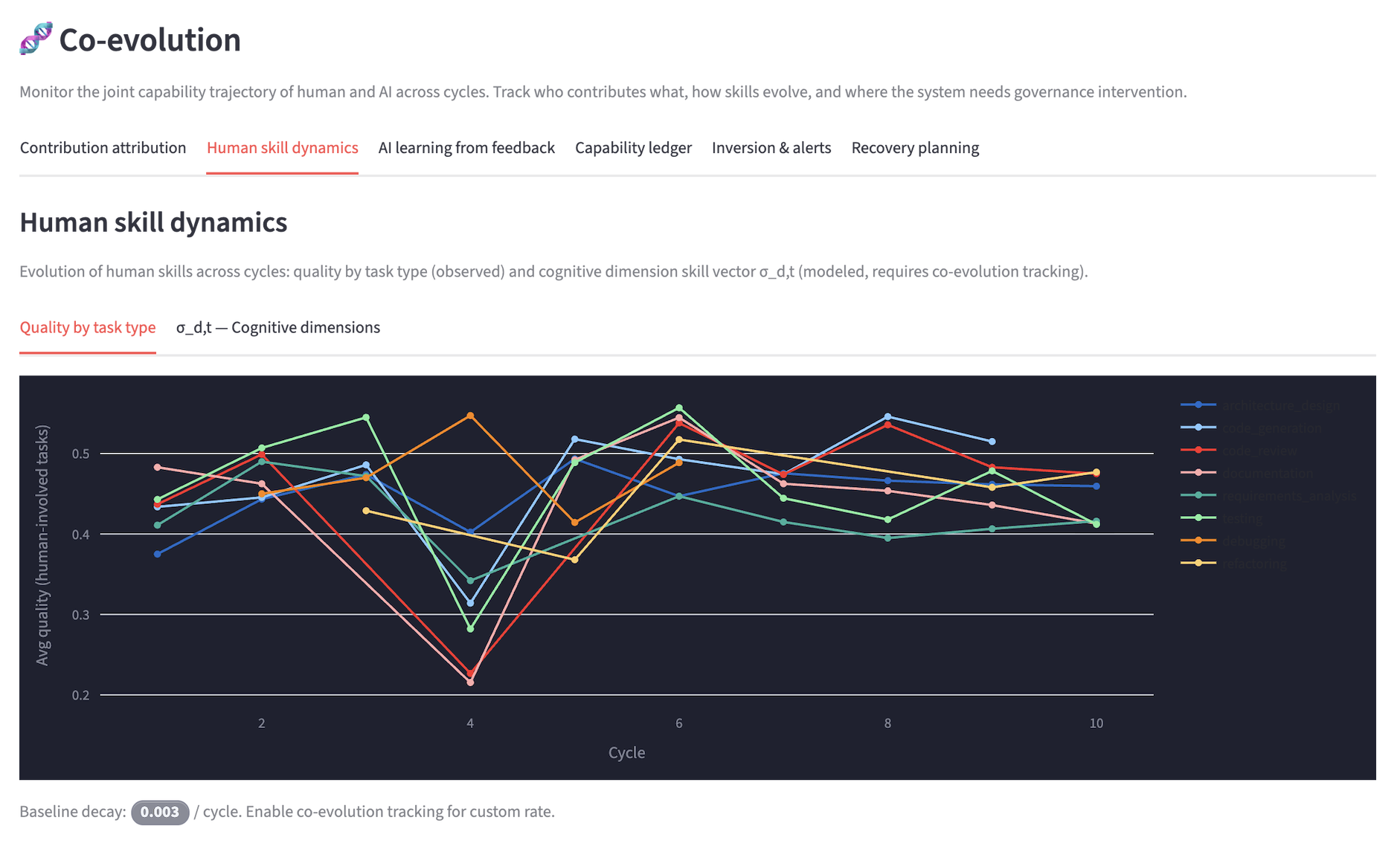}{%
Recipe~3 setup and entry into the \ui{Co-evolution} workflow for a LinUCB
simulation run, ending on the \ui{Human skill dynamics} task-type trajectory
view.\label{fig:r3-overview}}

\screenshot[0.97\linewidth]{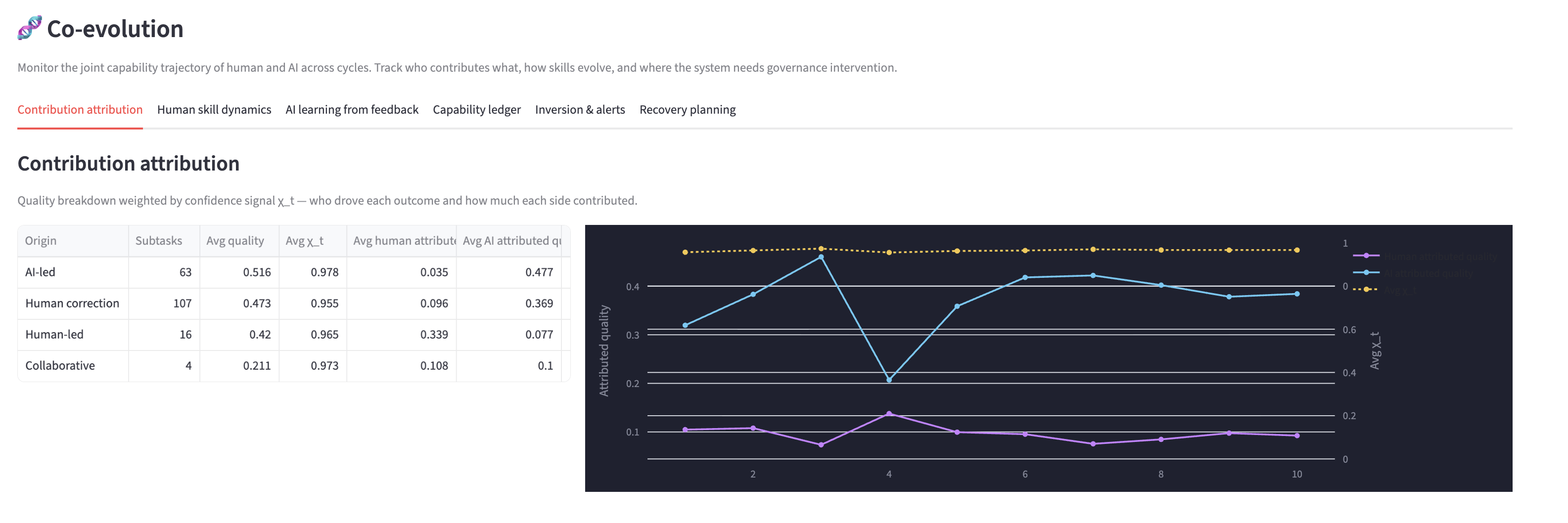}{%
\ui{Co-evolution} tab --- \ui{Contribution attribution} subtab, showing the
breakdown of outcome quality across AI-led, human-correction, human-led, and
collaborative origins, together with their cycle-level trajectories.
\label{fig:r3a}}

\screenshot[0.97\linewidth]{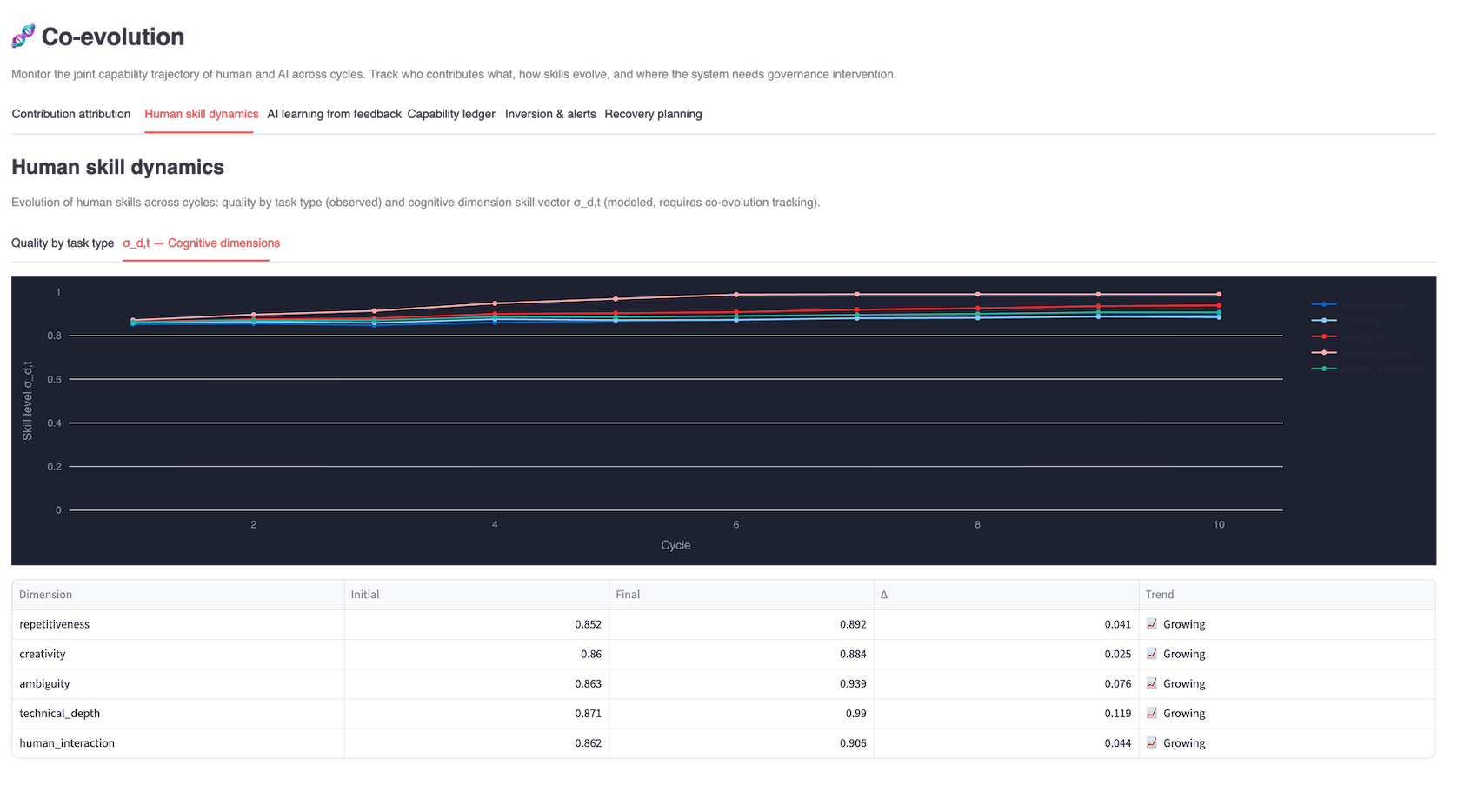}{%
\ui{Co-evolution} tab --- \ui{Human skill dynamics} subtab in
\term{Cognitive dimensions} view, showing modeled human skill trajectories by
dimension and the initial/final comparison table.
\label{fig:r3b}}

\screenshot[0.97\linewidth]{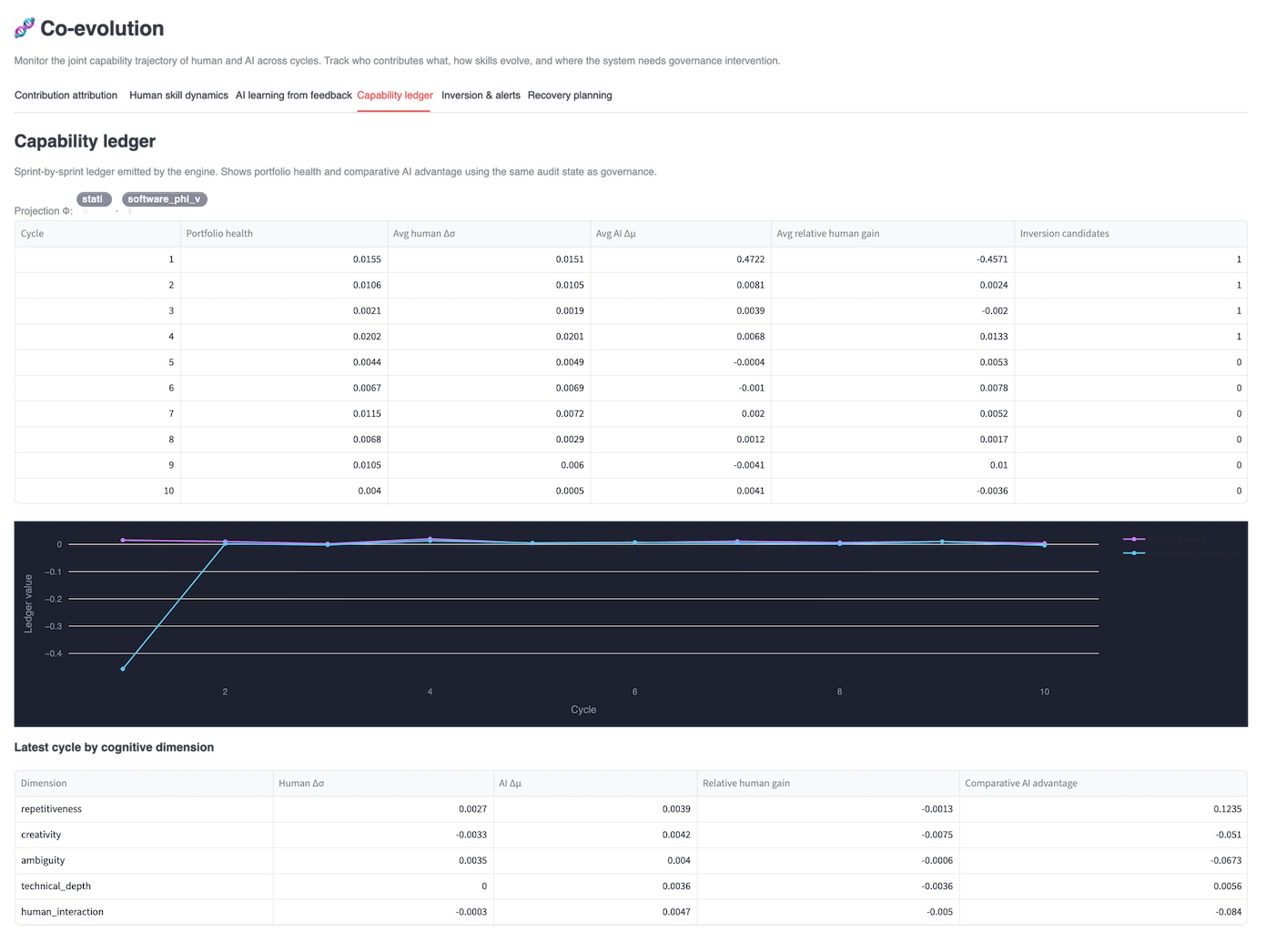}{%
\ui{Co-evolution} tab --- \ui{Capability ledger} subtab, showing sprint-level
ledger values, inversion-candidate tracking, and the latest per-dimension
comparative AI advantage snapshot.\label{fig:r3c}}

\screenshot[0.97\linewidth]{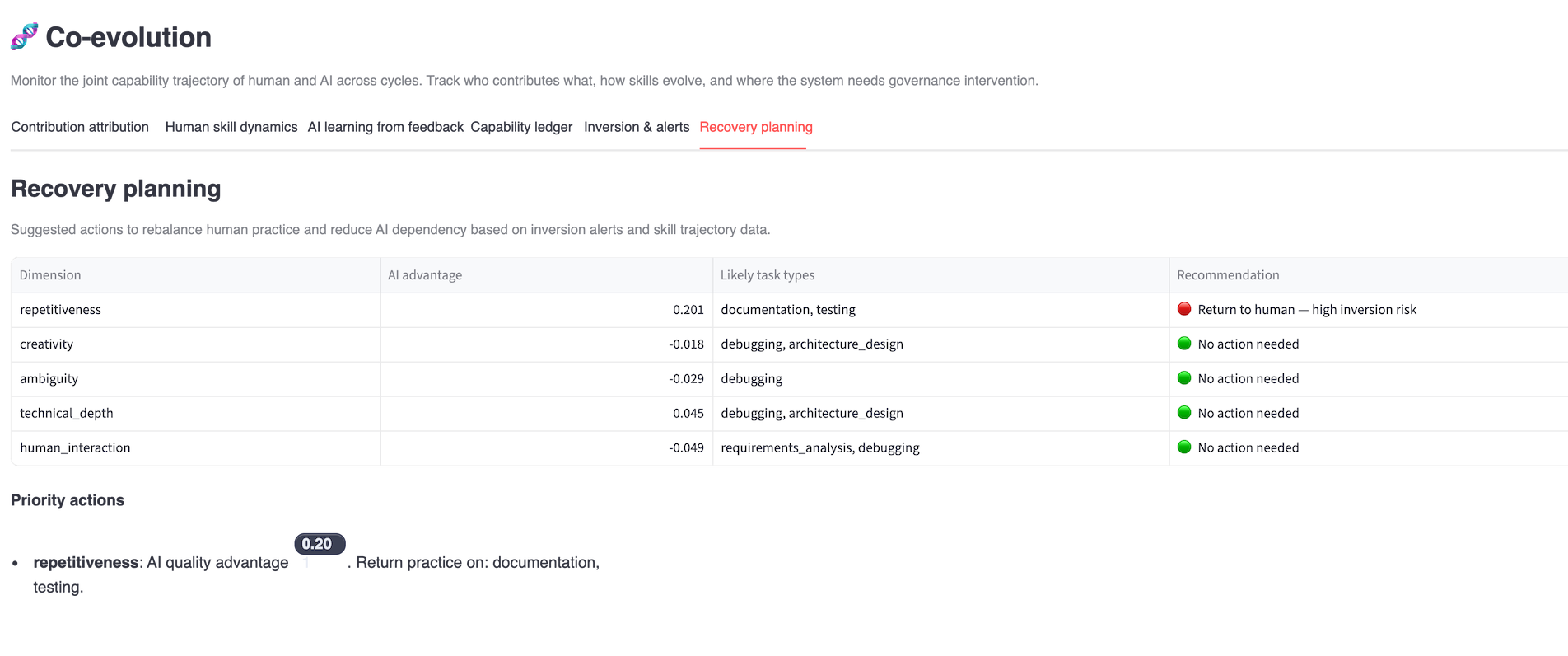}{%
\ui{Co-evolution} tab --- \ui{Recovery planning} subtab, mapping flagged
dimensions to likely task types and concrete human-rebalancing actions.
\label{fig:r3d}}

\screenshot[0.97\linewidth]{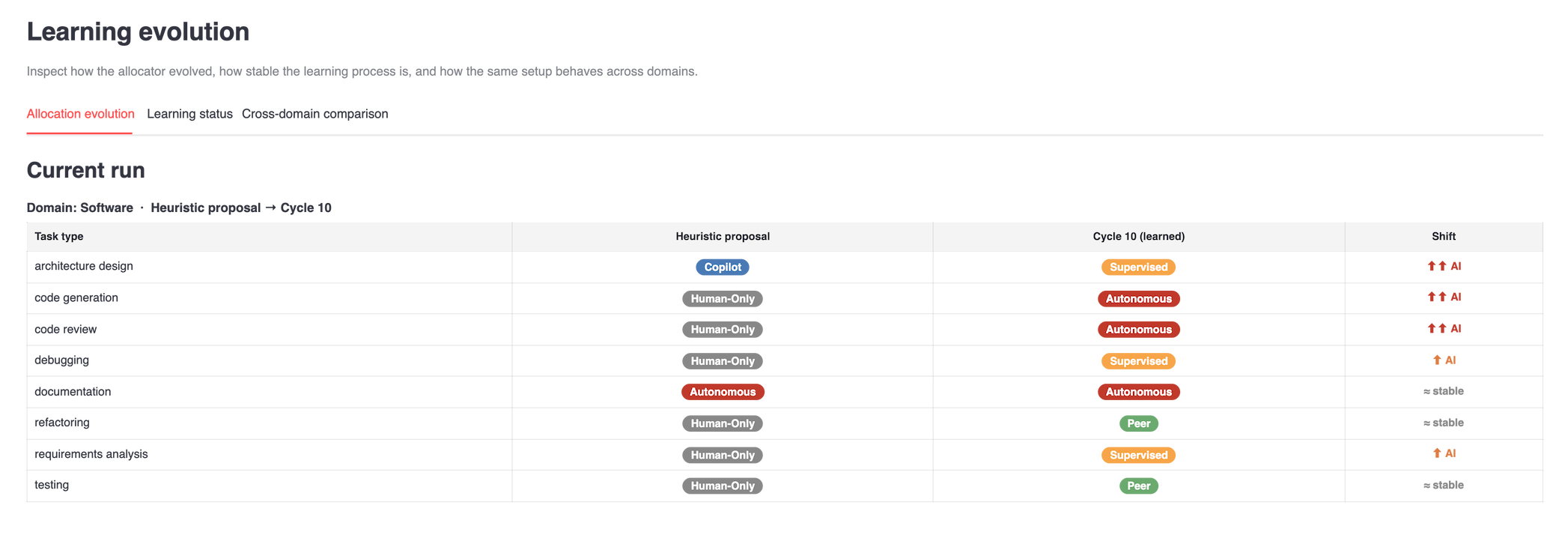}{%
\ui{Learning evolution} tab --- \ui{Allocation evolution} subtab, comparing
the initial heuristic proposal with the learned mode allocation by task type.
\label{fig:r3e}}

\begin{infobox}
\textbf{What to look for.} Treat Recipe~3 as a progression from diagnosis to
action. First verify whether attribution shocks are temporary
(Figure~\ref{fig:r3a}) or persistent. Then check whether the cognitive
dimensions are actually declining (Figure~\ref{fig:r3b}) rather than assuming
that a short-term quality dip implies deskilling. After that, use the ledger
and recovery panels (Figures~\ref{fig:r3c}--\ref{fig:r3d}) to decide whether
the observed AI advantage is operationally acceptable or requires rebalancing.
Finally, confirm in \ui{Learning evolution} (Figure~\ref{fig:r3e}) that LinUCB
has learned a task-type allocation pattern that is interpretable rather than
arbitrary.
\end{infobox}

\paragraph{Recipe~4 --- Track a persistent worker across sessions.}
\textit{Question:} how is a specific worker evolving across multiple
simulation sessions, and what fatigue or trust risks are projected in the
next planning horizon?

\begin{enumerate}
  \item Complete at least one simulation run for the target domain and
        scenario (e.g.\ \ui{Manufacturing / New Product Ramp-Up}, Thompson
        Sampling, 15~cycles, Operator profile).
  \item Open the \ui{Live Twin} tab. Assign a stable \ui{Worker ID}
        (e.g.\ \ui{rampup-op-01}), select the matching domain and profile,
        and enable \ui{Pre-sprint forecast}. Run the twin session. The tool
        stores the worker's fatigue level, skill state, trust score, and
        learning trace so that later sessions and forecasts start from the
        same accumulated state instead of from a clean slate.
  \item Read the resulting twin state before moving to planning.
        Figure~\ref{fig:r4a} shows the persistent worker snapshot for
        \ui{rampup-op-01}: a 4-sprint session with average quality
        $0.486$ and average fatigue $0.554$. The sprint table makes the
        intra-session pattern visible: fatigue drops to $0.192$ in
        sprint~2, then climbs back to $0.619$ and $0.714$ in sprints~3--4,
        while quality remains in a narrower band between $0.455$ and
        $0.532$.

        This is the point of the saved state: it records not only the final
        level, but also the shape of the recent trajectory that future
        forecasts will inherit.
  \item Check the governance panels in the same view. In
        Figure~\ref{fig:r4a}, \ui{Actionable alerts} reports no critical or
        warning events, but three informational alerts and
        \ui{100\% Model health}. That means the worker is not yet in an
        emergency state, but the twin is already surfacing signals worth
        monitoring.

        The \ui{Drift monitor} below then sharpens that reading:
        it marks \ui{Drift detected\allowbreak\ ---\allowbreak\ recalibration
        recommended}, with a quality error of $0.005$ and a larger fatigue
        error of $0.089$.

        Interpret this asymmetry carefully. The twin is still tracking quality
        well, but its fatigue projection is diverging enough from observations
        to justify recalibration before using the state for longer-horizon
        planning.
  \item Open the \ui{Planning} tab. Set a horizon of 6~sprints, select
        3~Monte Carlo samples, and associate the same Worker ID. Run the
        forward projection.
  \item Read the horizon chart in the same order the interface presents it.
        Figure~\ref{fig:r4b} starts with the alert badges: fatigue reaches the
        critical threshold at sprint~20, trust stays healthy, and AI share
        peaks at $100\%$, flagging a deskilling risk.

        Then read the chart:
        the orange fatigue trajectory rises from $0.042 \pm 0.007$ at
        sprint~17 to $0.337 \pm 0.043$ at sprint~19, spikes to
        $0.991 \pm 0.016$ at sprint~20, and only later falls back to
        $0.503 \pm 0.106$ and $0.174 \pm 0.111$. Because the chart also
        shows the warning line at $0.70$, the critical line at $0.90$, and
        the Monte Carlo uncertainty band around the fatigue curve, the user
        can see both \emph{when} the risk materialises and how uncertain the
        projection remains around that point.
  \item Use the table under the chart to confirm whether the alert is local
        or systemic. In Figure~\ref{fig:r4b}, projected quality stays in a
        relatively narrow interval from $0.401$ to $0.531$, and projected
        trust remains pinned at $1.000 \pm 0.000$ across the full horizon,
        so the dominant short-term problem is not generalized collapse but a
        fatigue spike concentrated at sprint~20.

        The \ui{Fatigue risk} column confirms this reading by marking
        sprint~20 as \ui{critical} while the other projected sprints remain
        \ui{ok}.
  \item Treat the two panels together as a causal chain. Figure~\ref{fig:r4a}
        explains why the forecast deserves attention: the worker already shows
        rising late-session fatigue and a non-trivial fatigue drift signal.
        Figure~\ref{fig:r4b} then extends that same saved state into the next
        six sprints and shows what happens if the operating conditions are not
        rebalanced in time.

        Read the pair as state diagnosis first, forward risk translation
        second. That sequence makes the logic of the recipe much easier to
        follow.
\end{enumerate}

\screenshot[0.97\linewidth]{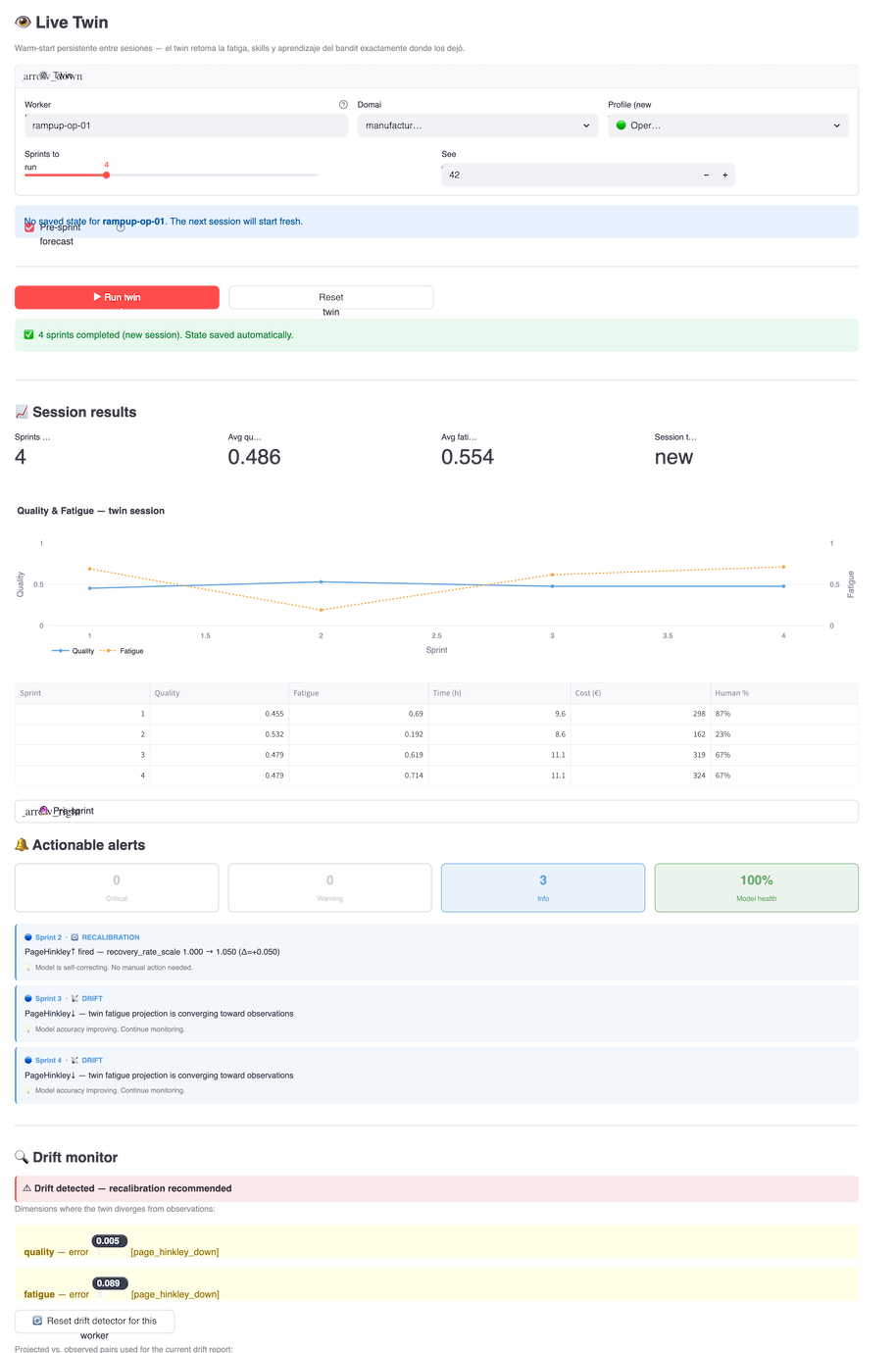}{%
Live Twin tab --- persistent worker state for \ui{rampup-op-01}, with
session metrics, alert summary, and drift monitor after repeated
sessions.\label{fig:r4a}}

\screenshot[0.97\linewidth]{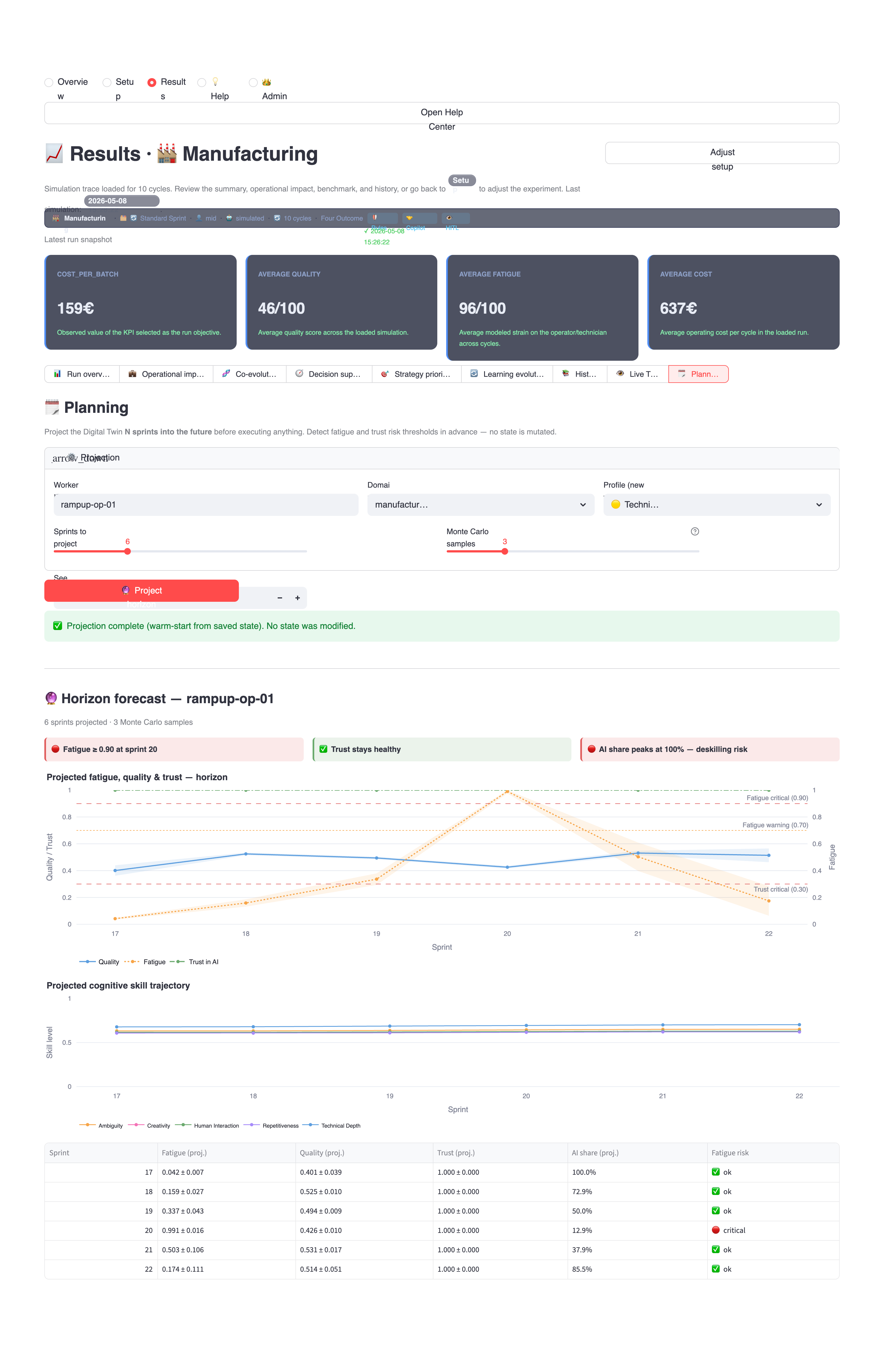}{%
Planning tab --- 6-sprint horizon forecast with fatigue threshold
crossing, trust status, deskilling alert, and Monte Carlo uncertainty
band around the projected trajectories.\label{fig:r4b}}

\begin{infobox}
\textbf{What to look for.} Read persistent-worker analysis in two passes.
First ask whether the twin state is trustworthy enough to project:
Figure~\ref{fig:r4a} should show stable model health and alerts that are
informative rather than catastrophic. Then ask whether the horizon forecast
isolates a specific risk mechanism or a general breakdown. In
Figure~\ref{fig:r4b}, trust and quality remain comparatively stable while
fatigue alone crosses the $0.90$ critical line, so the correct conclusion is
not ``the worker is failing'' but ``the current operating pattern is likely to
produce a fatigue spike unless the plan is rebalanced before sprint~20.''
\end{infobox}

Advanced benchmark-suite operation, task-level deployment prescription, and
allocator-selection heuristics are collected in Appendix~\ref{app:advanced-ops}
so that the main text can stay focused on representative workflows.

\section{How to Turn Results into Decisions}
\label{sec:interpretation}

\sectionrule

HAAS Studio is most useful when its outputs are treated as a decision workflow
rather than as disconnected charts. The core practical question is always the
same: \emph{given what the tool is showing, what should the user do next?}
This section organizes that question into recurring decision patterns,
action-oriented reading rules, common mistakes, and a compact decision matrix.

\begin{infobox}
\textbf{Interpretation rule.} First ask \emph{did the run or strategy pass the
contract?} Then ask \emph{did it also survive the responsible screen and
watchlist?} Only after those two filters should the user compare efficiency or
act on the transition plan.
\end{infobox}

\subsection{Decision patterns by user intent}

In practice, most user goals fall into one of five patterns:
\begin{itemize}
  \item \textbf{Single-run diagnosis.} Use \ui{Run overview} and
        \ui{Operational impact} when the immediate goal is to understand whether
        one specific configuration behaved as intended.
  \item \textbf{Strategy selection.} Use \ui{Decision support} and
        \ui{Strategy priorities} when the goal is to choose one strategy that is
        both effective and deployable under the active contract.
  \item \textbf{Deployment specification.} Use \ui{Deployment decision},
        \ui{Deployment planning}, and the watchlist when the goal is not merely
        to rank strategies, but to derive a task-level operating model.
  \item \textbf{Worker sustainability.} Use \ui{Co-evolution},
        \ui{Live Twin}, and \ui{Planning} when the main question is not only
        current performance, but whether the present posture is sustainable over
        repeated sessions.
  \item \textbf{Policy adjustment.} Use the watchlist, corrective actions,
        replay surfaces, and saved-run comparison when the recommendation is
        conditional or non-deployable and the user needs to understand what to
        change.
\end{itemize}

These five patterns cover most real uses of the tool. A reader who identifies
their intent early will reach the relevant surfaces faster and avoid mixing
diagnosis, comparison, and deployment interpretation in the same reading pass.

\subsection{From signal to action}

The most reliable way to convert outputs into decisions is to move through the
same ordered filter every time:
\begin{enumerate}
  \item \textbf{Confirm feasibility.} Check whether the run or strategy meets
        the active contract. If it fails here, do not interpret ranking or
        allocation elegance as deployability.
  \item \textbf{Check responsibility.} A feasible strategy may still fail the
        responsible screen or produce a watchlist signal that makes deployment
        unsafe or unstable.
  \item \textbf{Inspect transition logic.} If the strategy is feasible and
        responsible, open the transition plan to see how the recommendation is
        distributed across task types rather than treating the result as a
        single scalar verdict.
  \item \textbf{Stress-test the recommendation.} Use \ui{History} and
        \ui{Strategy priorities} to see whether the recommendation remains
        acceptable when priorities or one operating assumption change.
  \item \textbf{Decide whether to accept, revise, or escalate.} Accept a stable
        deployable strategy; revise a conditional one; escalate to corrective
        actions, policy overrides, or a new benchmark if no strategy remains
        acceptable.
\end{enumerate}

In action terms, the most common branches are:
\begin{itemize}
  \item if a strategy passes the contract and the responsible screen, move to
        task-level deployment planning;
  \item if a strategy passes the contract but fails the watchlist or
        responsibility filters, treat it as conditionally useful and inspect
        the problematic task types before accepting it;
  \item if the recommendation is conditional, compare saved variants and use
        sensitivity views before freezing a deployment posture;
  \item if no strategy is feasible, return to the contract and corrective
        actions before changing allocators or domains.
\end{itemize}

\begin{infobox}
\textbf{Five-step decision rule.} Confirm contract compliance; check
responsibility and watchpoints; inspect the transition plan; stress-test the
recommendation in \ui{History} or \ui{Strategy priorities}; only then accept or
revise the operating model.
\end{infobox}

\subsection{Common interpretation errors}

The same misreadings appear repeatedly in practice:
\begin{itemize}
  \item \textbf{Confusing benchmark winner with deployment recommendation.}
        The first is the best raw performer; the second is the best option that
        survives the active contract and governance screen.
  \item \textbf{Reading high AI share as a success signal.} A high automation
        share can still create fatigue, deskilling, trust, or governance
        problems.
  \item \textbf{Treating the oracle counterfactual as an operational target.}
        The oracle is a diagnostic ceiling, not a deployable policy.
  \item \textbf{Ignoring the watchlist because the top-level recommendation is
        green.} A strategy can be deployable and still sit close to unstable or
        undesirable boundaries.
  \item \textbf{Interpreting one good run as robustness.} Robustness should be
        checked in \ui{History}, \ui{Strategy priorities}, or benchmark suites.
  \item \textbf{Changing too many inputs at once.} If domain, profile,
        allocator, and contract all change together, the comparison is hard to
        interpret.
\end{itemize}

Two confusions deserve special emphasis. First, \emph{green} does not mean
\emph{optimal}; it means \emph{acceptable under the current decision frame}.
Second, \emph{first in rank} does not mean \emph{first to deploy}; the whole
point of the governed workflow is to preserve that distinction.

\subsection{Practical heuristics}

The following heuristics are useful when the result is not immediately clear:
\begin{itemize}
  \item \textbf{Repeat a benchmark} when the top two strategies are close and
        small priority changes invert the ranking.
  \item \textbf{Change the allocator} when the learning status remains unstable,
        the regret curve does not flatten, or the scenario is clearly
        contextual/non-stationary and the current strategy is not adapting.
  \item \textbf{Adjust guardrails} when every strategy fails feasibility for
        the same reason, especially if the thresholds are tighter than the
        operating reality the user is trying to model.
  \item \textbf{Use \ui{Live Twin} and \ui{Planning}} whenever a strategy looks
        good in one run but may create medium-horizon fatigue, trust, or
        deskilling exposure for a persistent worker.
  \item \textbf{Use policy rules or task locks} when the same task type
        repeatedly appears in the watchlist or systematically diverges from what
        the user considers an acceptable operating mode.
\end{itemize}

A useful rule of thumb is to change only one major decision dimension at a
time: allocator, contract, scenario, or worker profile. Otherwise, the user
often learns that the result changed, but not \emph{why} it changed.

\subsection{Decision matrix}

Table~\ref{tab:decision-matrix} summarizes the most common signals produced by
the tool, how they should be interpreted, where they should be verified, and
what the next action should usually be.

{\small
\begin{longtable}{@{}>{\raggedright\arraybackslash\bfseries}p{0.20\linewidth}>{\raggedright\arraybackslash}p{0.22\linewidth}>{\raggedright\arraybackslash}p{0.22\linewidth}>{\raggedright\arraybackslash}p{0.28\linewidth}@{}}
\caption{Decision matrix: how to turn common HAAS Studio outputs into next actions.}
\label{tab:decision-matrix}\\
\toprule
Observed signal & What it usually means & Where to verify it & Recommended next action \\
\midrule
\endfirsthead
\toprule
Observed signal & What it usually means & Where to verify it & Recommended next action \\
\midrule
\endhead
\midrule
\multicolumn{4}{r}{\small Continued on next page} \\
\endfoot
\bottomrule
\endlastfoot
Top strategy is deployable and stable & The current operating frame admits a
clear recommendation & Decision support, Strategy priorities, History & Move to
transition planning and define the operating specification \\

Top strategy is conditional & The strategy is useful, but depends on load,
contract, or profile assumptions & Deployment decision, watchlist, History &
Stress-test one variable at a time and document the condition before adoption \\

No feasible strategy & The contract is stricter than the current strategy set
can satisfy & Operational impact, Decision support, Corrective actions & Review
the contract, guardrails, or scenario assumptions before changing the whole
model \\

Feasible but not responsible & The benchmark result is acceptable on KPI terms
but unsafe on co-evolution or human-sustainability terms & Responsible screen,
watchlist, Co-evolution, Planning & Treat the strategy as non-deployable until
the flagged risk is mitigated \\

Strong ranking instability under priorities & The recommendation depends heavily
on value weighting & Strategy priorities, History & Keep more than one strategy
as viable and make the priority trade-off explicit \\

Deskilling alert with good productivity & The current operating posture may be
efficient but harmful to long-run human capability & Co-evolution, watchlist,
Planning & Reduce autonomy for the affected task types or add task-level
constraints \\

Persistent divergence from the oracle on one task type & The allocator may be
misassigning a specific class of tasks & Learning evolution, replay, oracle
views & Re-examine that task type and consider policy overrides or allocator
change \\
\end{longtable}
}

\paragraph{Mini-example.}
Suppose Strategy~A ranks first in the benchmark but fails the responsible
screen because it pushes fatigue and deskilling exposure too close to the
configured limit, while Strategy~B ranks second and remains feasible,
responsible, and stable under moderate changes in priorities. The correct
reading is not that Strategy~A is the best choice. The correct reading is that
Strategy~A is the efficiency ceiling under the current benchmark, while
Strategy~B is the candidate deployment option because it survives the full
decision workflow.

\section{Study Protocols and Analytical Evidence}
\label{sec:evidence}

\sectionrule

The configuration space of HAAS Studio---combining domain, scenario, allocator,
governance contract, and human profile---admits a large number of distinct study
protocols. The companion manual documents five representative cases that jointly
exercise the main features of the tool; Table~\ref{tab:cases} summarizes them
from the perspective of this paper. The following subsections describe three of
these cases from an analytical reading perspective: what question motivated each
one, and what the tool's surfaces reveal in response.

\subsection{Representative case-study protocols}

\begin{table}[H]
\centering
\small
\caption{Representative built-in study protocols documented for HAAS Studio.}
\label{tab:cases}
\begin{tabularx}{\linewidth}{@{}>{\bfseries}lYp{2.8cm}@{}}
\toprule
Case & Main purpose & Key features \\
\midrule
Software / Maintenance / UCB1 &
  Produce a full recommendation for a maintenance team under quality and cost
  constraints. &
  Recommendation workflow, contract reading, transition plan. \\
Manufacturing / Quality Crisis / UCB1 &
  Benchmark strategies in a safety- and quality-sensitive production context. &
  Four-outcome benchmark, watchlist, stability validation. \\
Software / Standard Sprint / LinUCB &
  Examine contextual allocation and co-evolution in a varied sprint. &
  LinUCB, Co-evolution, Learning evolution, Planning. \\
Manufacturing / New Product Ramp-Up / Thompson &
  Inspect deskilling risk and longitudinal worker evolution in a ramp-up phase. &
  Live Twin, Planning, reduced contracts, deskilling focus. \\
Healthcare / Alarm Fatigue / Discounted-UCB &
  Explore safety-oriented allocation under repeated false-positive alerts. &
  Healthcare domain, temporal adaptation, clinical guardrails. \\
\bottomrule
\end{tabularx}
\end{table}

For a tool paper, these protocols matter because they are effectively curated
demonstration scripts. They show that the artifact is not just a configurable
engine but also a guided experimental environment with repeatable walkthroughs.

\subsection{Software Maintenance: recommendation under guarded quality}

The \emph{Software / Maintenance} protocol asks a practical question: how much
routine work can be absorbed by AI without breaking a quality floor or losing
decision traceability? In the documented baseline, the user selects the
\emph{Maintenance} scenario, a \emph{Mid} developer profile, the \ui{UCB1}
allocator, a \emph{Quality First} objective, explicit guardrails over
\ui{cost\_per\_feature} (target $\leq$\,€190), minimum quality (0.72), and
maximum fatigue (0.45), and runs in \ui{Simulation + Benchmark} mode so that
both the operational trace and the \ui{Decision support} tab are populated.

From a tool-paper perspective, this case is important because it exercises the
full decision path:
\begin{enumerate}
  \item the setup wizard captures the operational question and contract;
  \item the run produces cycle-level traces of quality, fatigue, and human/AI
        distribution;
  \item the results tabs show whether documentation, testing, and refactoring
        move toward AI-led or shared execution;
  \item the decision layer compares strategies and distinguishes efficiency from
        deployment acceptability;
  \item the final output is a task-level transition plan rather than a single
        scalar score.
\end{enumerate}

This illustrates the intended use of HAAS Studio: not a binary verdict on AI
adoption, but a graded answer at the task level.

Figure~\ref{fig:sm-ranking} makes that distinction concrete. The efficiency
winner is \ui{AI-only baseline}, with a lead time of $14.6833$~h, average
quality $0.544$, average fatigue $0.101$, and average cost \texteuro$128.78$.
Yet it falls to responsible rank~9 with a responsible score of $0.3962$. By
contrast, \ui{UCB1 + policies off} is only efficiency rank~7, with a slower
lead time of $20.8271$~h, higher fatigue of $0.737$, and higher cost of
\texteuro$618.33$. Still, it rises to responsible rank~1 and is the row marked
as both \ui{Acceptable} and \ui{Reasonable}.

The point is therefore not ``which strategy is fastest?'' It is ``which
strategy remains admissible once efficiency is filtered through the
contract?''.

\emph{Which parts of the workflow can move first} is answered by the
transition plan. Figure~\ref{fig:sm-transition} shows that the recommended
changes are not uniform.

\ui{architecture\_design} is the clearest candidate for movement. It shifts
from \ui{Human Only} to \ui{Autonomous}, with $\Delta$quality $+0.161$,
$\Delta$cost $-$\texteuro$113.34$, $\Delta$time $-1.37$~h, support~12, and
similarity $0.79$. By contrast, \ui{documentation}, \ui{testing},
\ui{refactoring}, and \ui{code\_review} already remain in \ui{Autonomous}.
Their observed and counterfactual evidence aligns, with best-observed quality
values of $0.617$, $0.589$, $0.631$, and $0.676$, respectively.

The same panel also shows where caution is still needed. \ui{debugging}
changes from \ui{Supervised} to \ui{Autonomous} with confidence $0.4617$.
\ui{requirements\_analysis} moves from \ui{Human Only} to \ui{Autonomous} with
confidence $0.4512$. \ui{code\_generation} stays \ui{Supervised} with the
lowest displayed confidence ($0.3882$). This is why the output is a transition
map rather than a blanket ``more AI'' recommendation.

\emph{Under what conditions} is answered by the contract. In
Figure~\ref{fig:sm-deployment}, Stage~1 starts from 11 benchmarked strategies,
Stage~2 leaves only 1 passing the acceptability screen, and Stage~3 issues the
responsible recommendation \ui{UCB1 + policies off} with a
\ui{CONDITIONAL} deployment signal rather than an unconditional go-ahead. The
recommended configuration is summarized with quality $50.0\%$, average fatigue
$0.737$, human participation $29.5\%$, deskilling risk $0.000\%$, shared-mode
use $43.9\%$, responsible score $68.7\%$, and $0.000$ governance violations.

Those values show exactly what the contract is tolerating and what it is still
watching.

\emph{With what residual risks} is answered by the same decision-support
workspace. Figure~\ref{fig:sm-deployment} does not hide the remaining
exposure. The recommendation is accompanied by watchpoints and by concrete
``How to reach DEPLOYABLE'' adjustments.

The panel suggests shifting 10 subtasks from \ui{Autonomous} to
\ui{Supervised} to increase human participation, 8 subtasks from
\ui{Human Only} to \ui{Copilot} to reduce fatigue, and 14 subtasks from
\ui{Autonomous} to \ui{Peer} to limit deskilling.

Residual risk is therefore made operational. The tool does not just rank a
strategy; it shows which specific collaboration-mode changes would be needed
to convert a conditional recommendation into a safer deployment posture.

\screenshot[0.97\linewidth]{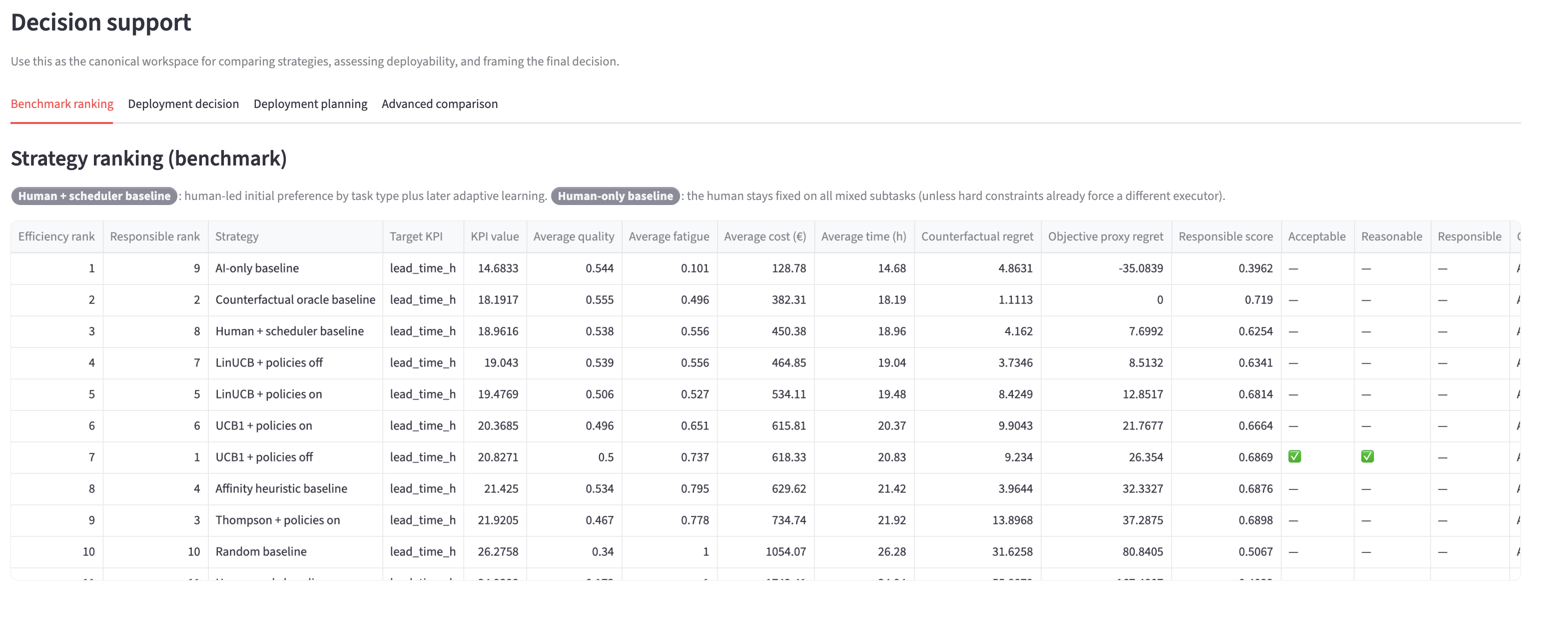}{%
Software / Maintenance --- \ui{Benchmark ranking} subtab: all strategies
ranked by efficiency and responsible criteria under the configured
contract.\label{fig:sm-ranking}}

\screenshot[0.97\linewidth]{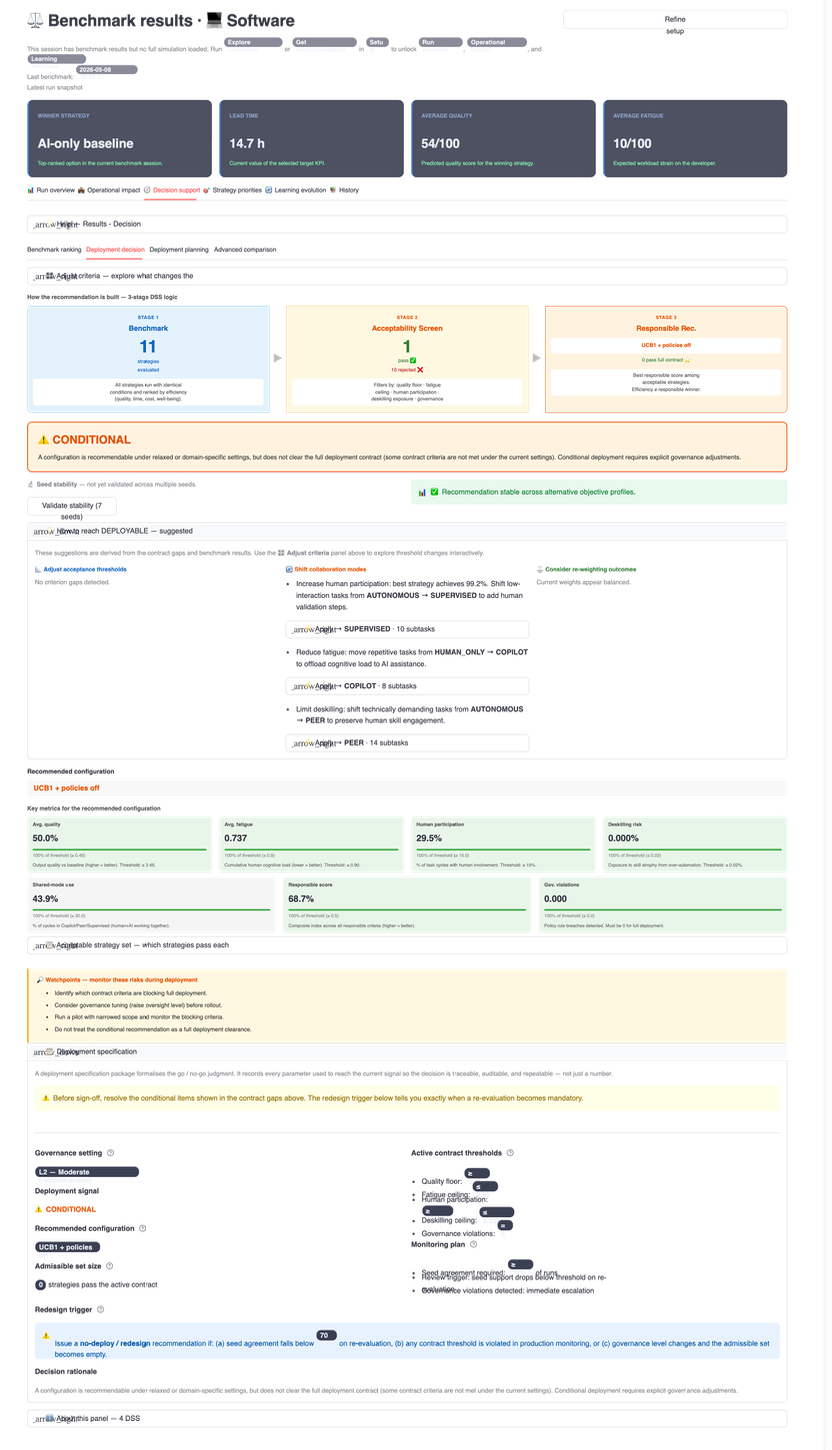}{%
Software / Maintenance --- \ui{Deployment decision} subtab: recommendation
card, acceptability screens, deployment signal, recommended configuration, and
monitoring plan for the selected strategy.\label{fig:sm-deployment}}

\screenshot[0.97\linewidth]{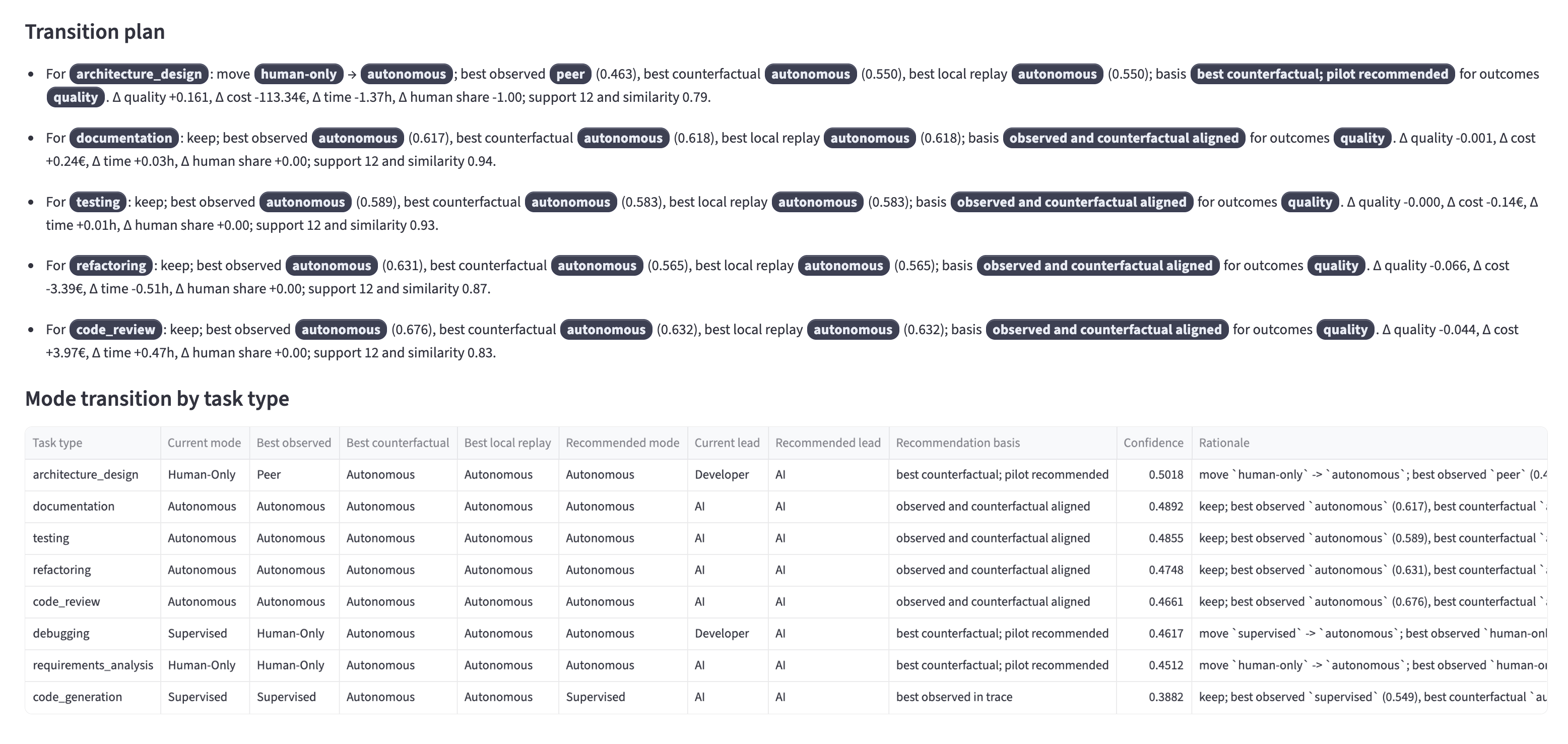}{%
Software / Maintenance --- \ui{Deployment planning} subtab: task-level
transition plan showing target collaboration mode, confidence, and recommended
action for each task type.\label{fig:sm-transition}}

\subsection{Manufacturing Quality Crisis: benchmark plus deployment reading}

The \emph{Manufacturing / Quality Crisis} protocol highlights a different
strength of the tool. The documented baseline uses the \emph{Quality Crisis}
scenario, an \emph{Engineer} operator profile, the \ui{UCB1} allocator, a
\emph{Four-outcome} objective, and guardrails over \ui{cost\_per\_batch}
(target $\leq$\,€220), minimum quality (0.78), and maximum fatigue (0.40),
running in \ui{Simulation + Benchmark} mode so that both the operational trace
and the \ui{Decision support} tab are populated.
The scenario mixes automatable activities such as visual inspection with
safety-critical tasks that must retain stronger human control, and the
objective is not merely to rank strategies by efficiency but to identify which
ones remain deployable under the full contract.

This case is especially useful for demonstrating three features:
\begin{itemize}
  \item \textbf{domain-specific structure}: the task catalog encodes real
        distinctions between inspection, diagnosis, and safety management;
  \item \textbf{governed recommendation}: under the default contract
        (quality $\geq 0.78$, fatigue $\leq 0.40$), the Quality Crisis scenario
        is likely to produce a \emph{non-deployable} or \emph{conditional}
        verdict rather than a clean approval---this is the intended result, and
        it shows the governance layer acting as a real filter rather than a
        cosmetic label; and
  \item \textbf{robustness workflow}: after the initial benchmark, the user
        re-runs with two or three different seeds to verify that the strategy
        ranking is stable and not an artifact of a single random draw; a second
        variant with tighter machine availability or a different company profile
        then checks whether the recommended strategy still holds under changed
        operational conditions.
\end{itemize}

This illustrates how HAAS Studio can support an operational question that is
common in practice: not merely which strategy is most efficient, but which one
remains defensible under quality and safety pressure. Figure~\ref{fig:manufacturing-quality-decision}
shows the benchmark and deployment reading for this case with the deployability
badge, acceptability screen failures, and the KPI watchlist visible, demonstrating
the governance layer acting as a real filter that produces non-deployable or
conditional verdicts rather than silently re-ranking strategies.

\screenshot[0.92\linewidth]{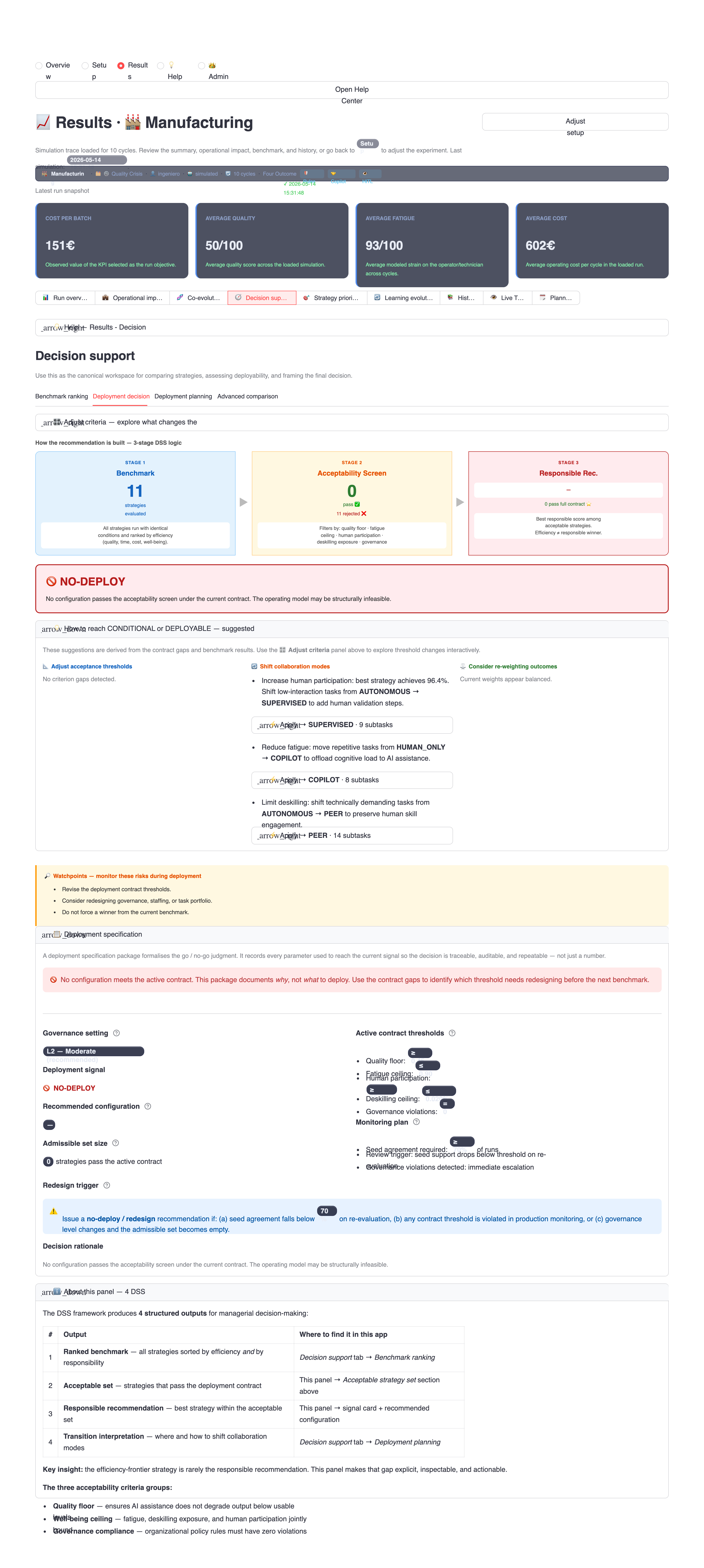}{%
Manufacturing / Quality Crisis case study. Recommended view: benchmark and
deployment reading under quality and safety constraints.\label{fig:manufacturing-quality-decision}}

The captured \ui{Deployment decision} screen should be read from top to bottom.
The three-stage DSS strip first shows the benchmark pool (11 evaluated
strategies), then the acceptability screen, where no strategy passes and all
benchmarked configurations are rejected, and finally the responsible
recommendation stage, which remains empty because there is no admissible
candidate. The red \ui{NO-DEPLOY} banner is therefore the central result: the
tool does not simply select the best-scoring strategy, but refuses deployment
when the active contract cannot be satisfied. The lower panels explain why the
rejection is actionable rather than opaque. The suggestion panel separates
possible interventions into threshold adjustment, collaboration-mode shifts,
and objective re-weighting; the yellow watchpoint panel warns the user not to
force a benchmark winner; and the deployment specification records the active
contract thresholds and the no-deploy decision. The KPI cards at the top keep
the operational drivers visible, including quality, fatigue, and
\ui{cost\_per\_batch}, so the rejection can be traced back to concrete
deployment criteria instead of to an unexplained ranking score.

\subsection{Advanced scenarios: co-evolution, deskilling, and prospective planning}

The more advanced documented cases broaden the picture. The
\emph{Software / Standard Sprint / LinUCB} protocol runs in \ui{Simulation}
mode and emphasizes the \term{Co-evolution} and \term{Learning evolution} tabs,
showing how contextual allocation can specialize by task type and how capability
balance evolves over time. The \emph{Manufacturing / New Product Ramp-Up /
Thompson Sampling} protocol also runs in \ui{Simulation} mode but requires
\ui{Live Twin} to be enabled, which activates the \term{Live Twin} and
\term{Planning} tabs and makes explicit the link between present allocation and
future worker sustainability. Finally, the \emph{Healthcare / Alarm Fatigue /
Discounted-UCB} case runs in \ui{Simulation} mode and shows that the same
artifact can express domain-specific risk logic in a clinical setting.

Taken together, these cases show that HAAS Studio is not limited to one
demonstration style. It supports:
\begin{itemize}
  \item pointwise recommendation,
  \item multi-strategy benchmark comparison,
  \item longitudinal worker monitoring (Figure~\ref{fig:advanced-coevolution}
        shows the Co-evolution or Learning evolution view with capability or
        allocation changes across cycles, illustrating how the tool tracks
        the human--AI capability balance over time), and
  \item forward-looking risk planning (Figure~\ref{fig:advanced-twin} shows
        the Live Twin or Planning view with worker state persistence or
        horizon projections, demonstrating that HAAS Studio operates as a
        multi-session longitudinal environment rather than a one-shot
        simulator).
\end{itemize}

\screenshot[0.92\linewidth]{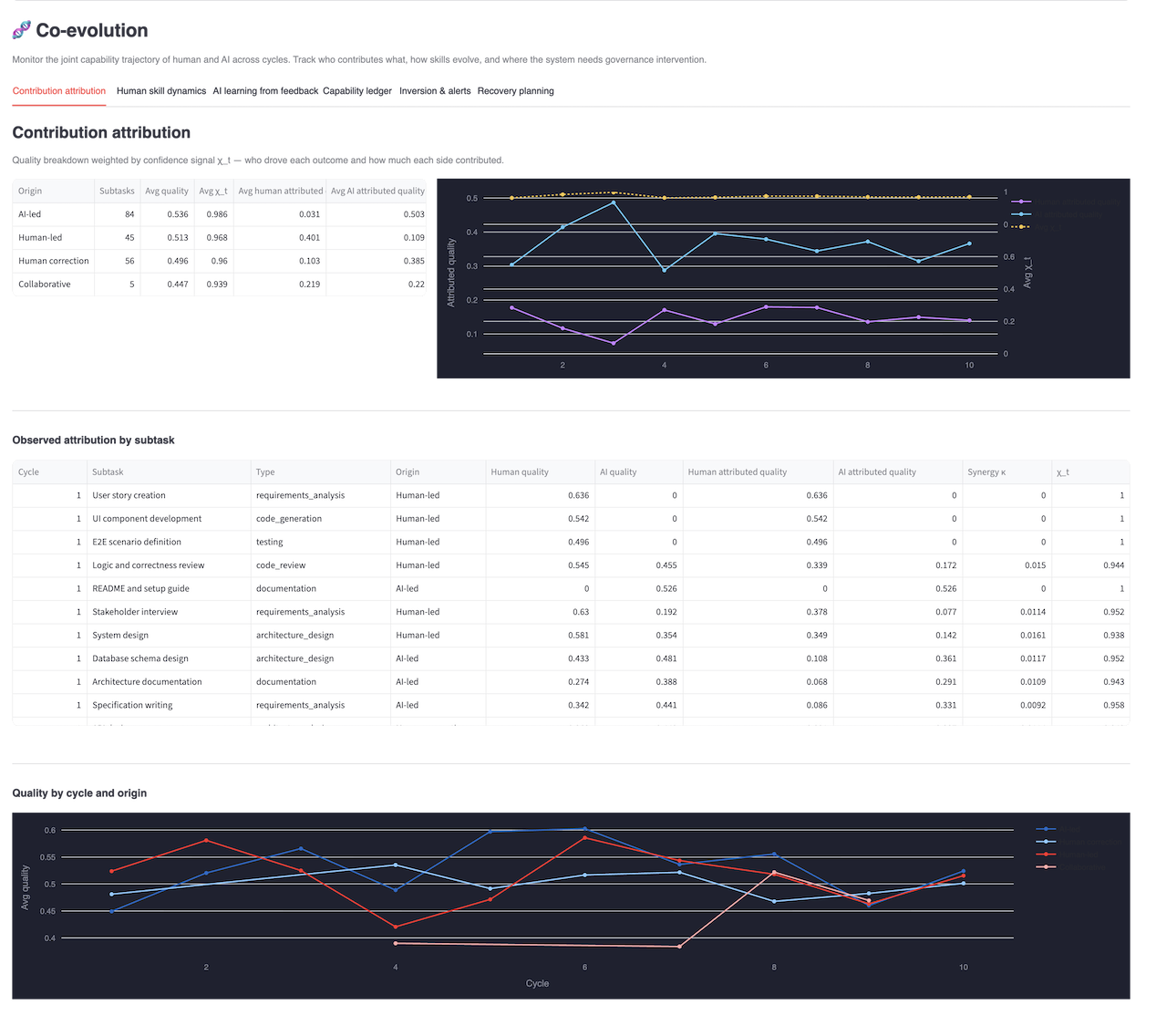}{%
Advanced analysis view. Recommended target: either \term{Co-evolution} or
\term{Learning evolution}, showing how capabilities or allocations change over
time.\label{fig:advanced-coevolution}}

Figure~\ref{fig:advanced-coevolution} shows the \ui{Co-evolution} tab in its
\ui{Contribution attribution} view. The upper summary table separates outcomes
by origin---AI-led, human-led, human correction, and collaborative---and reports
how much quality is attributed to each side after applying the attribution
confidence signal. The adjacent trend chart makes the same distinction temporal:
the AI-attributed and human-attributed quality traces can be compared cycle by
cycle while the confidence signal remains visible. The detailed table below
then expands the aggregate view to individual subtasks, exposing the cycle,
task type, origin, human quality, AI quality, attributed quality, synergy score,
and confidence value. Finally, the quality-by-origin chart at the bottom shows
whether one origin is consistently stronger or whether performance alternates
across cycles. This screen is therefore the evidence surface for the
co-evolution claim: the tool does not only report a final KPI, but records who
contributed to each outcome and whether the collaboration pattern changes over
time.

\screenshot[0.92\linewidth]{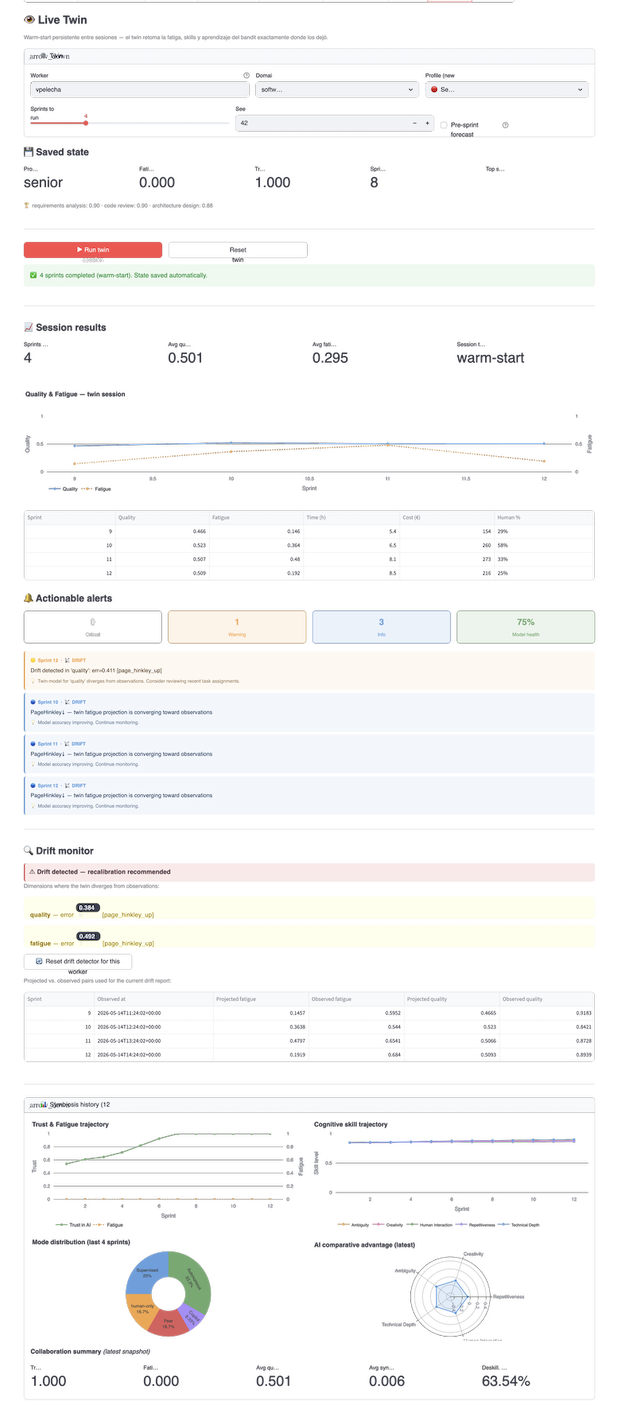}{%
Longitudinal analysis view. Recommended target: either \term{Live Twin} or
\term{Planning}, showing worker state persistence or horizon projections.\label{fig:advanced-twin}}

Figure~\ref{fig:advanced-twin} shows the \ui{Live Twin} view for a persistent
worker. The top controls identify the worker, domain, profile, sprint horizon,
and seed, while the saved-state cards summarize the worker state carried across
sessions rather than recomputed from a single run. The session section reports
the latest warm-start execution, including average quality, average fatigue,
session type, and the quality--fatigue trajectory over the simulated sprints.
The central value of the screen is the longitudinal governance layer: actionable
alerts classify detected conditions by severity, and the drift monitor flags
when the twin no longer matches observations closely enough and recalibration is
recommended. The lower history panel then summarizes trust and fatigue,
cognitive-skill trajectory, recent mode distribution, AI comparative advantage,
and aggregate collaboration metrics such as average synergy and deskilling
exposure. This is the visual evidence that HAAS Studio maintains worker state
across sessions and can reason about future sustainability, not only about the
last simulated batch.

\section{Availability and Implementation}
\label{sec:implementation}

\sectionrule

\subsection{Availability and access}

HAAS Studio is implemented in Python~3 using Streamlit as the UI framework and
is publicly available as a web application at
\url{http://3.248.226.75:8501}.

The application is accessed through the provided web address rather than
through a local installation process. Access is controlled by the application
administrator: users must either request access or be registered in the system,
and they can start using the tool once permission has been granted.

For normal use, the operational path is:
\begin{enumerate}
  \item access the web application at the provided address;
  \item request access or register a user account;
  \item log in once the application administrator has granted access;
  \item select an execution mode in the setup wizard;
  \item run a simulation, benchmark, or combined session;
  \item inspect the populated analysis surfaces.
\end{enumerate}

The implementation includes the required backend components for simulation
execution, benchmarking, and analysis, together with reproducibility scripts for
benchmark suites and deskilling runners.

The default \texttt{simulated} backend is sufficient for the full decision
workflow documented in this manual. External AI backends are optional
extensions, not prerequisites for using the tool.

\subsection{Layered architecture}

The artifact is organized around a small number of separable layers, shown in
Table~\ref{tab:architecture}. This structure is consistent with the broader
system documentation and helps keep the simulation, governance, and interface
concerns distinct.

\begin{figure}[H]
\centering
\resizebox{\linewidth}{!}{%
\begin{tikzpicture}[
  font=\small,
  node distance=0.48cm,
  layer/.style={draw=haasnavy, rounded corners=2pt, thick, align=center,
    text width=11.0cm, minimum height=0.72cm, fill=haasblue!6},
  support/.style={draw=haasgreen, rounded corners=2pt, thick, align=center,
    text width=5.2cm, minimum height=0.72cm, fill=haasgreen!8},
  arrow/.style={-{Latex[length=2.1mm]}, thick, draw=haasnavy},
  supportarrow/.style={-{Latex[length=2.1mm]}, thick, dashed, draw=haasgreen}
]
\node[layer] (dashboard) {\texttt{dashboard/}: Streamlit UI, setup wizard, analysis tabs, reports};
\node[layer, below=of dashboard] (simulation) {\texttt{simulation/}: run engine, KPI services, oracle traces, Live Twin, Planning};
\node[layer, below=of simulation] (core) {\texttt{core/}: immutable models, policy engine, allocators, human agent, collaboration decisions};
\node[layer, below=of core] (config) {\texttt{config/}: reward weights, guardrail thresholds, KPI catalog, calibration defaults};
\node[layer, below=of config] (domains) {\texttt{domains/}: task catalogs, scenarios, worker profiles, domain-specific KPI vocabulary};
\node[support, below=0.65cm of domains, xshift=-3.05cm] (storage) {\texttt{storage/}: SQLite runs, histories, worker state};
\node[support, below=0.65cm of domains, xshift=3.05cm] (scripts) {\texttt{scripts/}: benchmark suites, ablations, deskilling studies};
\draw[arrow] (dashboard) -- (simulation);
\draw[arrow] (simulation) -- (core);
\draw[arrow] (core) -- (config);
\draw[arrow] (config) -- (domains);
\draw[supportarrow] (simulation.south west) -- (storage.north);
\draw[supportarrow] (core.south east) -- (scripts.north);
\node[font=\scriptsize, align=center, below=0.1cm of storage] {persistence};
\node[font=\scriptsize, align=center, below=0.1cm of scripts] {reproducibility};
\end{tikzpicture}%
}
\caption{Layered implementation architecture. User interaction remains in the
Streamlit layer, while simulation, allocation, configuration, domain catalogs,
persistence, and reproducibility scripts remain separately auditable.\label{fig:architecture-diagram}}
\end{figure}
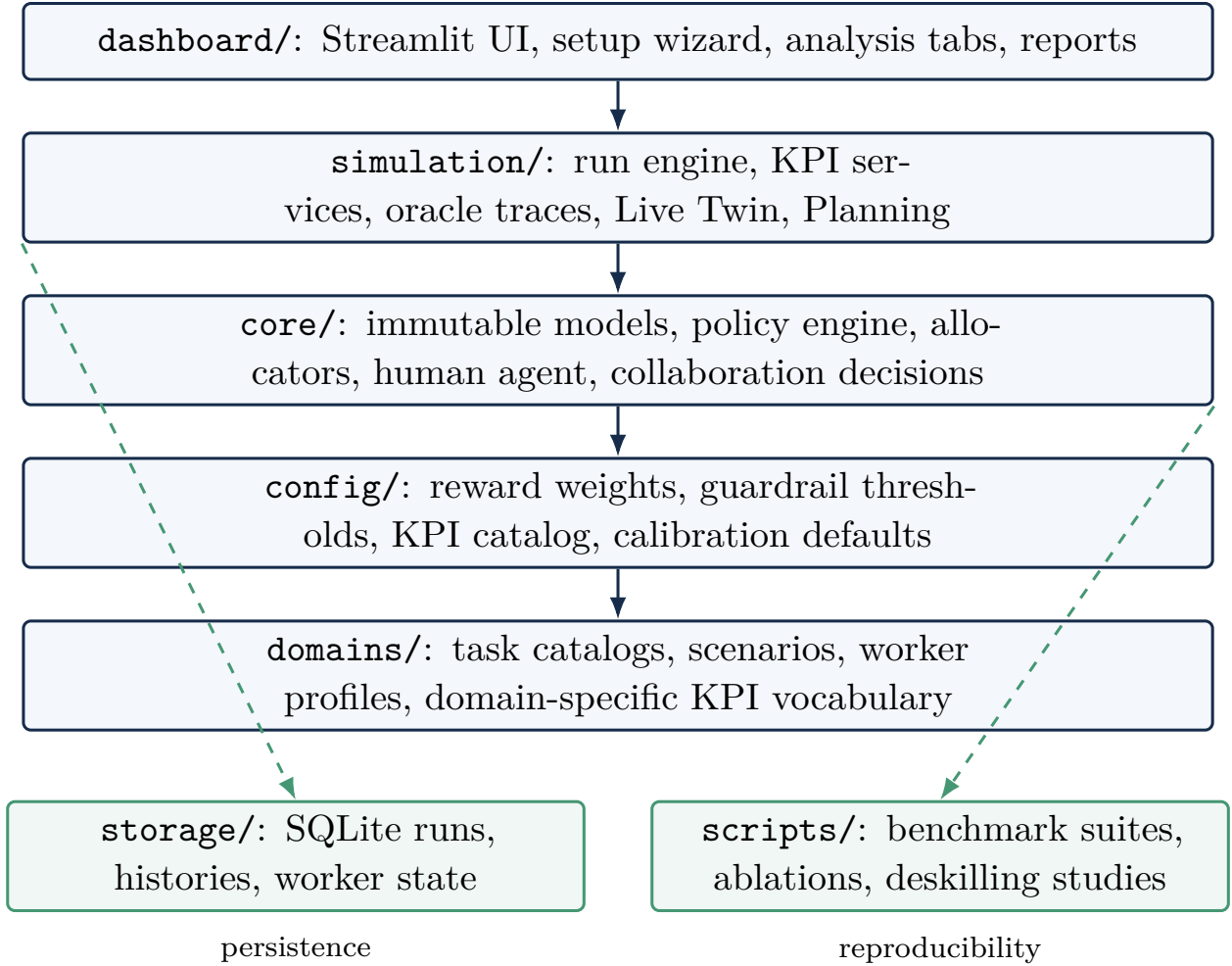

\begin{table}[H]
\centering
\small
\caption{High-level architecture of HAAS Studio.}
\label{tab:architecture}
\begin{tabularx}{\linewidth}{@{}>{\ttfamily\small}lY@{}}
\toprule
\textbf{Layer} & \textbf{Responsibility} \\
\midrule
dashboard/ & Streamlit user interface, setup wizard, tabs, reports, guided and expert modes. \\
simulation/ & Execution engine, cycle logic, KPI services, Digital Twin, forecasting. \\
core/ & Immutable models, human agent, allocators, policy engine, collaboration logic. \\
config/ & Settings, weight calibration, policy thresholds, KPI catalog. \\
domains/ & Scenario packs, task catalogs, human profiles, domain-specific KPIs. \\
storage/ & SQLite persistence for runs, histories, and worker state. \\
\bottomrule
\end{tabularx}
\end{table}

The internal model relies on immutable Pydantic objects, which makes state
transitions explicit and easier to audit. This is especially important in a tool
where a recommendation may later need to be justified from stored traces.

\subsection{Models, state, and reproducibility}

All state transitions are implemented through immutable Pydantic
models with \texttt{frozen=True}; mutation is performed exclusively
via \texttt{model\_copy(update=\{...\})}. The five-mode
\texttt{CollaborationMode} enum uses \texttt{(str, Enum)} to enable
direct serialization and string comparison across the stack.
Reproducibility is ensured through explicit random seeds on all
stochastic components. The tool has no external AI-service
dependency in its default \texttt{simulated} backend; real AI
backends (\texttt{claude\_cli}, \texttt{codex\_cli}, \texttt{groq},
\texttt{anthropic}, \texttt{gemini}, \texttt{ollama}) are optional
extensions that do not affect the simulation core.

\section{Operational Checklist}
\label{sec:checklist}

\sectionrule

Before running a study or making a recommendation, use the following checklist.

\subsection{Before running}
\begin{itemize}
  \item confirm the execution mode (\ui{Simulation}, \ui{Benchmark}, or
        \ui{Simulation + Benchmark});
  \item make the contract explicit: quality floor, cost target, fatigue cap,
        and any hard task restrictions;
  \item decide whether the goal is diagnosis, comparison, deployment
        recommendation, or longitudinal worker monitoring;
  \item keep only one major change between comparable runs whenever possible.
\end{itemize}

\subsection{After running}
\begin{itemize}
  \item check contract compliance before reading the ranking;
  \item inspect the watchlist and responsible screen before treating a strategy
        as deployable;
  \item confirm that the task-level transition plan is consistent with the
        observed task pool;
  \item save the run to \ui{History} if it may be used for comparison later.
\end{itemize}

\subsection{Before accepting a deployment recommendation}
\begin{itemize}
  \item verify that the recommendation survives both feasibility and
        responsibility filters;
  \item confirm that no critical task type is flagged as unstable or
        overdelegated;
  \item check whether the recommendation remains acceptable under small changes
        in priorities or contract settings;
  \item if \ui{Live Twin} is active, inspect whether the worker state remains
        sustainable over the next planning horizon.
\end{itemize}

\section{Limitations and Correct Use}
\label{sec:limitations}

\sectionrule

HAAS Studio has several deliberate scope limitations that are worth making
explicit. First, the simulation model is parameterized rather than empirically
calibrated: the cognitive profiles, reward functions, and fatigue dynamics are
grounded in the literature but not fitted to observed human-AI collaboration
data. The tool should therefore be read as a structured decision-support and
comparison environment, not as a predictor of exact field performance. Second,
the current co-evolution model captures six layers of adaptation but does not
model team-level dynamics; workers are simulated individually, without peer
effects or organizational hierarchy. Third, the domain packs (Software,
Manufacturing, Healthcare) are representative rather than exhaustive; their
task catalogs encode domain knowledge that should be reviewed by domain experts
before use in a specific deployment context. Finally, benchmark outputs remain
conditional on the active reward profile, guardrail configuration, and scenario
catalog, so recommendations should be interpreted as model-based operating
guidance within the configured decision frame rather than as universal policy
claims~\cite{iso42001,nistairmf}.

\section{Conclusion}
\label{sec:conclusion}

\sectionrule

This document presented HAAS Studio, a simulation and decision-support
tool for Human-AI work allocation. Its value lies not in introducing
a new algorithm, but in packaging allocation logic, governance
constraints, co-evolution monitoring, and forward-looking planning
into a single environment that can be configured, run, and interpreted
before any real deployment.

The tool is meant to help organizations make more defensible and
auditable decisions about how AI assistance should enter a workflow:
where to keep human control, where to allow shared execution, and
where higher automation may be acceptable under explicit guardrails.

The eleven task-oriented recipes---the first four presented in
Section~\ref{sec:workflow} and the remaining seven collected in the
appendices---are designed to make the tool usable by readers who want
not only to understand the artifact, but also to see how its outputs
should be read in practice.

\appendix

\section{Advanced Operational Recipes}
\label{app:advanced-ops}

This appendix collects the operational recipes omitted from the main body in
order to keep the paper focused on representative workflows.

\paragraph{Recipe~5 --- Compare strategies across reward profiles.}
\textit{Question:} if the organisation changes what it values most---quality,
balanced outcomes, or efficiency---does the same strategy still rank near the
top, or was it only the winner because one reward profile favored it?

This recipe tests whether a recommendation is robust to changes in priorities.
The scenario and guardrails stay fixed; only the reward profile changes. If the
same strategy remains strong under several profiles, it is a more defensible
candidate for deployment. If it only wins under one profile, it should be read
as a profile-specific choice rather than as a generally robust strategy.

\begin{figure}[H]
  \centering
  \fbox{\includegraphics[
    width=\dimexpr 0.97\linewidth - 2\fboxsep - 2\fboxrule\relax,
    height=0.85\textheight,
    keepaspectratio]{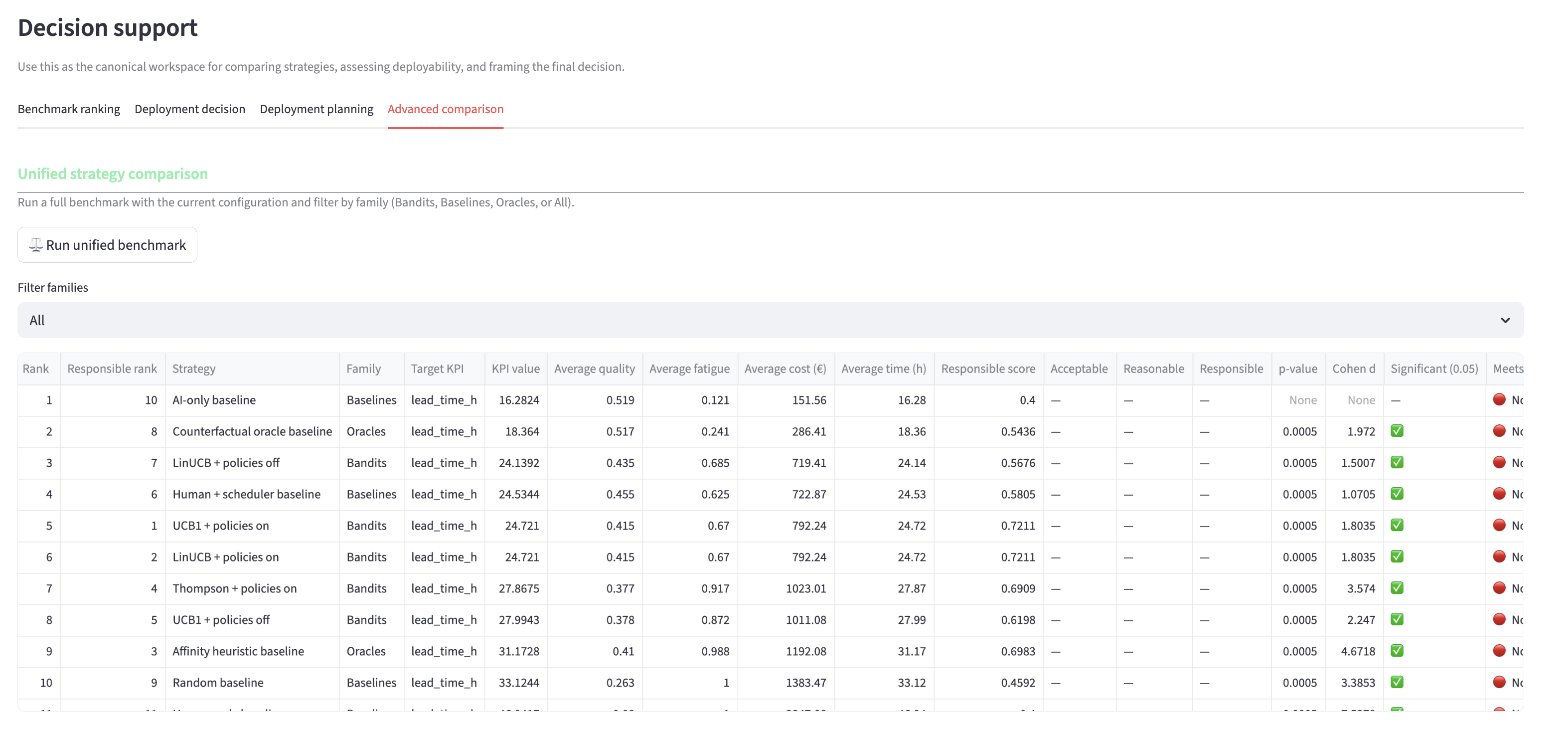}}%
  \caption{\ui{Advanced comparison} panel after running
  \ui{Compare strategies}, with the family filter set to \ui{All} so
  that bandit, baseline, and oracle strategies are visible together.}
  \label{fig:r5}
\end{figure}

\begin{enumerate}
  \item Decide on the scenario and guardrail configuration before starting.
        Keep them identical across all runs in this recipe so that any
        ranking differences are attributable to the reward profile, not to
        context changes.
  \item In the wizard, set \ui{Objective} to \ui{Compare strategies}.
        Set the reward profile to \ui{Quality First} and run the benchmark.
        Save the run to \ui{History} before proceeding.
  \item Change the reward profile to \ui{Four Outcome} and run again.
        Save to \ui{History}.
  \item Change the reward profile to \ui{Efficiency} and run again.
        Save to \ui{History}.
  \item Open \ui{Decision support} and navigate to \ui{Benchmark ranking}.
        Note which strategies appear in the top~3 under the current profile.
        Then open \ui{History} and repeat for the other two saved runs.
        Strategies that rank highly under all three profiles are robustly
        good; strategies that only dominate under one profile are
        context-sensitive.
  \item Open the \ui{Advanced comparison} panel. Use the family filter
        to compare \ui{Bandits}, \ui{Baselines}, and \ui{Oracles} separately.
        Figure~\ref{fig:r5} shows this panel with all families visible,
        illustrating the performance gap between adaptive strategies and
        fixed baselines and the distance from the oracle ceiling.
  \item To verify that the top-ranked strategy is stable and not an
        artefact of a single run, open \ui{Decision support} →
        \ui{Deployment decision} and use the \ui{Validate stability (7 seeds)}
        button. A winner with low seed agreement should not be recommended
        for deployment.
  \item Export the results using the download buttons in the \ui{Run overview}
        → \ui{Reports} subtab.
\end{enumerate}

Figure~\ref{fig:r5} is the cross-family view used in this recipe. The table
places fixed baselines, adaptive bandits, and oracle references in the same
ranking, so the reader can see whether the selected strategy is genuinely
robust or only attractive under the current reward profile. In this captured
run the cost-oriented ranking favours the \ui{AI-only baseline}, while the
counterfactual oracle and the LinUCB/UCB1 variants remain visible as
comparators. This is precisely the diagnostic point of the recipe: the
recommendation must be interpreted together with the value profile that
generated the ranking.

\begin{infobox}
\textbf{What to look for.} A strategy is considered \term{robust} if it
ranks in the top~3 under at least two of the three reward profiles and
passes the responsible screen. Strategies that only dominate under a
specific reward profile should be treated as locally optimal, not
generally deployable. Use the \ui{Validate stability (7 seeds)} button in \ui{Deployment decision}
to confirm that the top-ranked strategy is consistent across runs, not an
artefact of a single initialisation.
\end{infobox}

\paragraph{Recipe~6 --- Prescribe a task-level mode assignment that meets
given operational criteria.}
\textit{Question:} once the organisation has fixed its minimum quality,
maximum fatigue, and cost target, which collaboration mode should each task
type use in production: \ui{Human Only}, \ui{Copilot}, \ui{Peer},
\ui{Supervised}, or \ui{Autonomous}?

This recipe turns benchmark results into an operating rulebook. The goal is not
to choose the best allocator in the abstract, but to produce a concrete table
that says how each task type should be handled in production. A good answer
looks like: requirements analysis stays \ui{Copilot}, routine documentation can
move to \ui{Autonomous}, and safety-critical review remains \ui{Human Only},
because that combination satisfies the quality, fatigue, and cost constraints.

\begin{figure}[H]
  \centering
  \fbox{\includegraphics[
    width=\dimexpr 0.97\linewidth - 2\fboxsep - 2\fboxrule\relax,
    height=0.85\textheight,
    keepaspectratio]{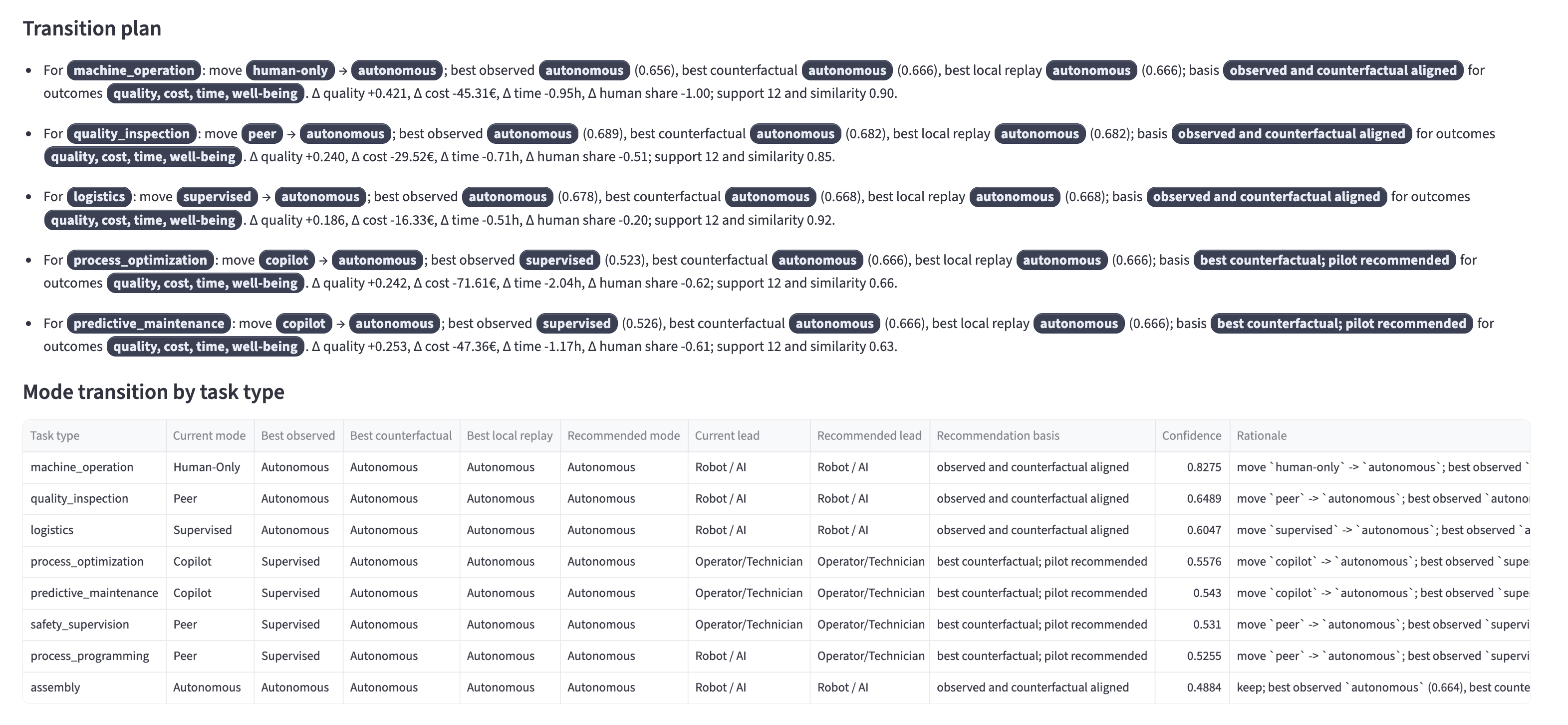}}%
  \caption{\ui{Deployment planning} view --- task-type mode assignment table
  showing the current mode, counterfactual best mode, recommended mode,
  recommendation basis, confidence, and rationale per task type.}
  \label{fig:r6a}
\end{figure}

\begin{figure}[H]
  \centering
  \fbox{\includegraphics[
    width=\dimexpr 0.97\linewidth - 2\fboxsep - 2\fboxrule\relax,
    height=0.85\textheight,
    keepaspectratio]{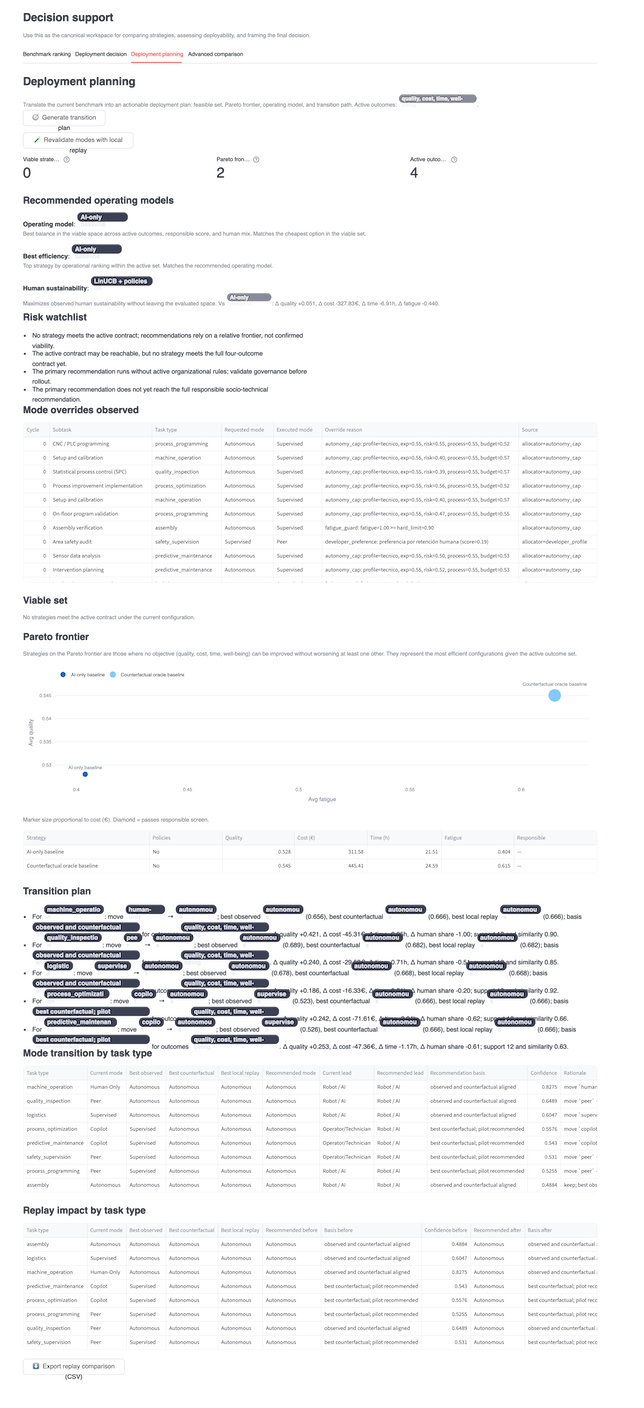}}%
  \caption{\ui{Deployment planning} after running \ui{Revalidate modes with
  local replay}, showing task-type mode recommendations before and after replay
  evidence.}
  \label{fig:r6b}
\end{figure}

\begin{enumerate}
  \item Define the criteria before opening the wizard. Write down, explicitly:
        the minimum acceptable quality score, the maximum fatigue level, the
        cost target per unit (feature, batch, or case), and any non-negotiable
        mode constraints for safety-critical or fully automatable tasks. These
        criteria will drive every subsequent configuration decision.
  \item In the wizard, set \ui{Objective} to \ui{Get recommendation} and
        enable \ui{Advanced} mode in the sidebar. This activates the full
        recommendation path, including the transition plan output.
  \item In \ui{Guardrails}, translate the criteria into contract parameters:
        set the quality floor, fatigue cap, and cost target as active
        constraints. For tasks that must remain human-controlled (e.g.\
        safety-critical inspection, clinical edge-case review), open the
        \ui{Subtask mode overrides} panel in the \ui{Guardrails} step and
        lock them to \ui{Human Only}. For tasks that are fully automatable
        (e.g.\ image preprocessing, routine report formatting), lock them to
        \ui{Autonomous}. Locked tasks are excluded from bandit learning and
        always execute in the assigned mode.
  \item Run the benchmark using \ui{Compare strategies}. The tool will
        evaluate all available allocators subject to the active contract,
        \emph{after} applying the locked assignments and governance guards.
  \item Open \ui{Decision support} and navigate to
        \ui{Deployment planning}. This view shows the recommended mode per
        task type derived from the benchmark trace --- it is the primary output
        of this recipe. Read each row as an operating rule: current mode,
        best observed mode, best counterfactual mode, best local replay mode
        when available, recommended mode, recommendation basis, confidence, and
        rationale.
  \item Cross-reference with the \ui{Risk watchlist}. Any task type flagged
        for deskilling exposure or instability should not be assigned the
        benchmark-recommended mode directly; check the \ui{Co-evolution} tab
        to verify that the mode preserves the relevant cognitive dimensions
        before confirming the assignment.
  \item To test sensitivity, change one criterion at a time and re-run:
        \begin{itemize}
          \item Tighten the quality floor by 0.05 and observe which task types
                shift from \ui{Supervised} to \ui{Copilot} or \ui{Peer}.
          \item Tighten the fatigue cap and observe which high-load tasks are
                demoted from AI-heavy modes.
          \item Raise the cost target and observe whether any task types
                currently assigned \ui{Copilot} can migrate to \ui{Supervised}
                without breaching quality.
        \end{itemize}
        Each variation should be saved to \ui{History} so that the
        mode-assignment tables can be compared side by side.
  \item In \ui{Deployment planning}, run \ui{Revalidate modes with local
        replay}. This replays representative subtasks for each observed task
        type and compares the mode suggested before and after replay evidence.
        Focus on rows where the recommendation is conditional or where the
        \ui{Best observed}, \ui{Best counterfactual}, and \ui{Best local replay}
        columns disagree. Figure~\ref{fig:r6b} shows this replay validation
        view: it reports the current mode, counterfactual best mode, replay
        best mode, final recommended mode, basis, and confidence for each task
        type, so the operating specification is grounded in local evidence
        rather than only in aggregate benchmark averages.
  \item Produce the final operating specification by combining: (a) the
        locked assignments from step~3; (b) the high-confidence assignments
        from \ui{Deployment planning}; (c) any manual overrides justified by
        the watchlist or replay analysis. Figure~\ref{fig:r6a} shows the
        resulting task-type mode assignment table with the current mode,
        counterfactual evidence, recommended mode, recommendation basis,
        confidence, and rationale per task type, which forms the concrete
        operating specification output of this recipe. Export it using the
        download buttons in the \ui{Run overview} → \ui{Reports} subtab.
\end{enumerate}

Figure~\ref{fig:r6a} shows the deployment-planning output used to build the
task-level operating specification. The transition plan lists task types such
as \ui{machine\_operation}, \ui{quality\_inspection}, and \ui{logistics}, and
states whether their current collaboration mode should be kept or moved toward
another mode. The \ui{Mode transition by task type} table then turns those
sentences into an auditable specification: for each task type it records the
current mode, observed evidence, counterfactual evidence, local replay evidence,
recommended mode, lead actor, recommendation basis, confidence, and rationale.
Rows such as \ui{process\_optimization} and \ui{predictive\_maintenance} are
especially useful because the observed best mode differs from the
counterfactual/local-replay best mode, making the recommendation conditional on
the evidence source rather than a simple aggregate winner.

Figure~\ref{fig:r6b} captures the same planning workspace after
\ui{Revalidate modes with local replay}. The upper part of the screen records
the governance context: no strategy satisfies the active contract, the Pareto
frontier is limited, and the risk watchlist warns that the operating model is
not yet directly deployable. The \ui{Mode overrides observed} table shows why
the raw recommendation cannot simply be applied: several autonomous requests
are executed as \ui{Supervised} because policy caps intervene. The lower
\ui{Replay impact by task type} table is the validation layer for the recipe;
it compares the recommendation before and after local replay and shows whether
the recommended mode, basis, or confidence changes once representative subtasks
are replayed. This is the evidence used to decide which assignments can enter
the operating specification and which must remain conditional.

\begin{infobox}
\textbf{What to look for.} The transition plan and deployment planning view
should converge on a mode assignment for each task type. Divergence between
them (e.g.\ transition plan suggests \ui{Supervised} while deployment
planning shows \ui{Peer} as stable) signals that the task type is sensitive
to load or profile --- treat it as conditional and document the condition in
the operating specification. A well-formed specification will have at most
one or two conditional assignments; if more than half the task types are
conditional, the criteria are likely too tight for the scenario and should
be relaxed before finalising.
\end{infobox}

\paragraph{Recipe~7 --- Identify which allocator fits the operational context.}
\textit{Question:} which learning allocator should be used for this scenario,
and how can we check that choice instead of guessing?

This recipe helps choose between \ui{UCB1}, \ui{Discounted-UCB}, \ui{LinUCB},
and \ui{Thompson Sampling}. The practical issue is simple: some scenarios are
stable, some change over time, and some contain very different task types. An
allocator that works well in one setting may be a poor fit in another. The
recipe first narrows the candidate allocators from the scenario properties, then
uses benchmark evidence to verify the choice.

\begin{enumerate}
  \item Characterise the scenario before running anything. Answer three
        questions:
        \begin{enumerate}[label=\alph*.]
          \item \textit{Is the reward environment stationary?} If the scenario
                involves changing conditions mid-run (e.g.\ \ui{Quality Crisis},
                \ui{New Product Ramp-Up}, \ui{Alarm Fatigue}), the reward
                distributions shift over time. If it involves stable, repetitive
                conditions (e.g.\ \ui{Maintenance}, \ui{Standard Production}),
                the environment is approximately stationary.
          \item \textit{Is there task-type diversity with observable features?}
                If the scenario has task types with meaningfully different
                cognitive profiles (e.g.\ \ui{Standard Sprint} mixing routine,
                creative, and high-ambiguity tasks), contextual information is
                available and can be exploited. If all tasks are similar in
                profile, contextual information adds little.
          \item \textit{What is the risk profile?} High-stakes or
                safety-sensitive scenarios require conservative exploration
                (low regret during learning), favouring algorithms with
                tighter confidence bounds.
        \end{enumerate}
  \item Use the answers to narrow the candidate set before benchmarking:
        \begin{itemize}
          \item \textbf{UCB1}: suited for stationary environments with
                moderate task diversity. Reliable baseline; converges
                predictably. Best starting point when no strong prior on the
                environment exists.
          \item \textbf{Discounted-UCB} ($\gamma \approx 0.90$): suited for
                non-stationary environments where recent outcomes should
                outweigh older ones. Use when conditions shift within a run
                (load spikes, fatigue accumulation mid-session, quality crises).
          \item \textbf{LinUCB} (contextual): suited for scenarios with
                diverse task types whose cognitive profiles differ substantially.
                The allocator conditions its estimate on the five-dimensional
                task profile, producing specialised assignments per task type.
                Requires more cycles to converge; favour runs of 10~cycles or
                more.
          \item \textbf{Thompson Sampling}: suited for non-stationary
                environments where uncertainty should drive exploration more
                aggressively. Performs well when the correct mode for a task
                type changes across cycles (e.g.\ ramp-up phases, recovery
                from governance demotions). More robust to sudden distributional
                shifts than Discounted-UCB.
        \end{itemize}
  \item Run \ui{Compare strategies} with the same scenario and guardrail
        configuration for each candidate allocator identified in step~2.
        Save each run to \ui{History} before changing the allocator.
  \item In \ui{Benchmark ranking}, compare the candidates on three axes:
        \begin{enumerate}[label=\alph*.]
          \item \textbf{Average reward} across seeds (efficiency).
          \item \textbf{Regret convergence speed} in \ui{Operational impact}
                (\ui{Trends and trade-offs}):
                how many cycles does each allocator need to reach
                $\bar{R}_C < 0.10$? A faster-converging allocator is
                preferable in short-run or high-cost exploration settings.
          \item \textbf{Stability across seeds}: expressed as the standard
                deviation of the reward distribution. A narrower distribution
                indicates the allocator is less sensitive to initialisation,
                which is a practical requirement for deployment.
        \end{enumerate}
  \item If the scenario is non-stationary, also inspect the
        \ui{Allocation evolution} panel in \ui{Learning evolution}. An
        allocator that fails to shift its mode distribution after a
        mid-run change (e.g.\ a fatigue spike or a quality drop) has not
        adapted to the distributional shift and should be replaced.
  \item Confirm the selection by running \ui{Compare strategies} on a
        second scenario in the same domain using the chosen allocator.
        A robust allocator should retain its performance advantage across
        scenarios, not only the one used for selection. Figures~\ref{fig:r7a}
        and~\ref{fig:r7b} split the \ui{Benchmark ranking} table into its
        left and right sides, because the full table is too wide to remain
        legible in a single paper figure.
\end{enumerate}

\screenshot[0.97\linewidth]{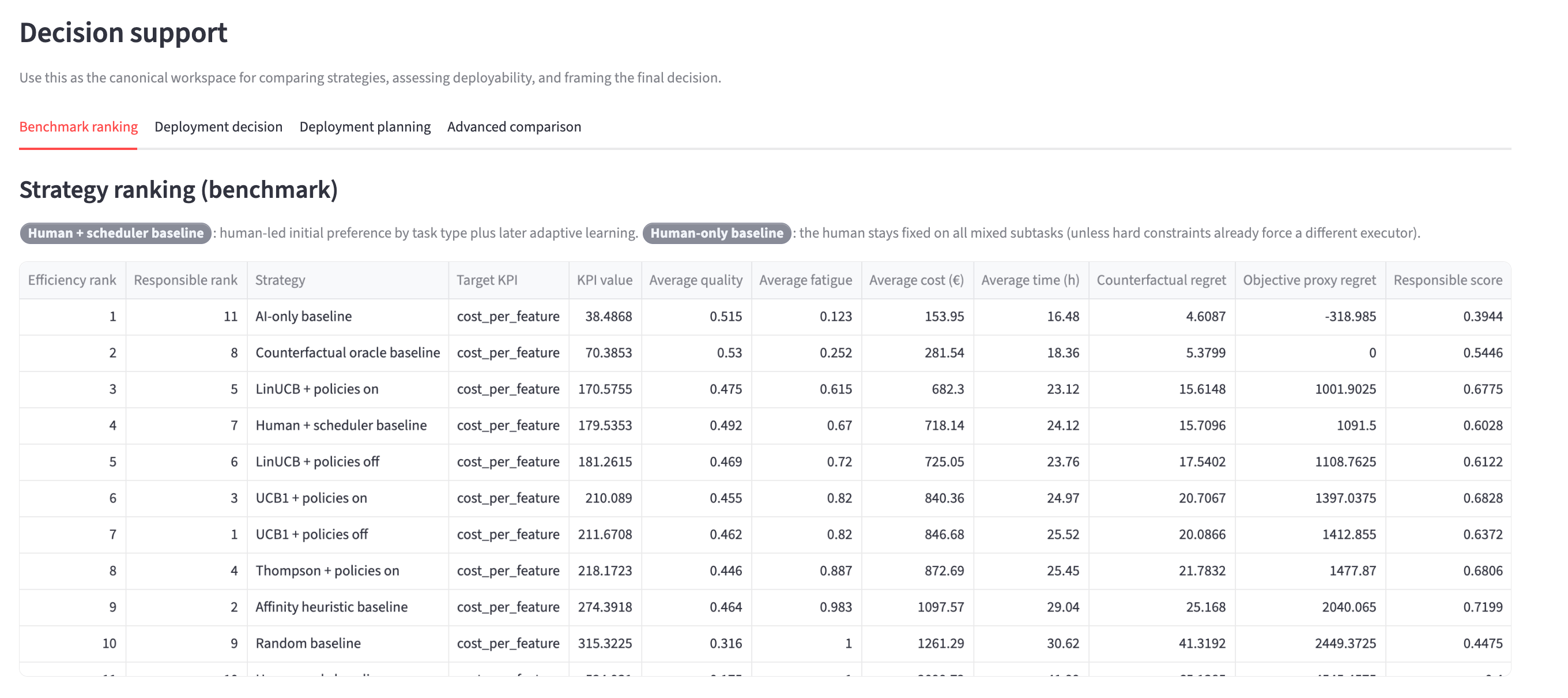}{%
Left side of the \ui{Benchmark ranking} table after running \ui{Compare
strategies} on Software / Standard Sprint. The visible columns show the
efficiency rank, responsible rank, strategy name, target KPI, KPI value,
average quality, fatigue, cost, time, regret, and responsible score.
\label{fig:r7a}}

\screenshot[0.97\linewidth]{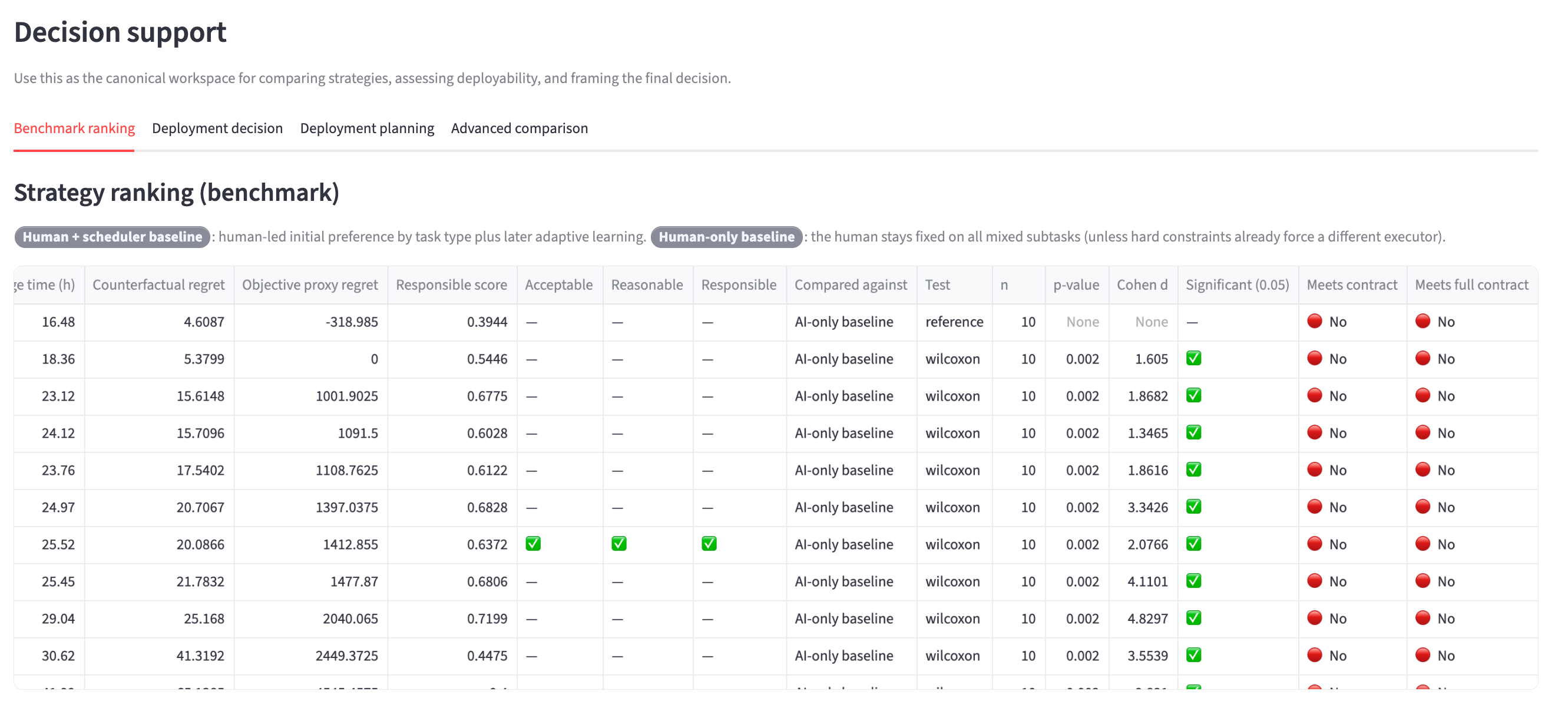}{%
Right side of the same \ui{Benchmark ranking} table. This continuation
shows the acceptability flags, responsible-recommendation flags,
statistical comparison against the reference strategy, sample size,
$p$-value, effect size, significance flag, and contract-compliance
columns.\label{fig:r7b}}

Taken together, Figures~\ref{fig:r7a} and~\ref{fig:r7b} make the allocator
selection auditable rather than impressionistic. Figure~\ref{fig:r7a} is
used to identify which allocator or baseline ranks best on the operational
objective and how far each candidate sits from the oracle or reference
baseline. Figure~\ref{fig:r7b} is then used to check whether that apparent
advantage is statistically supported and whether the candidate satisfies the
deployment contract. In this example, the left side shows that several
learning strategies can be compared against fixed baselines on cost, time,
quality, fatigue, and regret, while the right side shows that a high-ranking
strategy still needs to be read through the governance columns: statistical
significance alone is not enough if the contract-compliance columns remain
negative.

\begin{infobox}
\textbf{Decision heuristic.} Start with \ui{UCB1} as the baseline in any
new scenario. If the environment is non-stationary, test \ui{Discounted-UCB}
and \ui{Thompson Sampling} in parallel. If the scenario has high task
diversity (three or more meaningfully different task types), add \ui{LinUCB}
to the comparison. Use the convergence speed and seed stability, not only
the average reward, as the final selection criteria. An allocator that wins
on average but shows high variance across seeds is not a reliable choice
for deployment.
\end{infobox}

\section{Analytical Recipes}
\label{app:analytical-recipes}

This appendix collects the analytical reading patterns associated with the
artifact's output surfaces.

\paragraph{Recipe~8 --- Assess whether human and AI are genuinely co-evolving
or merely substituting.}
\textit{Question:} are the human and the AI becoming a better team over time,
or is the AI simply taking over work that the human used to do?

This recipe distinguishes useful collaboration from plain substitution. A run
can look efficient because the AI is doing more work, but that does not prove
that the human-AI pair is improving. The evidence to look for is whether
collaborative outcomes grow, whether synergy improves, and whether the human
keeps or loses capability in important cognitive dimensions.

\begin{figure}[H]
  \centering
  \fbox{\includegraphics[width=\dimexpr 0.97\linewidth - 2\fboxsep - 2\fboxrule\relax,
        height=0.85\textheight, keepaspectratio]{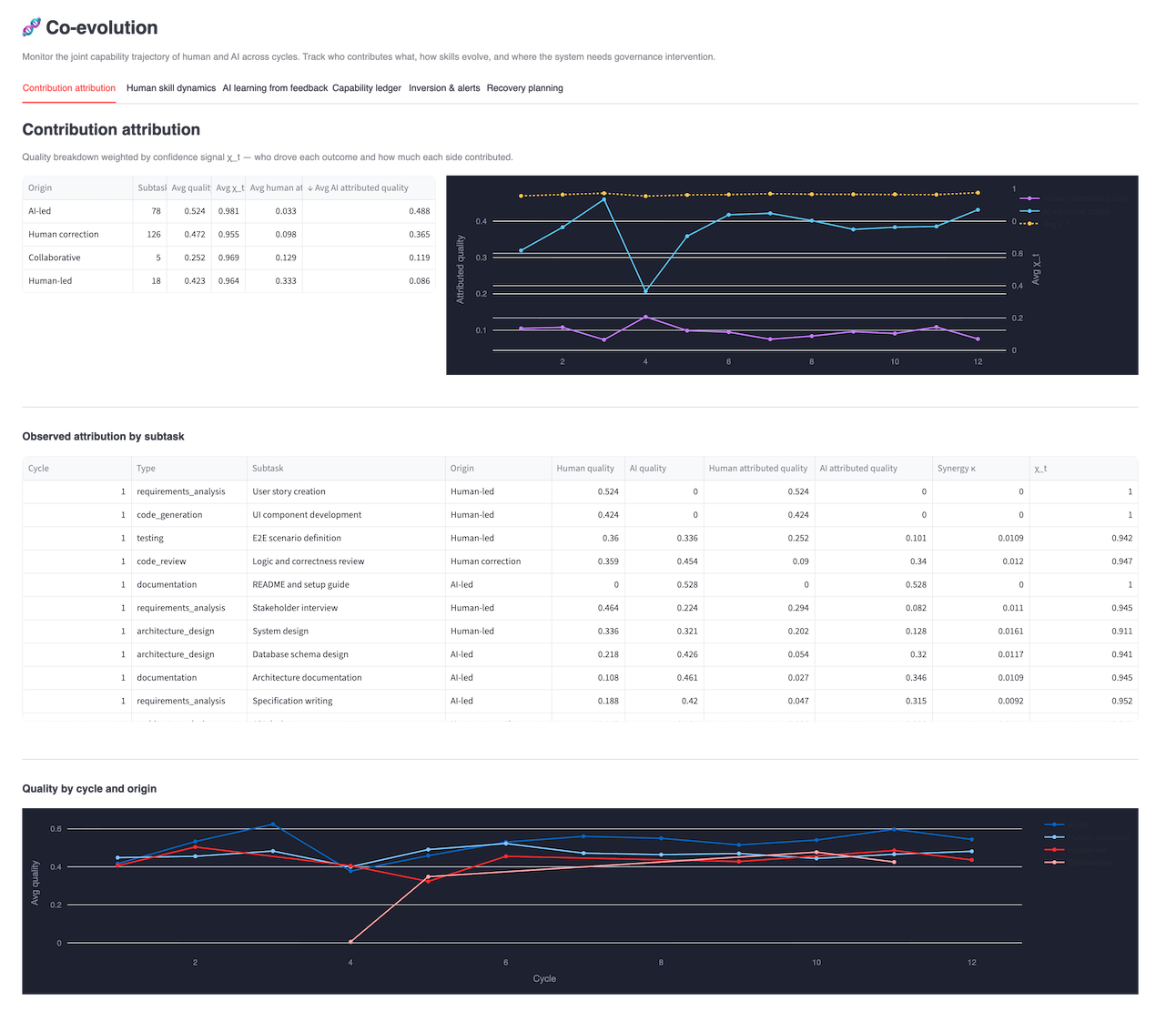}}%
  \caption{Co-evolution tab --- contribution attribution chart showing the
  proportion of human-led, AI-led, and collaborative outcomes across
  cycles, with the synergy score $\kappa_t$ overlaid.\label{fig:r8a}}
\end{figure}

\begin{figure}[H]
  \centering
  \fbox{\includegraphics[width=\dimexpr 0.97\linewidth - 2\fboxsep - 2\fboxrule\relax,
        height=0.85\textheight, keepaspectratio]{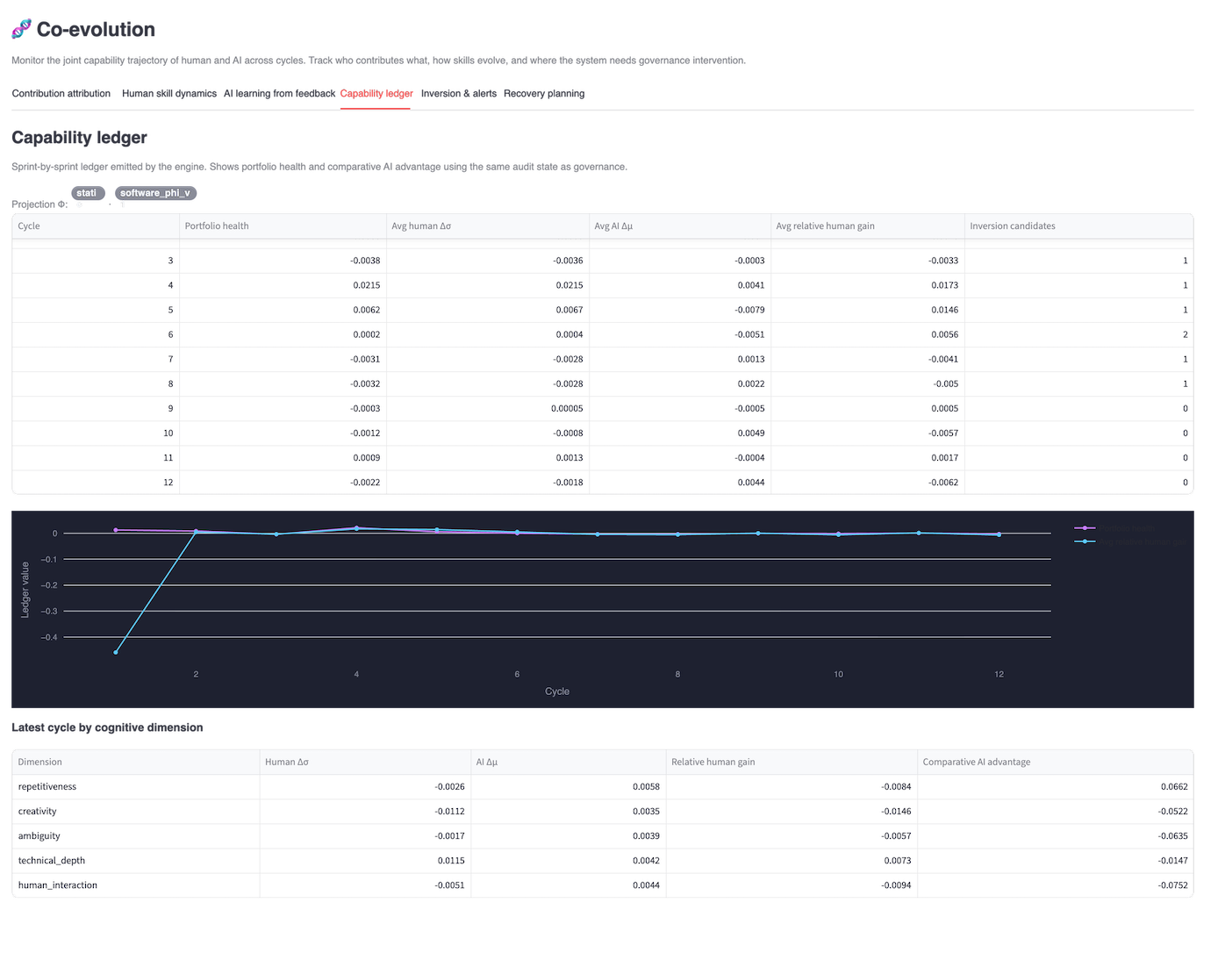}}%
  \caption{Co-evolution tab --- capability ledger showing comparative advantage
  scores ($\Delta_{\text{adv},d}$) per cognitive dimension, with human
  and AI advantage columns and any inversion alert
  flagged.\label{fig:r8b}}
\end{figure}

\begin{enumerate}
  \item Run a benchmark of at least 10~cycles with the \ui{Co-evolution
        horizon} set to \ui{balanced}. Use a scenario with meaningful task
        diversity (e.g.\ \ui{Software / Standard Sprint} or \ui{Manufacturing
        / New Product Ramp-Up}).
  \item Open the \ui{Co-evolution} tab and navigate to
        \textbf{Contribution attribution}. Figure~\ref{fig:r8a} shows the
        contribution attribution chart with the proportion of human-led,
        AI-led, and collaborative outcomes across cycles, and the synergy
        score $\kappa_t$ overlaid, making the substitution-vs-co-evolution
        distinction directly readable. Check whether the split shifts across
        cycles: pure substitution shows a monotonic increase in AI-led outcomes
        and a corresponding decrease in human-led outcomes with no growth in
        the collaborative category; co-evolution shows a growing collaborative
        share alongside stable human-led capacity. In the captured example,
        the summary table reports 78 \term{AI-led} subtasks, 126 \term{Human
        correction} subtasks, only 5 \term{Collaborative} subtasks, and
        18 \term{Human-led} subtasks, so the run should be read as
        supervision-heavy rather than strongly co-evolutionary.
  \item Navigate to \textbf{AI learning from feedback}. The synergy score
        $\kappa_t$ measures whether collaborative-mode outcomes exceed what
        either agent would have achieved alone. A value above 1.0 confirms
        synergy; a value near 1.0 indicates additive co-production; a value
        below 1.0 indicates that collaboration is underperforming individual
        execution (a signal of coordination friction or mode misassignment).
  \item Navigate to \textbf{Capability ledger}. Figure~\ref{fig:r8b} shows
        the capability ledger with comparative advantage scores
        ($\Delta_{\text{adv},d}$) per cognitive dimension and human and AI
        advantage columns visible alongside any inversion alerts, enabling
        the reader to see at a glance which agent dominates each dimension.
        In the \ui{Comparative AI advantage} column, positive values indicate
        an AI edge in that dimension and negative values indicate that the
        human remains stronger. Check
        for two patterns:
        \begin{itemize}
          \item \textit{Healthy specialisation}: humans dominate in high-ambiguity
                and high-interaction dimensions; AI dominates in repetitive and
                low-ambiguity dimensions. Each agent operates in its zone of
                advantage.
          \item \textit{Capability inversion}: the AI has gained a strong
                negative score in a dimension where the human was previously
                the leader. This signals that human practice has been crowded
                out, not that the AI has genuinely become superior.
        \end{itemize}
  \item Open \textbf{Inversion \& alerts}. Any flagged inversion should be
        treated as a governance signal, not a performance gain. Cross-reference
        with the skill trajectory in \textbf{Human skill dynamics}: a
        declining skill curve in the same dimension confirms that the inversion
        is driven by under-practice, not by a genuine shift in relative
        capability.
  \item Navigate to \textbf{Recovery planning}. If inversions are present,
        the tool generates corrective-action suggestions (e.g.\ reassigning
        specific task types back to \ui{Copilot} or \ui{Human Only} for a
        fixed number of cycles). Implement the suggestion with the highest
        impact score first.
  \item To verify improvement, repeat from step~2 after implementing the
        corrections. $\kappa_t$ should increase and the flagged inversions
        should clear within 3--5~cycles.
\end{enumerate}

Figure~\ref{fig:r8a} corresponds to a 12-cycle software run and gives a
concrete example of the distinction made in the recipe. The attribution
summary is not balanced: \term{AI-led} outcomes are the strongest on average
quality (0.524) and account for 78 subtasks, while \term{Human correction}
accounts for 126 subtasks with average quality 0.472. By contrast, the
\term{Collaborative} category appears only 5 times and has the weakest average
quality (0.252). The line chart tells the same story over time: AI-attributed
quality dominates most cycles, while human-attributed quality stays much lower.
Read against the recipe, this is evidence of effective oversight and AI
execution, but only weak evidence of genuine co-evolution because the
collaborative share remains small.

Figure~\ref{fig:r8b} adds the capability-side interpretation. At the
portfolio level, the ledger shows 1--2 inversion candidates from cycles~3 to~8
and then 0 inversion candidates from cycle~9 onward, which suggests that the
early inversion pressure is later stabilised rather than compounded. In the
latest-cycle dimension table, the clearest AI edge appears in
\term{repetitiveness} (\ui{Comparative AI advantage} = 0.0662), while
\term{creativity} (-0.0522), \term{ambiguity} (-0.0635),
\term{technical depth} (-0.0147), and \term{human interaction} (-0.0752)
remain on the human side. That pattern is consistent with healthy
specialisation: the AI is strongest on routine work, but the human retains the
advantage on the more ambiguous and interaction-heavy dimensions. Together,
Figures~\ref{fig:r8a} and~\ref{fig:r8b} show a run that is capability-safe by
the end of the horizon, yet still closer to substitution-plus-supervision than
to strong collaborative lift.

\begin{infobox}
\textbf{Synergy vs.\ substitution.} The clearest diagnostic is the joint
trajectory of $\kappa_t$ and the collaborative share. If $\kappa_t > 1$
and the collaborative attribution share is growing, the system is
co-evolving. If $\kappa_t \approx 1$ and AI-led attribution is growing at
the expense of both human and collaborative categories, the system is
substituting. The capability ledger will confirm which cognitive dimensions
are at risk.
\end{infobox}

\paragraph{Recipe~9 --- Select the right reward objective and contract for
the operational priorities.}
\textit{Question:} how should the reward objective and contract thresholds be
set so that the recommendation matches what the organisation actually cares
about?

This recipe is about translating priorities into configuration. If quality is
non-negotiable, the reward profile and contract should make low-quality
strategies fail even if they are fast or cheap. If efficiency matters more, the
tool should be allowed to prefer lower-cost strategies as long as the minimum
contract is still satisfied. The recipe compares profiles and thresholds to
check whether the final recommendation really reflects the stated priorities.

\begin{enumerate}
  \item Before running any benchmark, state the priorities explicitly in
        order of importance (e.g.\ quality first, then fatigue, then cost).
        This prevents post-hoc rationalisation of a result produced by the
        default weights.
  \item Run \ui{Compare strategies} three times on the same scenario
        (Software / Maintenance) and guardrails, using \ui{Thompson Sampling}
        as the allocator and changing only the reward profile each time: first
        \ui{Quality First}, then \ui{Four Outcome}, then \ui{Efficiency}.
        Save each run to \ui{History} before changing the reward profile.
        Use Thompson Sampling rather than UCB1: UCB1 converges to autonomous
        or copilot modes and does not assign enough risky tasks to shared
        modes (PEER or SUPERVISED) to pass the responsible screen; Thompson
        Sampling explores collaborative modes more broadly and is the
        allocator most likely to produce diamond markers on the Pareto
        frontier.
  \item In \ui{Decision support}, compare the strategy rankings across
        the three profiles. Note which strategies appear in the top~3 under
        all profiles (robustly good) and which only appear under one profile
        (context-sensitive).
  \item After each run, open \ui{Decision support} $\to$ \ui{Deployment
        planning} (not \ui{Deployment decision}) and scroll to the
        \ui{Pareto frontier} section. Figures~\ref{fig:r9a}, \ref{fig:r9b}, and~\ref{fig:r9} show the three
        Pareto frontiers, one per profile.
        Each point represents a non-dominated strategy: no other strategy
        beats it simultaneously on quality \emph{and} fatigue.
        Strategies that also pass the \ui{responsible screen} (deskilling
        risk, human participation, and workload constraints all within
        threshold) are shown as \textbf{diamonds}; the rest are circles.
        Panel~(a) (\textit{Quality First}) concentrates three strategies in
        a compact, high-quality band (avg.\ quality 0.51--0.55) at moderate
        fatigue, none passing the responsible screen because quality
        pressure forces high AI share.
        Panel~(b) (\textit{Four Outcome}) expands the frontier horizontally
        across a wider fatigue range (0.55--0.89), reflecting the balanced
        weight across all four outcome dimensions; four strategies appear but
        none clears the responsible-screen threshold under this profile.
        Panel~(c) (\textit{Efficiency}) shows the most spread frontier, with
        one strategy (\textsc{UCB1\,+\,policies on}) earning a diamond
        marker: cost pressure pushes the allocator toward shared
        collaboration modes (PEER / SUPERVISED) that satisfy the
        deskilling-risk and human-participation constraints.
        Identify the frontier point in the panel that corresponds to the
        stated priority order from step~1.
  \item Navigate to \ui{Strategy priorities}. Use the priority sliders to
        encode the stated priority order numerically (e.g.\ quality weight
        0.6, fatigue weight 0.25, cost weight 0.15). Observe how the ranking
        changes. If the top-ranked strategy under the slider weights is not
        on the Pareto frontier, the priorities contain an internal
        inconsistency that should be resolved before proceeding.
  \item Test contract sensitivity. Run the same benchmark twice: once with
        the full contract (all four outcome dimensions active) and once with
        a reduced contract (e.g.\ quality and well-being only, dropping cost
        and time as hard constraints). Compare the size of the feasible set
        and the top recommendation under each contract. A reduced contract
        typically expands the feasible set; if the top recommendation changes
        substantially, the cost and time constraints are binding --- they
        should be kept in the full contract.
  \item Set cost sensitivity. Return to the \ui{Setup} wizard and adjust
        the human and AI hourly cost inputs, then re-run \ui{Compare
        strategies}. Observe at what cost ratio the ranking between the top
        two strategies inverts. If the inversion happens at a ratio that is
        within realistic variation for the organisation, treat both strategies
        as viable and document both in the deployment specification.
\end{enumerate}

\screenshot[0.97\linewidth]{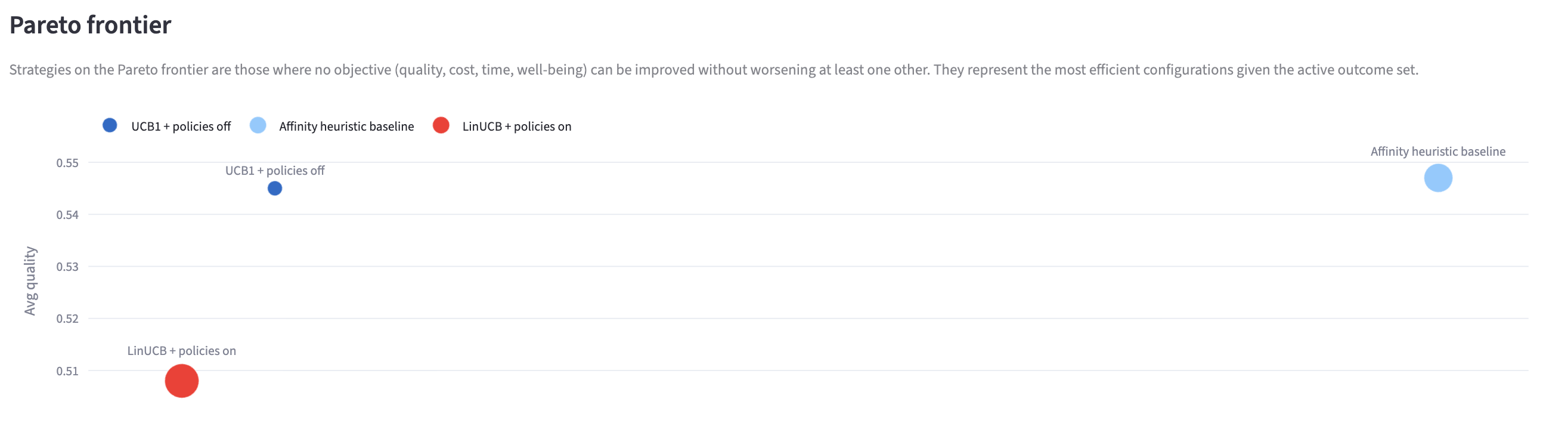}{%
  Pareto frontier --- \textbf{Quality First} profile (Software /
  Maintenance, Thompson Sampling, 7~seeds). Three strategies form a
  compact frontier at high quality (0.51--0.55) and moderate fatigue.
  No strategy clears the responsible screen (all circles) because the
  quality objective drives the allocator toward AI-heavy modes.
  Marker area is proportional to cost~(\texteuro).\label{fig:r9a}}

\screenshot[0.97\linewidth]{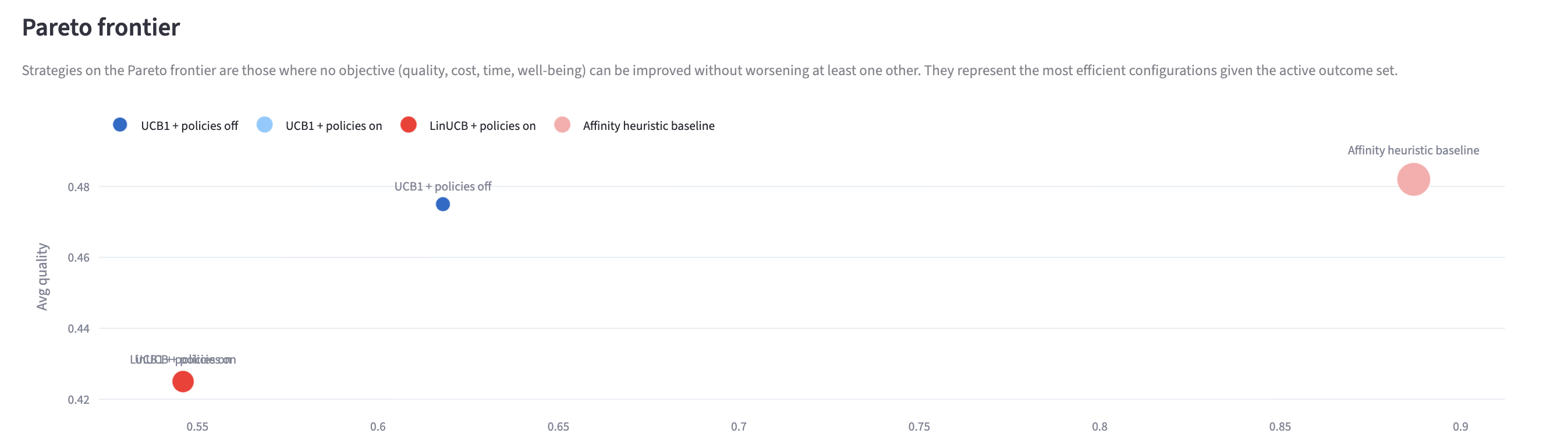}{%
  Pareto frontier --- \textbf{Four Outcome} profile (same scenario and
  seeds as Fig.~\ref{fig:r9a}). Four strategies span a wider fatigue
  range (0.55--0.89) as the balanced objective trades well-being against
  quality and cost. No diamond: no strategy clears the responsible
  screen under this profile.\label{fig:r9b}}

\screenshot[0.97\linewidth]{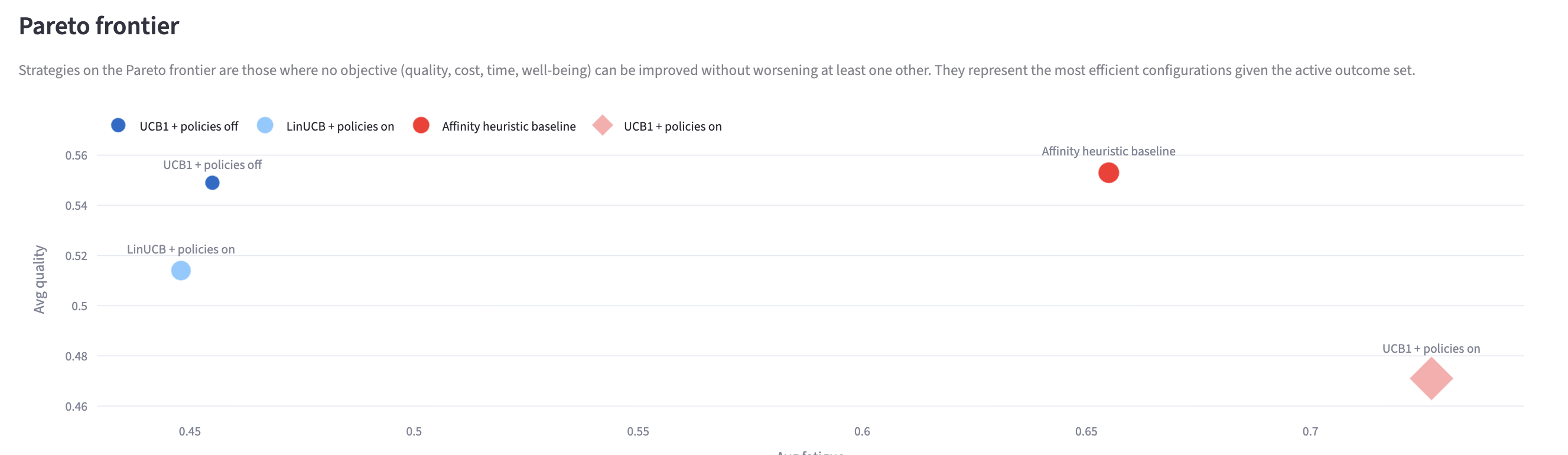}{%
  Pareto frontier --- \textbf{Efficiency} profile (same scenario and
  seeds as Fig.~\ref{fig:r9a}). The frontier is the broadest of the
  three profiles (quality 0.47--0.55). \textsc{UCB1\,+\,policies on}
  earns a diamond~($\diamond$): cost pressure guides the allocator
  toward PEER and SUPERVISED collaboration modes, which satisfy the
  deskilling-risk and human-participation thresholds of the responsible
  screen. Marker area is proportional to cost~(\texteuro).\label{fig:r9}}

\begin{infobox}
\textbf{What to look for.} A recommendation is stable if the same strategy
appears in the top~3 under all three reward profiles and under both the full
and reduced contracts. A recommendation that only survives under a specific
reward profile or a tight contract is locally optimal, not generally
deployable. The Pareto frontier should guide the choice, not the efficiency
ranking alone.
\end{infobox}

\paragraph{Recipe~10 --- Verify that governance guardrails are correctly
calibrated.}
\textit{Question:} are the guardrails firing at the right time, or are they too
strict, too weak, or too expensive in performance terms?

This recipe checks whether the governance layer is calibrated. A useful guard
fires when fatigue, deskilling, trust, or RLHF risk becomes real. A guard that
fires constantly may be blocking otherwise acceptable strategies; a guard that
never fires may be too loose. The recipe compares guarded and relaxed runs to
measure both the safety value and the efficiency cost of each guard.

\begin{enumerate}
  \item Run a single simulation (not a benchmark) with all four guards
        active (fatigue, deskilling, trust, RLHF) on Manufacturing /
        Quality Crisis, UCB1, 10~cycles, fatigue cap~38/100. Record the
        average reward and total governance events shown in the
        \ui{Run overview} headline card.
  \item Open \ui{Decision support} and navigate to
        \ui{Deployment planning}. Scroll past the Pareto frontier to the
        \textbf{Mode overrides observed} section. Figure~\ref{fig:r10}
        shows this table for a Manufacturing / Quality Crisis run with
        fatigue cap~0.38 and UCB1. Two guard types appear: the
        \texttt{fatigue\_guard}, which fires when the human fatigue
        metric reaches the warning limit (e.g.\ \texttt{fatigue=0.87
        >= warn\_limit=0.70}) or the hard limit
        (\texttt{fatigue=1.00 >= hard\_limit=0.90}), demoting the
        requested mode one step toward higher human involvement; and the
        \texttt{autonomy\_cap}, which fires when the operator profile
        (here \texttt{tecnico}) sets a ceiling on autonomous delegation
        based on experience, risk, process, and budget signals. In both
        cases the executed mode is more conservative than the requested
        mode --- \ui{Autonomous} becomes \ui{Supervised}, and
        \ui{Supervised} becomes \ui{Peer} --- preventing the allocator
        from compounding fatigue or exceeding the delegation ceiling.
        Count the firing frequency per guard type over the full run;
        this is the primary evidence for calibration diagnosis.
  \item Diagnose over-constraint. If a guard fires more than once every
        2~cycles, the corresponding threshold may be set too tight. Symptoms:
        the strategy is frequently demoted mid-run; quality is high but well
        above the floor; cost is well above target despite the efficiency
        allocator winning the benchmark.
  \item Diagnose under-constraint. If a guard never fires across 5+ seeds,
        the threshold may be set too loose. Symptoms: fatigue trend is rising
        but no guard activates; deskilling exposure metric is near the
        threshold but no demotion occurs.
  \item For each guard that fired frequently, return to the \ui{Setup}
        wizard, go to the \ui{Guardrails} step, and disable that specific
        guard (available in \ui{Advanced} mode). Re-run the benchmark with
        the same configuration. Measure the delta in average reward between
        the guarded and unguarded runs. This is the \textit{efficiency cost}
        of the guard. If the cost is below 5\%, the guard is well-calibrated.
        If the cost is above 15\%, the threshold should be relaxed slightly
        and the run repeated.
  \item Adjust thresholds incrementally. Increase the fatigue cap by 0.03,
        or raise the deskilling exposure threshold by 0.05. Re-run. If the
        guard now fires at a frequency of one event per 3--4~cycles, the
        threshold is calibrated. Save the run to \ui{History} to preserve the
        calibration evidence.
  \item Verify that relaxing a guard does not breach the operational
        contract. After each threshold adjustment, check \ui{Operational
        impact} to confirm that the quality floor and cost target are still
        met. A guard is correctly calibrated when it fires rarely enough to
        allow the allocator to learn freely, but reliably enough to prevent
        sustained exposure to fatigue or deskilling.
\end{enumerate}

\screenshot[0.97\linewidth]{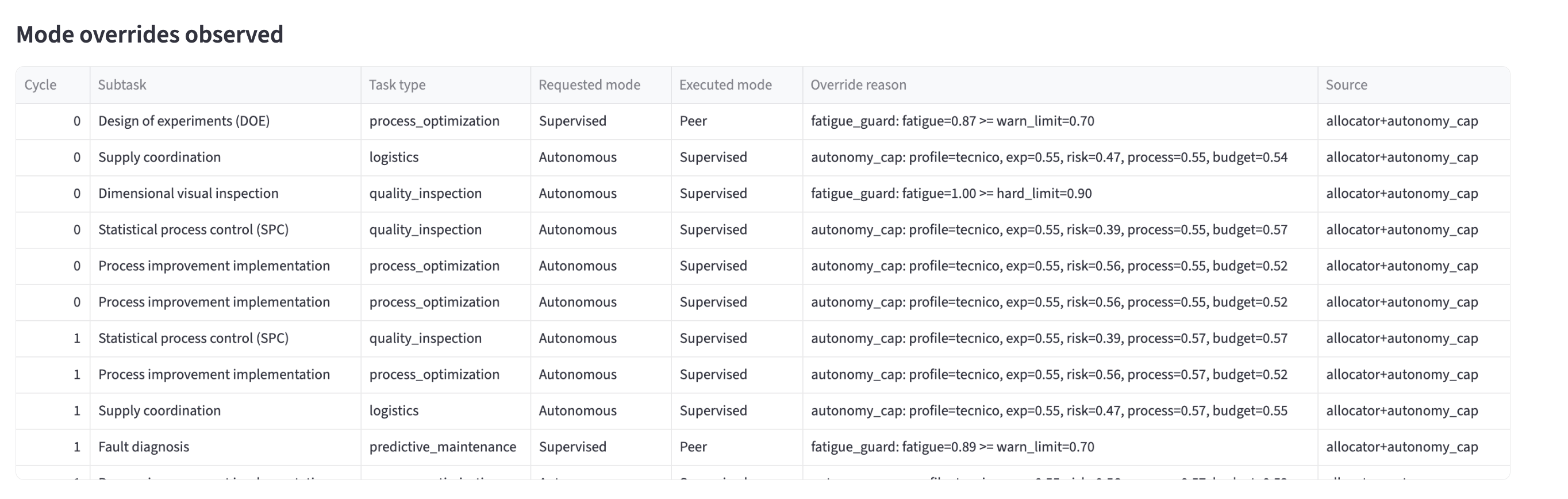}{%
\ui{Decision support} --- \ui{Deployment planning}, \textbf{Mode
overrides observed} table. Manufacturing / Quality Crisis, UCB1,
10~cycles, fatigue cap~0.38. Each row is one guard-firing event showing
the cycle number, subtask, requested collaboration mode (what the
allocator chose), executed mode (after the guard override), and the
override reason. Two guard types are active: \texttt{fatigue\_guard}
(fires at the warning limit \texttt{fatigue >= 0.70} and hard limit
\texttt{fatigue >= 0.90}, demoting Supervised~$\to$~Peer and
Autonomous~$\to$~Supervised respectively) and \texttt{autonomy\_cap}
(fires when the \texttt{tecnico} operator profile cap is exceeded based
on experience, risk, process, and budget signals, demoting
Autonomous~$\to$~Supervised). The \texttt{Source} column confirms both
guards are applied jointly (\texttt{allocator+autonomy\_cap}).\label{fig:r10}}

\begin{infobox}
\textbf{Key distinction.} Guards are not performance penalties; they are
safety mechanisms. Removing a guard that produces a large efficiency gain
signals not that the guard was wrong, but that the system was relying on
over-delegation to achieve performance. The correct response is to adjust
the threshold, not to disable the guard.
\end{infobox}

\paragraph{Recipe~11 --- Measure how far the current strategy is from the
theoretical optimum.}
\textit{Question:} how far is the chosen strategy from the best outcome that
would have been possible with perfect hindsight, and is that gap shrinking over
time?

This recipe uses the oracle counterfactual as a diagnostic ceiling. The oracle
is not a deployable strategy: it knows, after the fact, which agent would have
done best on each subtask. Comparing the real allocator with that ceiling shows
where value is being lost, which task types are most often misassigned, and
whether learning is reducing the gap across cycles.

\begin{enumerate}
  \item Run any benchmark that includes the \ui{Oracle counterfactual
        baseline} in the strategy set (this baseline is automatically
        included in all benchmark modes). The oracle assigns every subtask to
        whichever agent --- human or AI --- would have produced the higher
        reward on that specific instance, using perfect hindsight.
  \item In \ui{Benchmark ranking}, locate the oracle baseline row.
        Figure~\ref{fig:r11b} shows the full ranking table for a
        Software / Standard Sprint benchmark. The
        \ui{Counterfactual oracle baseline} appears at efficiency
        rank~2 with KPI~18.364 (lead time in hours) and zero
        counterfactual regret (by definition, the oracle has no
        regret against itself). The best learned strategy (\ui{AI-only
        baseline}, rank~1) achieves a lower lead time of 16.021~h but
        carries a responsible score of~0 and does not pass the
        governance filters. Among responsible strategies,
        \ui{LinUCB + policies off} (efficiency rank~3, responsible
        rank~1) is the closest deployable option to the oracle ceiling,
        with a counterfactual regret of~23.4 --- meaning it forfeits
        roughly 23 lead-time units per run relative to perfect
        hindsight. This gap is the quantified opportunity cost of
        operating under uncertainty and governance constraints.
  \item Open \ui{Operational impact} and navigate to
        \ui{Trends and trade-offs}. Figure~\ref{fig:r11a} shows the
        full cycle-by-cycle evolution panel, with the
        \textbf{Regret convergence} chart at the bottom: green bars
        show per-cycle regret and the yellow line is the cumulative
        regret $R_C = \sum_s \text{regret}_s$. The heuristic
        (warm-up) phase and the adaptive learning phase are marked by
        the shaded background. The curve rises steeply in the first
        three cycles, inflects around cycle~4--5, and flattens toward
        the end of the run, indicating that the allocator has converged
        on most of the learnable signal. A curve that continues to rise
        steeply beyond cycle~8 indicates that the environment is too
        non-stationary for the allocator to converge within the run
        length.
  \item Compute the normalized regret $\bar{R}_C = R_C / R_{\text{oracle}}$.
        Values below 0.10 at the end of the run indicate that the allocator
        is capturing more than 90\% of the theoretical value. Values above
        0.20 indicate significant room for improvement --- either from a
        different allocator, a longer exploration phase, or better guardrail
        calibration.
  \item Navigate to \ui{Decision support} and open the \textbf{Risk
        watchlist}. Look for entries flagged as
        \textit{distance large between observed mode and counterfactual
        optimum}: these identify the specific task types where the allocator
        is systematically diverging from what the oracle would have chosen.
        These are the highest-priority targets for manual policy intervention.
  \item Use \textbf{Local replay} in the Advanced comparison panel to
        re-examine individual subtask decisions for the flagged task types.
        For each replayed decision, the tool shows the reward that would have
        resulted from each alternative mode. If the oracle mode consistently
        outperforms the allocator's choice by a large margin, consider locking
        that task type to the oracle-preferred mode using a policy rule
        (see Recipe~6, step~3).
  \item To verify whether the gap is closing, compare regret across three
        time windows: cycles 1--4 (exploration), cycles 5--8 (convergence),
        and cycles 9+ (exploitation). The regret rate (slope of $R_C$) should
        be highest in the first window and close to zero in the third. If the
        slope in the exploitation window is still non-trivial, increase the
        run length and re-examine.
\end{enumerate}

\screenshot[0.97\linewidth]{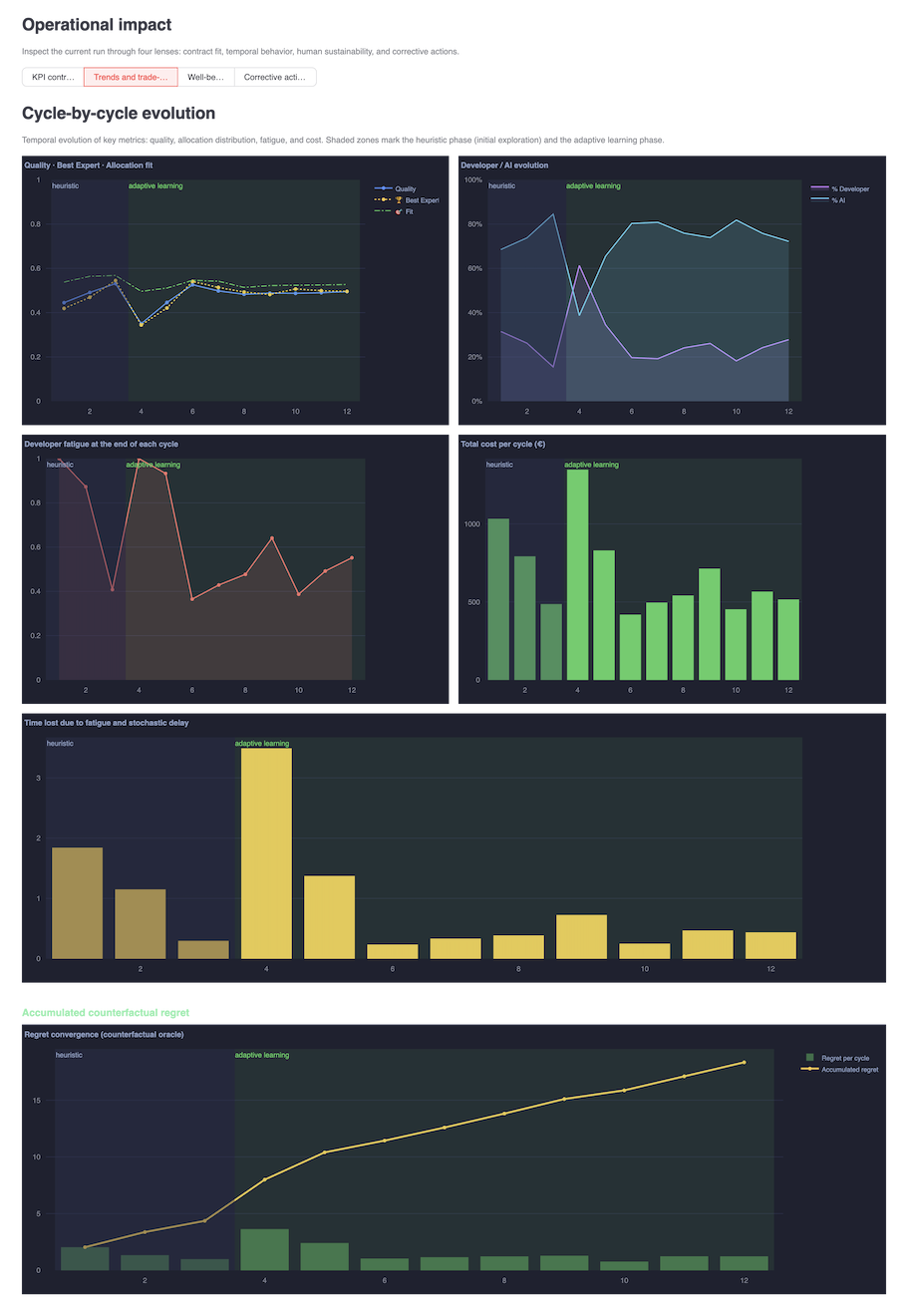}{%
\ui{Operational impact} --- \ui{Trends and trade-offs} subtab,
showing the full cycle-by-cycle evolution panel for a Software /
Standard Sprint run. The \textbf{Regret convergence (counterfactual
oracle)} chart at the bottom plots per-cycle regret (green bars) and
accumulated regret $R_C$ (yellow line). The heuristic warm-up phase
and adaptive learning phase are distinguished by background shading.
The steep initial rise flattens progressively from cycle~4 onward,
indicating allocator convergence. The remaining panels show quality
and AI evolution (top row), developer fatigue and cost per cycle
(middle row), and time lost due to fatigue and stochastic delay
(bottom row).\label{fig:r11a}}

\screenshot[0.97\linewidth]{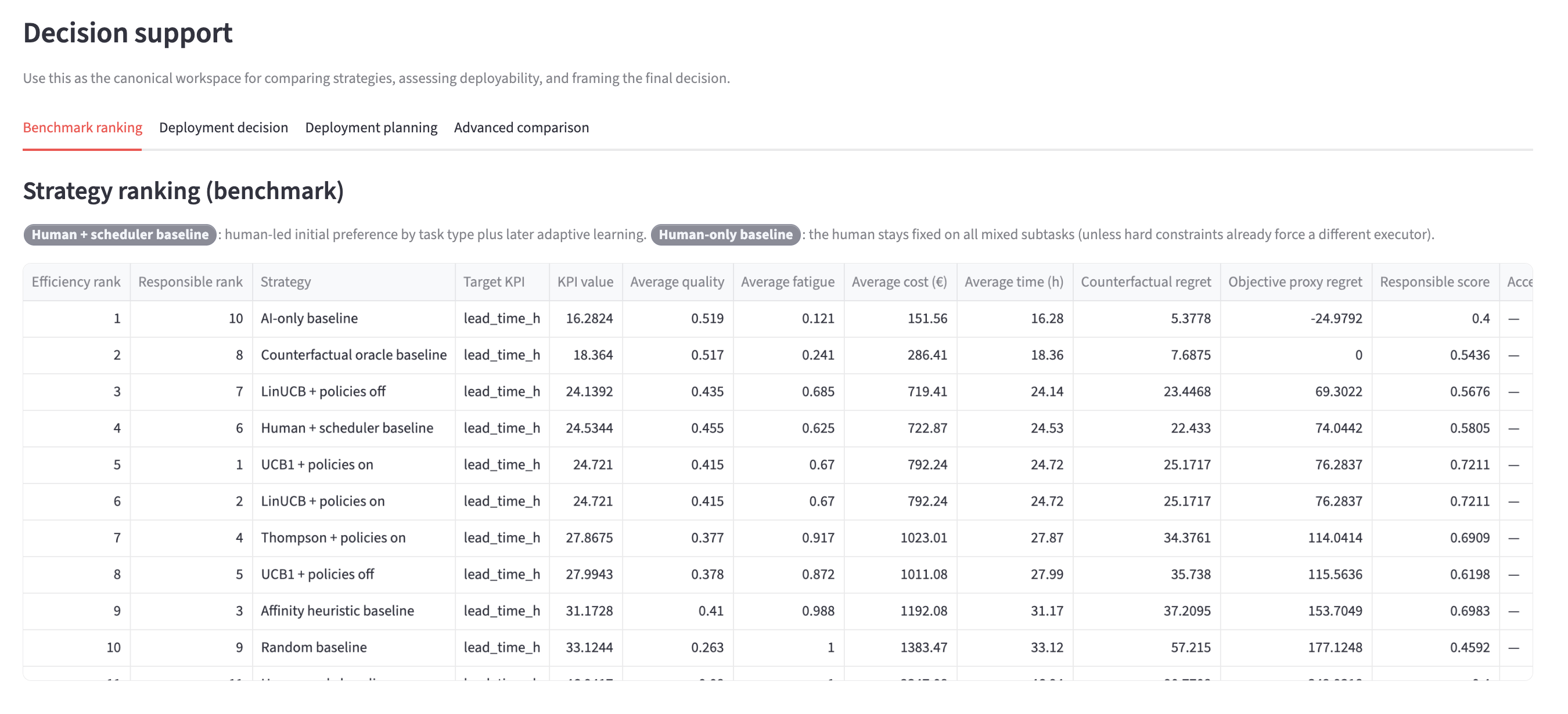}{%
\ui{Decision support} --- \ui{Benchmark ranking} table for the same
run. The \ui{Counterfactual oracle baseline} (efficiency rank~2, KPI
18.364~h lead time, counterfactual regret~0) sets the theoretical
ceiling. The best learned strategy (\ui{AI-only baseline}, rank~1,
KPI 16.021~h) achieves a lower lead time but carries zero responsible
score and fails governance filters. The closest responsible and
deployable option (\ui{LinUCB + policies off}, responsible rank~1) has
a counterfactual regret of~23.4 --- the quantified opportunity cost
of learning under uncertainty and governance
constraints.\label{fig:r11b}}

\begin{infobox}
\textbf{What to look for.} The oracle baseline is a useful diagnostic
ceiling, not a deployment target --- it is unreachable in practice because
it requires knowledge of future outcomes. A learned strategy that achieves
$\bar{R}_C < 0.10$ at cycle~8 is operating near the practical optimum for
the scenario. The most actionable output of this recipe is the watchlist
of task types with large counterfactual divergence: these are candidates
for policy overrides or allocator replacement, not for guardrail adjustment.
\end{infobox}

\section*{Acknowledgements}

The author thanks the colleagues and reviewers who provided
feedback on earlier drafts of this work. Development of HAAS
Studio was carried out within the research activities of the
VRAIN institute at Universitat Polit\`ecnica de Val\`encia.
This work was developed with the support of the Spanish Ministry
of Science and Innovation under Project PRODIGIOUS
PID2023-146224OB-I00. The funders had no role in study design,
data collection and analysis, decision to publish, or preparation
of the manuscript.

\bibliographystyle{plain}
\bibliography{haas_studio}

\end{document}